%% file: Gravity-Edge-Modes-3.tex
\documentclass[11pt,showlabels]{article}

\usepackage{authblk}
\usepackage{amsmath}
\usepackage{amsfonts}
\usepackage{amssymb}
\usepackage{mathrsfs}
\usepackage{slashed}
\usepackage{color}
\usepackage{mathtools}
\usepackage{marvosym}
\usepackage{amssymb}
\usepackage{dsfont}
\usepackage{slashed}
\usepackage{fullpage}
\usepackage{cite}
\usepackage{accents}
\usepackage[colorlinks=true,linkcolor=blue,citecolor=magenta,linktocpage=true]{hyperref}
\usepackage{titlesec}
\titleformat*{\section}{\normalsize\bfseries}
\titleformat*{\subsection}{\normalsize\bfseries}
\titleformat*{\subsubsection}{\normalsize\bfseries}
\usepackage{xcolor}
\usepackage{subcaption}
\DeclareMathAlphabet{\bbvar}{U}{BOONDOX-ds}{m}{n}

\makeatletter
\renewcommand{\@dotsep}{10000}
\makeatother

\input Macro.tex 
\begin{document}

\title{\Large{\textbf{\sffamily Edge modes of gravity - III:\\ Corner simplicity constraints}}}
\author{\sffamily Laurent Freidel$^1$, Marc Geiller$^2$, Daniele Pranzetti$^{1,3}$}
\date{\small{\textit{
$^1$Perimeter Institute for Theoretical Physics,\\ 31 Caroline Street North, Waterloo, Ontario, Canada N2L 2Y5\\
$^2$Univ Lyon, ENS de Lyon, Univ Claude Bernard Lyon 1,\\ CNRS, Laboratoire de Physique, UMR 5672, F-69342 Lyon, France\\
$^3$Dipartimento di Fisica, Universit\`a degli Studi di Udine,
via delle Scienze, 208, I-33100 Udine, Italy\\
}}}

\maketitle

\begin{abstract}
In the tetrad formulation of gravity, the so-called \textit{simplicity constraints} play a central role. They appear in the Hamiltonian analysis of the theory, and in the Lagrangian path integral when constructing the gravity partition function from topological BF theory. We develop here a systematic analysis of the corner symplectic structure encoding the symmetry algebra of gravity, and perform a thorough analysis of the simplicity constraints. Starting from a precursor phase space with Poincar\'e and Heisenberg symmetry, we obtain the corner phase space of BF theory by imposing \textit{kinematical constraints}. This amounts to fixing the Heisenberg frame with a choice of position and spin operators. The \textit{simplicity constraints} then further reduce the Poincar\'e symmetry of the BF phase space to a Lorentz subalgebra. This picture provides a particle-like description of (quantum) geometry: The internal normal plays the role of the four-momentum, the Barbero--Immirzi parameter that of the mass, the flux that of a relativistic position, and the frame that of a spin harmonic oscillator. Moreover, we show that the corner area element corresponds to the Poincar\'e spin Casimir. We achieve this central result by properly splitting, in the continuum, the corner simplicity constraints into first and second class parts. We construct the complete set of Dirac observables, which includes the generators of the local $\mathfrak{sl}(2,\mathbb{C})$ subalgebra of Poincar\'e, and the components of the tangential corner metric satisfying an $\mathfrak{sl}(2,\mathbb{R})$ algebra. We then present a preliminary analysis of the covariant and continuous irreducible representations of the infinite-dimensional corner algebra. Moreover, as an alternative path to quantization, we also introduce a regularization of the corner algebra and interpret this discrete setting in terms of an extended notion of twisted geometries.
\end{abstract}

\thispagestyle{empty}
\newpage
\setcounter{page}{1}

\hrule
\tableofcontents
\vspace{0.7cm}
\hrule


\newpage

\section{Introduction}

We have recently proposed in \cite{Edge-Mode-I,Edge-Mode-II} a new local holographic perspective on quantum gravity. The aim is to study the notion of corner symmetry algebra in gravity, with the expectation that understanding its representation theory will reveal universal features of quantum gravity \cite{Donnelly:2016auv}. The associated symmetry group at the corner surface $S$ has the semi-direct product structure $\text{Diff}(S)\ltimes G^S$, where $G$ is a Lie group which depends on the formulation of gravity under consideration, and $G^S$ denotes the set of maps $S\to G$. This result can be understood by systematically decomposing the symplectic structure of various formulations of gravity into a universal bulk piece, parametrized by the canonical ADM pair, and a corner contribution whose explicit form depends on the formulation being studied. In the tetrad formulation of gravity with non-vanishing Barbero--Immirzi parameter, the corner symplectic structure contains the internal normal to the foliation $\n^I$ as a dynamical variable as well as the corner coframe field, and in this case the symmetry group contains a factor $G=\SL(2,\C)\times\SL(2,\R)_\para$. We have shown in \cite{Edge-Mode-II} that the corner $\sll(2,\R)_\para$ algebra, which is generated by the tangential corner metric components, is at the origin of the discreteness of the corner area element. This illustrates the non-trivial semi-classical physical information encoded in the corner symplectic structure.

Here we continue our analysis of tetrad gravity by focusing on the so-called \textit{simplicity constraints}. These constraints appear in the Hamiltonian analysis of the Einstein--Cartan--Holst action \cite{Peld_n_1994,BarroseSa:2000vx}, and also play a central role in the construction of the spin foam regularizations of the gravitational path integral \cite{Alexandrov:2011ab,Perez:2012wv}. In spin foam models, one writes gravity as a topological BF theory supplemented by the simplicity constraints, ensuring that the $B$ field can be written as the wedge product of frame fields. The challenge is then to consistently implement these constraints in the quantum theory. This is a notoriously subtle issue since, under the spin foam quantization map which assigns  Lie algebra elements to the discrete $B$ field (which corresponds to integrals of the $B$ field along 2-dimensional surfaces and replaces the continuous Poisson-commuting \footnote{In the following we may sometimes omit to specify whether the commutation relations are at the level of the classical phase space, that is in terms of Poisson brackets, or at the level of a Hilbert space, that is in terms of commutator brackets, as this should be clear from the different notation used, namely $\{\cdot,\cdot\}$ for the former and $[\cdot,\cdot]$ for the latter. In section \ref{sec:discretization} where more brackets are introduced we provide a list of notation.} bulk $B$ field of the classical phase space
\footnote{In the standard spin foam treatment the smearing on a codimension-2 discrete surface is introduced as a proxy for the quantization of the continuous bulk $B$ field, although one never introduces a continuous classical boundary phase space explicitly. We have pointed out in detail in \cite{Edge-Mode-II} the inconsistency, both at the classical and quantum levels, of naively identifying corner variables with the pullback of bulk fields (see in particular Section 7 there); solving this inconsistency represented the main reason to introduce edge modes variables.}), the simplicity constraints become non-Poisson-commuting \textit{with themselves}. More precisely, the discrete $B$ field is assumed to generate a Lorentz algebra and the simplicity constraints appear as a proportionality between the boost and rotation generators \cite{Engle:2007qf,Freidel:2007py,Engle:2007wy}. One of the central issues of this approach is the fact that these constraints break the internal Lorentz symmetry down to the rotation subgroup.

In our framework, the non-commutativity of $B$ field is naturally implemented in the continuum, without the need for any discretization. This is done by shifting the viewpoint from the bulk symplectic structure to the corner one, which allows us to perform a rigorous treatment of the simplicity constraints. Furthermore, this reveals a new type of geometrical structure related to a particular parametrization of the Poincar\'e algebra. In particular, we show that the internal normal is part of the phase space and it becomes {\it dynamical}, in the sense that it has nontrivial Poisson brackets with other phase space variables. At the quantum level this means that it is promoted to a quantum operator with new quantum numbers. This crucial ingredient, which was missed or not fully exploited in  previous analyses, allows us to restore Lorentz symmetry even after imposing the simplicity constraints.

More precisely, we show that the BF corner phase space can be obtained by imposing \textit{kinematical constraints} on a larger (precursor) phase space exhibiting Poincar\'e--Heisenberg symmetry. These kinematical constraints correspond to a fixing of the Heisenberg frame and a choice of position and spin operators. We then explain how the \textit{simplicity constraints} reduce the Poincar\'e symmetry of the BF phase space to a Lorentz subalgebra. This gives rise to a particle-like description of (quantum) geometry, where the internal normal plays the role of the 4-momentum, the Barbero--Immirzi parameter that of the mass, the flux that of a relativistic position, and the frame that of a spin harmonic oscillator. Importantly, the corner area corresponds to the Poincar\'e spin Casimir. This means that, using the internal normal $\n^I$ and the Lorentz generator $\JJ_{IJ}$, one can construct a relativistic spin generator $\S^I$ which satisfies a relativistic invariant $\su(2)$ algebra. Its Casimir is proportional to the area element. In other words we have 
\be
\S^I =\frac12 \epsilon^{IJKL} \JJ_{JK}\n_L,
\qquad 
\S^I \S_I = \beta^2 q,
\ee
where $\beta^{-1}$ is the Barbero--Immirzi parameter and $q$ the determinant of the metric on $S$. The relativistic spin $\S^I$ is the gravitational analog of the Pauli--Lubanski vector. The relation between the area element and the relativistic spin gives us another proof that the area spectrum is quantized. Moreover since $\S^2$ is a relativistic invariant this also reconciles the area discreteness with internal Lorentz symmetry invariance.

One of our main results is to give an explicit split, in the continuum, of the corner simplicity constraints into first and second class parts, and then identify a complete set of Dirac observables. They are given by the generators of the local $\mathfrak{sl}(2,\mathbb{C})$ subalgebra of Poincar\'e, and the components of the tangential corner metric satisfying an $\mathfrak{sl}(2,\mathbb{R})$ algebra. Another central result  is to show that all the Casimirs involved in this construction are related to the corner area element. More precisely, we have that
\be
C_{\SL(2,\RR)}=-\beta^2q,\q C_{\SU(2)}=\beta^2q,\q C_{\SL(2,\mathbb{C})}^{(1)}=(\beta^2-1)q,\q C_{\SL(2,\mathbb{C})}^{(2)}=-2\beta q,
\ee
where $\beta$ is the (inverse) Barbero--Immirzi parameter, $q$ the determinant of the corner metric and $\SU(2)$ refers to the Poincar\'e spin subgroup generated by $\S^I$. This provides another important example of the quantum algebraic  information encoded in the corner symmetry algebra. It suggests that gravity (expressed here in terms of tetrads and using the simplicity constraints) picks out the representations of $\text{Diff(S)}\ltimes G^S$ which satisfy these balance equations. This also gives a new  perspective on one of the key insights of LQG: It shows that the Barbero--Immirzi parameter provides a mass gap from the Poincar\'e perspective, which is really an area gap from the perspective of quantum geometry. This gap is regularizing these representations.

The plan of the paper is the following. In Section \ref{sec:BF} we introduce some key ingredients of the BF formulation of gravity, and recall the main result of \cite{Edge-Mode-II} about the bulk and corner decomposition of the BF symplectic potential. In Section \ref{sec:Bound-PS} we use this result to extend the  phase space by adding a set of edge modes living at the corner of the space-like hypersurface, which allows us to restore internal Lorentz gauge invariance. We then introduce the precursor Poincar\'e--Heisenberg phase space, together with a set of kinematical constraints which reduce this phase space to that of BF theory. This is done by rewriting the edge modes in terms of Dirac observables with respect to these kinematical constraints. They correspond to the internal normal, the boost generator, and the tangential coframe. The latter can be repackaged as a spin generator, a 2-dimensional tangential metric $q_{ab}$, satisfying an $\sll(2,\RR)$ algebra, and an angle $\theta$ which turns out to be conjugate to the area element. This angle plays an important role in the reconstruction of twisted geometries. The boost and spin generators define a set of Lorentz generators $\JJ_{IJ}$, which together with the internal normal form an elemental corner Poincar\'e algebra. This is one of the most important results of the paper. Next, we introduce the corner  simplicity constraints and show that they form a second class system already at the classical and continuum level.

The study of this algebra of simplicity constraints is the main topic of Section \ref{sec:simp-algebrs}. There, we identify and separate the second class part of the simplicity constraints from the first class component. We rewrite the second class pair in terms of holomorphic and anti-holomorphic constraints. This allows us to clarify a confusion often met in the spin foam literature about imposition of second class constraints \`a la  Gupta--Bleuler. In particular, we show how a \textit{strong} imposition of the holomorphic component is equivalent to the minimization of the associated master constraint in the quantum theory, by verifying a consistency condition which is usually only implicitly assumed. With this analysis at hand, we can identify 9 corner  Dirac observables given by the Lorentz generators and the tangential metric components. Of these, only 7 are the independent observables, and the missing geometrical information is encoded in the angle $\theta$. In this way we recover the 8 corner physical degrees of freedom forming the set 
$(\JJ_{IJ},q_{ab},\theta)$, and we establish that the corner symmetry group is given by $\SL(2,\mathbb{C})^S \times \SL(2,\mathbb{R})^S_\para \times \text{U}(1)^S_\para$.

In Section \ref{sec:symbreaking} we analyze in more details the structure of the Poincar\'e algebra we have discovered. We explain how this allows us to reconcile internal Lorentz invariance with the imposition of the simplicity constraints and the discreteness of the area spectrum. We also provide an understanding of the standard LQG picture in the time gauge from the point of view of this more general covariant framework.

Section \ref{sec:discretization} sets the stage for the quantization of the corner symmetry algebra, which will be developed further in subsequent papers of the series. We show how we can consider smooth representations of the corner symmetry algebra labelled by a choice of measure on the sphere, and give one example of such a representation. We also show how, by using piecewise smearing functions, we can recover a regularized discrete subalgebra  that bears resemblance with the algebra studied in LQG. We  provide a prescription to regularize the Casimir operators, as well as the tangential metric operators. With this structure at hand, we then elucidate the interpretation of the new corner geometrical data in terms of a generalization of the notion of twisted geometry. This also gives us the chance to clarify further the geometrical origin of the edge modes through their relation to the bulk variables, and explain in particular their relationship with holonomies.

A final discussion is presented in Section \ref{sec:conclusions}. The three appendices \ref{App:Brackets}, \ref{App:simplicity-sec}, and \ref{App:Poinc}, contain the explicit derivations of many Poisson brackets used in the main text.

\section{Preliminaries}\la{sec:BF}

We start with a brief review of the BF formulation of gravity, and recall the decomposition of its symplectic potential obtained in \cite{Edge-Mode-II}. This decomposition of the potential is the starting point of our analysis.

\subsection{Conventions}

We consider a spacetime $M$ and introduce at each point a coframe field $e^I=e_\mu{}^I\rd x^\mu$, with inverse $\hat{e}_I=e_I{}^\mu\pa_\mu$, such that the spacetime metric is given by $g_{\mu\nu}=e_\mu{}^Ie_\nu{}^J\eta_{IJ}$, with $\eta_{IJ}=\rm{diag}(-,+,+,+)$ the internal Lorentzian metric. We consider a foliation in terms of codimension-1 space-like slices $\Sigma$, such that $M=\Sigma\times\mathbb{R}$, with unit normal form $\un=n_\mu\rd x^\mu$ satisfying $g^{\mu\nu}n_\mu n_\nu=-1$. We denote by $\hat{n}=n^\mu\pa_\mu$ the outward pointing normal vector, by $\epsilon=\sqrt{|g|}\rd^4 x$ the volume form on $M$, and $\teps=-\sqrt{|\tilde{g}|}\rd^3x$ such that $\epsilon=\un\wedge\teps$ is the induced volume form on $\Sigma$. In what follows the tilde will always denote quantities pulled back to $\Sigma$. We also introduce an internal normal $n^I$ such that
\be\label{form}
\un=e^In_I,\q\hat{n}\ip e^I=n^I,\q n^2=-1,
\ee
where we used the interior product notation $\hat{n}\ip\alpha=n^\mu\alpha_\mu$ and $n^2=n^In^J\eta_{IJ}$. The normal 1-form and the internal normal allow us to introduce the tangential coframe field
\be
\te_\mu{}^I\coloneqq e_\mu{}^I+n_\mu n^I,
\ee
which is both tangential and horizontal in the sense that $\te^I n_I=0$ and $\hat{n}\ip\te=0$. The induced metric on $\Sigma$ is then given by
\be
\tilde{g}_{\mu\nu}\coloneqq\te_\mu{}^I\te_\nu{}^J\eta_{IJ}=g_{\mu\nu}+n_\mu n_\nu.
\ee
We define the duality map $\as$ acting on Lie algebra-valued functionals as
\be
\as M_{IJ}=\f{1}{2}{\eps_{IJ}}^{KL}M_{KL},\q\as^2=-1,
\ee
and the cross-product $(M\times N)_I\coloneqq\tilde\eps_{IJK}M^J\wedge N^K$, where $\teps_{IJK}\coloneqq\epsilon_{IJKL}n^L$. Finally, for two vector-valued objects $M^I$ and $N^I$ we will denote $M\cdot N=M^IN^J\eta_{IJ}$. Other useful relations are found in Appendix A of our companion paper \cite{Edge-Mode-II}.

Since we are interested in the corner symmetries, which are independent of any boundary conditions one may specify on a time-like or null hypersurface in $M$, we consider the case where all the space-like hypersurfaces $\Sigma$ meet at an entangling 2-sphere $S$, which we call the corner.

\subsection{BF formulation of gravity}

Topological BF theory is defined by the bulk action
\be\label{bulk action}
S_\BF=\int_MB_{IJ}\wedge F^{IJ},
\ee
where $F^{IJ}=\rd\omega^{IJ}+\omega^I{}_K\wedge\omega^{KJ}$ is the curvature of the Lorentz connection 1-form $\omega^{IJ}$, and $B^{IJ}$ is a Lie algebra-valued 2-form. The BF pre-symplectic potential is
\be\la{Theta-BF}
\Theta_\BF\coloneqq\int_\Sigma B_{IJ}\wedge\delta\omega^{IJ}.
\ee
As  shown in \cite{Edge-Mode-II}, using the internal normal one can decompose the pull-back of $B_{IJ}$ to $\Sigma$ in terms of a boost 2-form $\tB^I$ and a spin 2-form $\tS^I$, according to
\be\label{B on Sigma}
B^{IJ}\stackrel{\Sigma}{=}-2\tB^{[I}n^{J]}+\teps^{IJ}{}_K\tS^K.
\ee
The boost and spin 2-forms are both tangential, i.e. $\tB^In_I=0=\tS^I n_I$. They can respectively be expressed as the cross product of a \textit{boost frame} $\tb^I$ and a \textit{spin frame} $\ts^I$ as
\be
\tB^I=\f{1}{2}(\tb\times\tb)^I\,,\q \tS^I=\f{1}{2}(\ts\times\ts)^I.
\ee
As explained in \cite{Edge-Mode-II}, this rewriting amounts to trading the 9 components of $\tB^I$ for the 9 components of $\tb^I$, and similarly for $\tS^I$ and $\ts^I$. Both frames are horizontal and tangential 1-forms.\footnote{A vector-valued form $\alpha^I$ is called horizontal if $\hat{n}\ip\alpha^I=0$, and tangential if $\alpha^In_I=0$.} We can also decompose the Lorentz connection as
\be\la{con-dec}
\omega^{IJ}\stackrel{\Sigma}{=}\tilde{\Gamma}^{IJ}-2\tK^{[I}n^{J]},
\ee
where $\tK^I$ denotes the horizontal component of $K^I\coloneqq\rd_\omega n^I$.

Using these decompositions of the $B$ field and the connection, one of the main results of \cite{Edge-Mode-II} was to rewrite the BF symplectic potential as a canonical bulk component plus a corner term parametrized by the corner canonical pairs $(\tB_I,n^I)$ and $(\ts_I,\ts^I)$. More precisely, we have $\Theta_\BF=\Theta_\BF^\Sigma+\Theta_\BF^S$, where the bulk component is given on-shell of the Gauss constraint by
\be
\Theta_\BF^\Sigma\simeq-\int_\Sigma\left(\tB_I\wedge\delta\tK^I-\tilde{\rd}_\Gamma\ts_I\wedge\delta\ts^I\right)-\delta\left(\f{1}{2}\int_\Sigma\ts_I\wedge\tilde{\rd}_\Gamma\ts^I\right),
\ee
and the corner component is
\be\label{symptom}
\Theta_\BF^S=\int_S\left(\tB_I\delta n^I-\f{1}{2}\ts_I\wedge\delta\ts^I\right).
\ee
This corner potential is the central object of study of this paper.

In order to go from BF theory to gravity we need to impose the \textit{simplicity constraints}. At the level of the action, they read
\be
B^{IJ}=(*+\beta)(e\wedge e)^{IJ},
\ee
and turn \eqref{bulk action} into the Einstein--Cartan--Holst (ECH) action. At the level of the decomposition in terms of boost and spin frames, they read
\be\la{simp-frames}
\tB^I=\tE^I=\f{1}{2}(\te\times\te)^I,\q\ts^I=\sqrt{\beta}\te^I.
\ee
This turns the BF pre-symplectic potential into the potential of ECH gravity \cite{Edge-Mode-II}. After the imposition of the simplicity constraints, the Gauss law in the bulk of the slice becomes
\be
\tilde{\rd}_\Gamma\te^I\simeq0,\q\tK^I\wedge\te_I\simeq0.
\ee
As explained in \cite{Edge-Mode-II} as well, the names of the BF frames is actually not relevant. What matters is that topological BF theory has \textit{two} frames while ECH gravity only has \textit{one}. In other words, the simplicity constraints identify the two frames of BF theory with the gravitational frame of ECH gravity. We can therefore choose for convenience a ``notational gauge'' in which we rename $(\tb^I,\ts^I)\to(\tb^I,\sqrt{\beta}\,\te^I)$. This is what we adopt below.

\section{Corner phase space}
\la{sec:Bound-PS}

When imposing the simplicity constraints, the bulk potential $\Theta_\BF^\Sigma$ becomes the gravitational bulk potential $\Theta_\ECH^\Sigma$, which as we have shown in \cite{Edge-Mode-II} coincides with the universal potential $\Theta_\GR$ of canonical gravity. Different formulations of gravity only differ by the form of the corner potential. Here we focus on that of BF theory, namely $\Theta_\BF^S$, which is the precursor before the simplicity constraints of the ECH corner potential $\Theta_\ECH^S$.

As explained in \cite{Edge-Mode-II}, the edge mode formalism amounts to extending the corner phase space by introducing new corner fields, which are a priori independent from the pull-back of the bulk fields to the corner. These corner fields, or edge modes, are introduced via a non-trivial potential $\bTh_\BF^S$. The goal of this paper is to study in great details the corresponding corner symplectic structure. With this edge mode potential we define the extended potential
\be
\Theta_\BF^\ext\coloneqq\Theta_\BF-\bTh_\BF^S,
\ee
with
\be\label{BF potential}
\bTh_\BF^S=\int_S\left(\BB_I\bd\n^I-\f{\beta}{2}\e_I\wedge\bd\e^I\right),
\ee
where $\bd\coloneqq\delta-\varphi^{-1}\delta\varphi$ is a horizontal variational derivative which depends on the edge mode field $\varphi$. This latter is a group element whose role is to ensure proper gluing of the bulk and corner fields. In fact, the extended potential is defined in such a way that if we set $\varphi=1$ and impose the strong (and naive) gluing condition $B_I\stackrel{S}{=}\BB_I$ and $e^I\stackrel{S}{=}\e^I$ between the bulk and corner fields, we get $\bTh_\BF^S=\Theta_\BF^S$, and therefore $\Theta_\BF^\ext$ reduces to the bulk piece $\Theta_\BF^\Sigma$.

This construction using the extended potential is such that gauge invariance is restored on the phase space. Moreover, it enables us to express gauge invariance at the corner as a continuity condition relating the pull-back of the bulk fields to the dressed edges modes. We refer the reader to Section 6.1 in \cite{Edge-Mode-II} and Section \ref{sec:twisted} below for more details. In short, we have that
\be\la{continuity}
B^I\stackrel{S}{=}\vphi^{I}{}_J\BB^J,\q
n^I\stackrel{S}{=}\vphi^{I}{}_J\n^J,\q
e_a{}^I\stackrel{S}{=}\vphi^{I}{}_J\,\e_b{}^J\,\rho^b{}_a,
\ee
where $(\vphi,\rho)$ is an element of the corner symmetry group, with $\vphi\in\SL(2,\C)^S$ and $\rho\in\SL(2,\R)^S$. These continuity equations are first class constraints which commute with the symmetry generators. They guarantee that the extended phase space preserves gauge invariance, i.e. that the canonical generator of gauge transformations is vanishing on-shell even if the gauge transformation is non-trivial at the corner. Restoration of gauge invariance still allows for non-vanishing corner symmetry charges. These are the charges of transformations rotating only the edge modes \cite{Edge-Mode-II,Donnelly:2016auv}.

In summary, although BF theory with the simplicity constraints is equivalent in the bulk to metric gravity, it has additional corner charges which give rise to a non-trivial representation of the corner local Lorentz algebra. This is detailed in \cite{Edge-Mode-II}. Here we will momentarily set aside the group elements by setting $\varphi=1$, and focus on $(\BB^I,\n^I,\e^I)$ in order to study the simplicity constraints. Notice that since these fields satisfy the 4 relations\footnote{We use $a,b,\dots$ to label indices $x=(x^1,x^2)$ on the corner surface $S$.}
\be\label{kinematical BF relations}
\BB^I\n_I=0=\e_a{}^I\n_I,\q\n^2=\n^I\n_I=-1,
\ee
the corner BF phase space parametrized by $(\BB^I,\n^I,\e^I)$ is 12-dimensional. In the following section we start by showing that the corner BF potential naturally descends from a corner potential with Poincar\'e and Heisenberg symmetry.

\subsection{Poincar\'e--Heisenberg corner symplectic structure}

The extended phase space of Einstein--Cartan--Holst gravity is obtained form the BF one after imposition of the bulk and corner simplicity constraints. It turns out that the corner simplicity constraints are a mixture of first and second class constraints, and as such need to be studied very carefully. This is the main goal of this paper. In order to achieve this, it will prove very convenient to introduce another corner phase space, called the Poincar\'e--Heisenberg\footnote{The rationale behind this name will become clear in Section \ref{sec:heisenberg} where we study the Heisenberg symmetry.} phase space, from which the BF corner phase space can be obtained after imposition of \emph{kinematical constraints}. These kinematical constraints ensure that we have \eqref{kinematical BF relations}, and like the simplicity constraints they contain both first and second class parts. Starting from the Poincar\'e--Heisenberg corner phase space, the gravitational one is obtained after imposing both the kinematical and simplicity constraints.

The Poincar\'e--Heisenberg corner phase space is parametrized by a vector-valued 2-form $\X^I$ on $S$, a vector-valued 1-form $\se^I=\se_a{}^I\rd x^a$ on $S$, and a vector-valued scalar $\n^I$ on $S$. These $4+4+8$ variables\footnote{The 1-form has a 2-dimensional index $a$ on $S$, and the internal index $I=0,1,2,3$ is for the moment not restricted.} $(\X^I,\n^I,\se^I)$ define a 16-dimensional phase space with symplectic potential given by
\be\label{poincare theta}
\bTh_\Poin^S=\int_S\left(\X_I\delta\n^I-\f{\beta}{2}\se_I\wedge\delta\se^I\right).
\ee
Assuming that all the variables $(\X^I,\n^I,\se^I)$ are independent, we have two canonical pairs $(\X_I,\n^I)$ and  $(\se_1{}^I,\se_2{}^I)$ on the corner, and the Poisson brackets are\footnote{Note that in order to write these brackets we should first convert the potential from forms to densities using $\X^I=\undertilde{\X}^I\rd^2x$ and $\se_I\wedge\delta\se^I=\se_a{}^I\delta\se_b{}^I\eta_{IJ}\eps^{ab}\rd^2x$. In what follows we allow ourselves an obvious and innocent abuse of notation and do not write the density explicitly. With this notation, one should simply recall that all objects appearing in brackets are densities.}
\be\la{basic poincare brackets}
\{\X^I(x),\n^J(y)\}=\eta^{IJ}\delta^2(x,y),
\q
\{\se_a{}^I(x),\se_b{}^J(y)\}=-\f{1}{\beta}\epsilon_{ab}\eta^{IJ}\delta^2(x,y),
\ee
where $\delta^2(x,y)$ is the density Dirac delta function on $S$. The gravitational phase space is obtained from this 16-dimensional phase space after imposing respectively the \textit{kinematical} and \textit{simplicity constraints}. We now study in details these two sets of constraints.

As we are about to see, an important role is played by the internal normal $n^I$, which becomes a field on phase space with non-commuting Poisson brackets. This normal has appeared previously in the literature, for example when writing down covariant first order boundary terms \cite{0264-9381-4-5-011,bianchi2012horizon,Bodendorfer:2013hla}, in extensions of LQG beyond the time gauge \cite{Alexandrov:2002br,Alexandrov:2007pq,Alexandrov:2008da,Alexandrov:2011ab,Gielen:2010cu}, and also in group field theory, where it serves as an extra kinematical structure to impose the simplicity constraints \cite{Baratin:2010wi,Baratin:2011tx}. In studies of LQG beyond the time gauge, where the internal normal plays an important role, the normal was however only considered from the point of view of the bulk phase space, where it commutes with the $B$ field and the tetrad components. It was only in later prescient work such as \cite{Wieland_2014,Wieland_2017,Wieland:2017zkf,Wieland:2017cmf} and studies of LQG in higher dimensions \cite{Bodendorfer_2013,Bodendorfer:2013jba,Bodendorfer:2013sja}, that the role of the normal in the boundary phase space was recognized, along with its non-commutativity with the tetrad components. Our approach relies heavily on this non-commutativity of the normal with $B$ on the corner, and we will eventually promote the normal to an operator at the quantum level.

\subsection{Kinematical constraints}\la{sec:kin-constraints}

The kinematical constraints\footnote{We use this terminology to differentiate them from the simplicity constraints that turn topological BF theory into gravity, thus introducing local degrees of freedom. The reader should not be misled by the distinction between kinematical and dynamical constraints usually introduced in the canonical analysis of the bulk (see also the discussion at the beginning of Section \ref{sec:simp-algebrs}).} are given by
\be\la{sec-con}
\n^2=\n_I\n^I=-1,\q\q\n_a\coloneqq\se_a{}^I \n_I=0.
\ee
The first kinematical constraint is simply the normalisation condition on $\n$. The second constraint is geometrically more interesting: it means that the pull-back of the form $\underline{\n}$ to $S$ vanishes. As we will see, this condition can be understood as the condition that the normal vector $\n^I$ is at rest with respect to $S$. These kinematical constraints correspond to 1 first class and 2 second class constraints since their algebra is
\be
\{\n_a(x),\n_b(y)\}=\f{1}{\beta}\epsilon_{ab}\delta^2(x,y),\q\q\{\n^2(x),\n_a(y)\}=0.
\ee
The  complete set of  Dirac observables that commute with these constraints are parametrizing 12 degrees of freedom (dof). These are given by the normal (3 dof), the boost operator (3 dof), and the tangential frame (6 dof), respectively 
\be\la{tse}
\n^I,\q\q\BB^I\coloneqq\X^I+\n^I(\X\cdot\n)+\beta(\se_a{}^I\n_b\epsilon^{ab}),\q\q\e_a{}^I\coloneqq\se_a{}^I+\n_a\n^I.
\ee
We have that $\BB^I $ and $\e_a{}^I$ are tangential observables satisfying $\BB^I\n_I=0=\e_a{}^I\n_I$, with $\n^2=-1$. We show explicitly in Appendix \ref{App-Boost} that these observables commute strongly\footnote{For first class constraints $C_I=0$ it is enough that Dirac observables $\CO$ commute weakly with the constraints, i.e. $\{C_I,\CO\}=\CO_I{}^J C_J$. However, for second class constraints it is required that Dirac observables commute strongly with the constraints, i.e. $\{C_I,\CO\}=0$.} with the kinematical constraints \eqref{sec-con}. The commutators of the boost operator and the tangential frame operator are 
\be\la{Eeee}
\{\BB^I(x),\e_a{}^J(y)\}=\e_a{}^I\n^J(x)\delta^2(x,y),\q\q
\{\e_a{}^I(x),\e_b{}^J(y)\}=-\f{1}{\beta}\epsilon_{ab}\tilde{\eta}^{IJ}\delta^2(x,y),
\ee
where $\tilde{\eta}^{IJ}\coloneqq\eta^{IJ}+\n^I\n^J$ is the tangential internal metric. As expected, these brackets, which are also derived in Appendix \ref{App-Boost}, correspond to the commutation relations of the BF corner potential
\be\la{TS1}
\bTh_\BF = \int_S\left(\BB_I\delta \n^I- \f{\beta}{2}\e_I\wedge\delta\e^I\right),
\ee
thereby showing that the kinematical constraints indeed reduce the Poincar\'e corner phase space to the BF one.

The information about the tangential coframe $\e_a{}^I$  can be conveniently rewritten in terms of a \textit{spin} operator (3 dof), a Lorentz-invariant 2-dimensional tangential metric (3 dof), and an angle $\theta$. The spin operator and the tangential metric are defined as
\be\la{Sq-def}
\S_I\coloneqq\f{\beta}{2}\epsilon_{IJKL}(\e\wedge\e)^{JK}\n^L,\q\q \qq_{ab}\coloneqq\e_a{}^I\e_b{}^J\eta_{IJ},
\ee
 The reason why we need an additional angle is because $\S_I$ and $q_{ab}$  are not independent: They satisfy the geometrical balance relation\footnote{
An explicit derivation is given by
\be
\S^2
&=\frac{\beta^2}{4} \teps_{IAB}  \teps^I{}_{JK}  \e_a{}^A\e_b{}^B  \e_{c}{}^J\e_{d}{}^K \eps^{ab} \eps^{cd}\cr
&=\frac{\beta^2}{4} (\eta_{AJ} \eta_{BK} -\eta_{AK}\eta_{BJ})  \e_a{}^A\e_b{}^B  \e_{c}{}^Je_{d}{}^K \eps^{ab} \eps^{cd}\cr
&=\frac{\beta^2}{2} (q_{ac}q_{bd}-q_{ad}q_{bc}) \eps^{ab} \eps^{cd}\cr
&=\beta^2q.\nn
\ee}
\be\label{Sdet}
\S^2=\beta^2 \qq ,
\ee
which relates the Poincar\'e spin Casimir to the determinant of the corner metric $\qq\coloneqq\det(\qq_{ab})$. As we are about to see, $ \qq $ is the Casimir of an $\sll(2,\mathbb{R})^S_\para$ algebra\footnote{We denote by $G^S$ the set of maps $S\to G$, which also forms a group. It is a 2-dimensional generalization of the loop algebra. Its infinitesimal generators form an ultra-local algebra, where ultra-local refers to the fact that the commutation relations only involves $\delta$ distributions and no derivatives of $\delta$ distributions.} \eqref{sl2R} commuting with the Poincar\'e generators. The geometrical balance equation therefore identifies two Casimirs, and it means that the spin of the elemental Poincar\'e algebra is given by the area element times $\beta$. One can check that $\S^I$, $q_{ab}$ and $\epsilon_{ab}$ are left invariant by the frame rotation
\be\la{framerot}
\e_a{}^I\mapsto\e_a{}^I(\theta)\coloneqq\cos\theta\,\e_a{}^I+\sin\theta\,\star\!\e_a{}^I,\q\q\star\e_a{}^I\coloneqq\f{q_{ab}\epsilon^{bc}\e_c{}^I}{\sqrt{ q }}\,,
\ee
with $\epsilon_a{}^b\coloneqq\epsilon_{ac}q^{cb}$. Here we have introduced a 2-dimensional notion of Hodge duality\footnote{This should not be confused with the internal 4-dimensional duality $\as$.} which defines a complex structure on $S$ since $\star^2=-1$. This duality is such that $\star\e_a{}^I\e_b{}^J\eta_{IJ}=\sqrt{ q }\epsilon_{ab}$ is the area form. The angle $\theta$, which represents the information in $\e_a{}^I$ not captured by $(\S_I,q_{ab})$, is conjugated to the area element as
\be\la{qtheta}
\beta\{\sqrt{ q }\,(x),\theta(y)\}=\delta^2(x,y).
\ee 
This can be seen by applying the shift $\e_a{}^I\mapsto\e_a{}^I(\theta)$ in the potential. Equivalently, one can just shift the variation of the frame as $\delta\e_a{}^I\mapsto\delta\e_a{}^I+\delta\theta\star\e_a{}^I$, and the symplectic potential becomes
\be
\bTh_\text{BF}^S\mapsto\bTh_\text{BF}^S+\beta\int_S\sqrt{ q }\,\delta\theta\,\rd^2x.
\ee
The commutation relation \eqref{qtheta} shows that $\theta$ is the continuum analog of the twist angle entering the definition of twisted geometries \cite{Freidel:2010aq,Freidel:2010bw}. We come back to this point in Section \ref{sec:twisted}.

The variables $(\S_I,q_{ab},\theta)$ encode the same information as the tangential frame $\e_a{}^I$. It is enlightening to investigate a little more the relationship between these quantities. The details are given in Appendix \ref{App-Spin}. The spin density $\S_I$ generates an $\su(2)^S$ ultra-local Lie algebra which preserves $\n^I$, as can be seen from the brackets
\bea\label{SS}
\{\S_I(x),\S_J(y)\}=-\teps_{IJ}{}^K\S_K(x)\delta^2(x,y),\q\q
\{\S_I(x),\n_J(y)\}=0\,,
\eea
where we denote $\teps_{IJK}\coloneqq\eps_{IJKL}\n^L$.
In the time gauge, where $\n^I=(1,0,0,0)$, this generator is the celebrated LQG flux generator which gets quantized in terms of spin network states \cite{Rovelli:1994ge, Ashtekar:1996eg}. Its commutation relation with the frame is given by
\be
\{\S^I(x),\e_{a}{}^J (y)\}=-\teps^{IJ}{}_K\e_a{}^K(x)\delta^2(x,y).
\ee
It also has the following non-trivial commutator with the boost operator:
\be\label{BS}
\{\S_I(x),\BB_J(y)\}=-\n_I\S_J(x)\delta^2(x,y).
\ee
Finally, it satisfies by definition the relations
\be\label{S1}
\S_I=\f{\beta}{2}(\e\times\e)_I,\q\q\teps_{IJ}{}^K\S_K=\beta(\e\wedge\e)_{IJ}.  
\ee
Focusing now on the boost pair $(\BB_I,\n^J)$, one can show that it satisfies the algebra
\begin{subequations}\label{BB}
\be
\{\BB_I(x),\BB_J(y)\}&=\big(\BB_I\n_J-\BB_J\n_I-\teps_{IJ}{}^K\S_K\big)(x)\delta^2(x,y),\\
\{\BB_I(x),\n^J(y)\}&=(\delta_I{}^J+\n_I\n^J)(x)\delta^2(x,y),\\
\{\n^I(x),\n^J(y)\}&=0.
\ee
\end{subequations}
As mentioned above, the tangential metric $q_{ab}$ generates an $\sll(2,\mathbb{R})^S_\para$ algebra
\be\la{sl2R}
\{\qq_{ab}(x),q_{cd}(y)\}=-\f{1}{\beta}\big(q_{ac}\epsilon_{bd}+q_{bc}\epsilon_{ad}+q_{ad}\epsilon_{bc}+q_{bd}\epsilon_{ac}\big)(x)\delta^2(x,y).
\ee
This corner  algebra $\sll(2,\mathbb{R})^S_\para$ associated with the tangential metric was first revealed in \cite{Freidel:2015gpa} and studied further in \cite{Freidel:2016bxd,Edge-Mode-II}. It adds a crucial element to the corner algebra that had been ignored up to now in most studies of quantum gravity. The metric $\qq_{ab}$ also commutes with the generators $(\BB_I,\S_I,\n_I)$. The tangential frame, however, transforms as a 2-dimensional vector under these generators, namely
\be
\{q_{ab}(x),\e_c{}^I(y)\}=\f{1}{\beta}\big(\epsilon_{ca}\e_b{}^I+\eps_{cb}\e_a{}^I\big)(x)\delta^2(x,y).
\ee
From these relations, we can establish as in \eqref{detqstar} that the area element generates the infinitesimal frame rotation \eqref{framerot}. Indeed, we have
\be\la{qec}
\beta\{\sqrt{ q }\,(x),\e_a{}^I(y)\}=\star\e_a{}^I(x)\delta^2(x,y)=\f{\delta\e_a^I(\theta)(x)}{\delta\theta(y)}. 
 \ee
This is a confirmation that the angle $\theta$ is conjugate to the area element. This result also means that the  Hodge duality transformation on the sphere can be represented as the commutator with the total area, i.e. $\beta\{\text{Ar}(S),\cdot\}=\star$. Using the Jacobi identity, this shows that the Hodge dual is compatible with the Poisson structure, i.e.
\be
\{\star\e_a{}^I(x),\star \e_b{}^J(y)\}=\{\e_a{}^I(x),\e_b{}^J(y)\},\q\q
\{\star\e_a{}^I(x),\e_b{}^J(y)\}=-\{\e_a{}^I(x),\star\e_b{}^J(y)\}.
\ee
Using \eqref{qec} and the fact that $\{\text{Ar}(S),q_{ab}\}=0$, it is immediate to establish that the dual frame field also transforms as a 2-dimensional vector, i.e.
\be
\{q_{ab}(x),\star\e_c{}^I(y)\}=\frac{1}{\beta}\big(\epsilon_{ca}\star\!\e_b{}^I+\eps_{cb}\star\!\e_a{}^I\big)(x)\delta^2(x,y).
\ee
For completeness, we can evaluate as in \eqref{*ee} the bracket of the frame with its dual. This gives
\be\label{*eemain}
\{\star\e_a{}^I(x),\e_b{}^J(y)\}=\f{1}{\beta\sqrt{ q }}\big(q_{ab}\tilde{\eta}^{IJ}-\star\e_{a}{}^J\star\!\e_{b}{}^I-\e_{a}{}^J\e_{b}{}^I\big)(x)\delta^2(x,y) .
\ee
Last, but not least, the balance equation \eqref{Sdet} leads to an alternative expression for the Hodge dual in the form
\be\la{*eS}
\star\e_a{}^I(x)=\beta\{\text{Ar}(S),\e_a{}^I(x)\}=\Big\{\int_S|\S|,\e_a{}^I(x)\Big\}=\frac{1}{|\S|}\big(\tilde{\eps}^{I}{}_{JK}{\S^J}\e_a{}^K\big)(x).
\ee
This identity is proven by a direct calculation in \eqref{*eSapp}.

This closes the study of the parametrization of the corner phase space once the kinematical constraints are imposed, and of the various Poisson bracket relations between these corner variables. We now turn to the study of the corner simplicity constraints.

\subsection{Simplicity constraints}

In addition to the 3 kinematical constraints \eqref{sec-con}, we have the corner \textit{simplicity constraints}, which relate the boost operator  to the spin operator. They read\footnote{This expression of the corner simplicity constraints follows immediately from \eqref{simp-frames} and  the continuity conditions \eqref{continuity}. A notion of corner simplicity constraints was previously exploited in \cite{Freidel:2019ofr}.}
\be\la{simpl-n}
\CC_I\coloneqq\BB_I-\f{1}{\beta}\S_I\stackrel{S}{=}0.
\ee
Since $\n^I\CC_I=0$, these correspond to 3 constraints. As shown in \eqref{CCapp}, these simplicity constraints satisfy the algebra
\be\la{CCbra}
\{\CC_I(x),\CC_J(y)\}=\left(\CC_I\n_J-\CC_J\n_I-\left(1+\f{1}{\beta^2}\right)\teps_{IJ}{}^K\S_K\right)(x)\delta^2(x,y).
\ee
We see that this algebra is first class when $\beta^2=-1$. This is expected since this choice corresponds to the self-dual formulation of gravity. In the next section we will study these simplicity constraints in great details, and show that when $\beta$ is real they can be split between 2 second class constraints and 1 first class constraint, in perfect analogy with the kinematical constraints \eqref{sec-con}. Therefore, in this case the simplicity constraints remove $4$ degrees of freedom from the 12 degrees of freedom of the previous subsection (after imposing the kinematical constraints), leaving us with 8 physical corner degrees of freedom.

We now want to identify the Dirac observables which describe these $8$ physical degrees of freedom. We already have that the tangential metric components $q_{ab}$ provide some of these observables since $\{\CC_I,q_{ab}\}=0$. It is then natural to look for the remaining ones among the Lorentz generators. One can use \eqref{SS}, \eqref{BS}, and \eqref{BB} to show that the generators defined as
\be\label{J1}
\JJ_{IJ}\coloneqq\BB_J\n_I-\BB_I\n_J+\teps_{IJ}{}^K\S_K
\ee
satisfy the $\sll(2,\mathbb{C})^S$ Lie algebra
\bea\la{PoincJJ}
\{\JJ_{IJ}(x),\JJ_{KL}(y)\}=\big(\eta_{JK}\JJ_{IL}+\eta_{IL}\JJ_{JK}-\eta_{IK}\JJ_{JL}-\eta_{JL}\JJ_{IK}\big)(x)\delta^2(x,y).
\eea
The generators $\JJ_{IJ}$ can also be interpreted as the components of total angular momentum, with $\n_I$ playing the role of momenta, $\mathsf L_{IJ}\coloneqq\BB_I\n_J-\BB_J \n_I$ playing the role of the angular momenta, and $\S_I$ being its spin component. It is important to note that together, the 10 generators $(\JJ_{IJ},\n_I)$ form a Poincar\'e algebra, where in addition to the Lorentz commutation relations \eqref{PoincJJ} we also have the brackets
\begin{subequations}\la{PoincJP}
\be
\{\JJ_{IJ}(x),\n_K(y)\}&=\big(\n_I\eta_{JK}-\n_J\eta_{IK}\big)(x)\delta^2(x,y),\\
\{\n_I(x),\n_J(y)\}&=0.
\ee
\end{subequations}
One can see that the crucial distinctive feature of our analysis which makes these structures available is that the internal normal is now part of the corner phase space.

The reason why the generators $\BB_I$ and $\S_I$ can be interpreted, respectively, as the covariant boost and spin components of a Lorentz algebra is that they can be written as
\be
\BB_I=\JJ_{IJ}\n^J,\q\q\S_I=*\JJ_{IJ}\n^J.
\ee
Since the generators $\JJ_{IJ}$ are Lorentz generators of a Poincar\'e algebra, below we will often call $\S_I$ the Poincar\'e spin. As expected, the total angular momentum components $\JJ_{IJ}$ generate Lorentz transformations on the corner phase space, i.e.
\be\la{JC}
\{\JJ_{IJ}(x),\mathsf{V}^K(y)\}=\big(\mathsf{V}_I\delta_J{}^K-\mathsf{V}_J\delta_I{}^K\big)(x)\delta^2(x,y),
\ee
where $V^I=\big(\BB^I,\S^I,\n^I,\e_a{}^I\big)$. The proof of this statement is given case by case in Appendix \ref{Lotrentzt}. This implies that the generators $\JJ_{IJ}$ represent weak Dirac observables, in the sense that they commute weakly with the simplicity constraints. Indeed, the bracket 
\be\la{JC}
\{\JJ_{IJ}(x),\CC^K(y)\}=\big(\CC_I\delta_J{}^K-\CC_J\delta_I{}^K\big)(x)\delta^2(x,y),
\ee
which is computed in \eqref{JC-app}, vanishes only when the simplicity constraints are imposed. Finally we have that the angle $\theta$ involved in the reconstruction of the frame commutes with all the constraints and is the last Dirac observable we are looking for. This shows that the algebra of weak Dirac observables is generated by $(\JJ_{IJ},q_{ab},\theta)$ and forms a subalgebra of $\mathfrak{sl}(2,\mathbb{C})^S\oplus\mathfrak{sl}(2,\mathbb{R})^S_\para\oplus\mathfrak{u}(1)^S_\para$. Indeed, this algebra is 10 dimensional, but there are however two balance relations among the Casimirs: The area matching condition \eqref{Sdet} and diagonal simplicity relation \eqref{diagsimp}, which leave us with 8 independent generators. These are the 8 Dirac observables that we were looking for.
 
This central result is in sharp contrast with the results obtained in the spin foam literature (see e.g. \cite{Perez:2012wv} and reference therein). There, one starts with the algebra\footnote{In LQG we only have access to a discrete analog of this algebra.} $\mathfrak{sl}(2,\mathbb{C})^S$, and the imposition of the simplicity constraints leads to a breaking of this Lorentz symmetry down to $\mathfrak{su}(2)^S$. Here we still have the full Lorentz algebra as our symmetry algebra, even after imposition of the simplicity constraints. In section \ref{sec:symbreaking} we review in more details the differences in symmetry breaking patterns between our analysis and the usual analysis of LQG.

Now, more care is needed in order to analyse the simplicity constraints $\CC_I$ since they contain second class components. It is therefore not enough to consider only weak observables, and we need to understand the nature of the strong Dirac observables. At this point, we can only identify three strong Dirac observables. Two are given by the two $\sll(2,\C)$ Casimirs
\begin{subequations}\la{QQtilde}
\be
\mathsf{Q}&=\f{1}{2}\JJ_{IJ}\JJ^{IJ}=\S^2-\BB^2,\\
\widetilde{\mathsf{Q}}&=\f{1}{2}*\JJ_{IJ}\JJ^{IJ}=-2\BB\cdot\S,
\ee
\end{subequations}
and the third one is the Poincar\'e spin Casimir $\S^2$. We need a deeper understanding of the simplicity constraints in order to promote $\JJ_{IJ}$ to strong Dirac observables. This can be done by properly identifying the first and second class components of the simplicity constraints, which we are now going to do.



\section{Algebra of simplicity constraints}
\la{sec:simp-algebrs}

In this section, we analyze in detail the algebra of simplicity constraints. Since there are three simplicity constraints $\CC_I$, not all of them are second class. Indeed, as we are about to see, only two are second class, while the other component is first class. To quantize the corner variables, we need to characterize the explicit splitting of the simplicity constraints into first and second class components. Such a split determines which constraints can be imposed strongly at the quantum level. To do so, we propose to use a Gupta--Bleuler imposition of the constraints \cite{Hasiewicz:1990xc, Kalau:1991jp}. This means that, given second class constraints $\CC_a$ and a kinematical metric $q_{ab}$, we need to find a splitting $\CC_a=(\CC_a^+,\CC_a^-)$ of these second class constraints into holomorphic and anti-holomorphic components such that:
\begin{enumerate}
\item[1)] $\quad\CC^-_a$ are first class,
\item[2)] $\quad\CC^-_aq^{ab}\CC^-_b=0$, and 
\item[3)] $\quad\CC_a^{-*}=\CC_a^+$. 
 \end{enumerate}
At the quantum level, we then impose strongly the holomorphic first class constraint $\widehat \CC^-_a|\Psi\rangle=0$, or alternatively (but less rigorously) use the master constraint $\mathcal{M}\coloneqq\CC_aq^{ab}\CC_b$. This allows us to construct the Dirac observables which can be used in the quantum theory for the proper construction of the corner Hilbert space.

The simplicity constraints have a long history and an intricate relationship with quantum gravity. They first appeared at the classical level in the work of Plebanski on the self-dual formulation of gravity \cite{Plebanski:1977zz,Capovilla:1991kx,Capovilla:1991qb,Obukhov:1996rb, Capovilla:2001zi}. In the Plebanski formulation, gravity is obtained from topological BF theory after imposition of the simplicity constraints. Their study and the proposal for their discretisation/quantisation has led to the creation of spin foam models \cite{Reisenberger:1996pu,Reisenberger:1997sk, Barrett:1997gw, Baez:1997zt, Markopoulou:1997wi, Freidel:1998pt,Barrett:1999qw,Livine:2001jt}, 
which provide a state sum representation of the path integral for quantum gravity. 
The main idea behind spin foam models is to start from a quantization of topological BF theory, and to then impose the simplicity constraints 
at the quantum level on the BF partition function. A key step in this construction was to understand that the simplicity constraints 
can be expressed as quadratic constraints on the discrete $B$ fields \cite{Reisenberger:1996ib,DePietri:1998hnx,Freidel:1999rr}.

This realization triggered a more in-depth study of the simplicity constraints in the continuum path integral. The Hamiltonian analysis of Plebanski theory was performed in \cite{Buffenoir:2004vx} (see also \cite{Alexandrov:2008fs}). It reveals that the primary simplicity constraints on the $B$ field lead to secondary constraints which also depend on the connection, implying that the complete set of canonical simplicity constraints is \emph{second class}. This poses a challenge for the understanding of spin foam quantization from a canonical perspective  \cite{Alexandrov:2006wt, Alexandrov:2011ab, Alexandrov:2007pq,Alexandrov:2008da,Alexandrov:2010pg,Alexandrov:2010un,Anza:2014tea}. Indeed, spin foams are thought of as a Lagrangian path integral, and as such focus only on the (primary) constraints on the $B$ field appearing in the Plebanski Lagrangian. This difference of treatment of the simplicity constraints in spin foams and canonical LQG is intimately related to the quest for a covariant formulation of LQG initiated by Alexandrov  \cite{Alexandrov:2001wt, Alexandrov:2002xc, Alexandrov:2002br}, which aims at the construction of a Lorentz covariant canonical connection and an explicit imposition of the secondary simplicity constraints in the spin foam path integral.

Focusing on the spin foam approach, Engle, Pereira and Rovelli realized \cite{Engle:2007uq, Engle:2007qf} that the 
discrete simplicity constraints used in the Barrett--Crane model were second class constraints, and that their strong imposition was responsible for the suppression of propagating degrees of freedom \cite{Alesci:2007tx}. This then led to the construction of a new family of spin foam models for quantum gravity, using path integral discretization \cite{Freidel:2007py} or canonical techniques \cite{Engle:2007wy}. The main technical feature of these constructions was to replace the discrete quadratic constraints by a set of discrete linear constraints involving an internal normal. These models were shown to possess the correct semi-classical limit \cite{Conrady:2008mk, Barrett:2009gg} and provided a new test ground for covariant quantum gravity amplitudes \cite{Rovelli:2010wq, Perez:2012wv}.

The quantization of the discrete simplicity constraints, known to be second class, was studied by many authors.
One can identify two types of studies: One involving canonical analysis and weak imposition of the constraints \cite{Livine:2007ya,Engle:2007wy,Engle:2007mu,Pereira:2007nh,Rovelli:2010ed,Ding:2010ye} and another one involving the use of coherent states \cite{Livine:2007vk, Freidel:2007py, Dupuis:2010iq,Dupuis:2011wy}. Despite this large literature dealing with the quantization of the simplicity constraints \cite{Perez:2012wv}, to the best of our knowledge only \cite{Wieland:2011ru,Speziale_2012} (which relies on a spinorial formulation) implement a clean split of the simplicity constraints between first and second class, and deal with the proper quantum implementation \`a la Gupta--Bleuler of the second class constraints.

In spite of all this work, a puzzle remains, which is the reconciliation between the continuum approaches and the discrete ones.
Indeed, although in both cases it is recognized that the simplicity constraints form a second class system, they however do so for different reasons. 
In the discrete approach, the geometrical simplicity constraints are second class due to the 
non-commutativity of the discrete $B$ field operator, and not because of the presence of secondary constraints (which as mentioned above are typically ignored in spin foam models) as in the continuum Hamiltonian analysis. Despite several attempts to understand the secondary simplicity constraints in the discrete framework \cite{Dittrich:2008ar,Dittrich:2010ey,Anza:2014tea} and proposals for their implementation in spin foams \cite{Alexandrov:2010pg,Alexandrov:2007pq,Alexandrov:2008da,Alexandrov:2011ab}, no conclusive resolution has been achieved yet.
The main difficulty in this task is reconciling the notion of a commutative continuous bulk $B$ field and the non-commutative discrete $B$ field.
Achieving this reconciliation via the introduction of edge mode operators was the purpose of \cite{Edge-Mode-II}. 

We are now in a position to revisit in details the implementation of the simplicity constraints in the continuum. 
First, as already pointed out in \cite{Edge-Mode-I} and explained more in detail in  \cite{Edge-Mode-II}, the covariant phase space formalism differs from the standard Dirac's algorithm of Hamiltonian analysis in that the former is an on-shell formalism and the  bulk dynamical content is taken into account by the fact that all the bulk equations of motion are imposed.  In particular, the secondary constraints are solved on-shell by the decomposition of the connection $\omega$ as in \eqref{con-dec} (see Section 3.5 of \cite{Edge-Mode-II}) and one is left with only the corner simplicity constraints \eqref{simpl-n} to impose.
Second, the introduction of the corner variables allows us to distinguish the bulk simplicity constraints form the corner simplicity constraints involving non-commutative variables. The key ingredient present in our setting, which enables us to perform the splitting of the constraints, is the existence of the coframe field as part of the phase space. Another notable difference between our analysis and the standard analysis is that the algebra of simplicity constraints obtained in \eqref{CCbra} is different from the constraint algebra studied in spin foam models. This is because the internal normal is now also part of the corner phase space. Although the general structure is similar, this leads to crucial differences at the classical and quantum level, which we are going to reveal and investigate.

Before proceeding, let us clarify how, in the locally holographic approach we are proposing \cite{Edge-Mode-I, Edge-Mode-II}, as a new path towards the quantization of gravitational degrees of freedom, the  bulk dynamical content is encoded at the boundary. More precisely, 
the bulk equations of motion translate to conservation laws for the corner charges (for the kinematical sector) and continuity equations relating the change of charges to the symplectic flux across a boundary representing the time development of the corner (for the dynamical sector). In the quantum theory then, the former set can   be implemented in terms of generalized intertwiners \cite{Freidel:2016bxd,Freidel:2019ees,Freidel:2019ofr}, while the latter in terms of
the fusion product for corner Hilbert
spaces. While the implementation of the dynamical content of the bulk clearly requires further investigation, we expect the corner symmetry group elements represented by the edge mode fields at the corner  (see Section 7 in \cite{Edge-Mode-II}) to play a crucial role in encoding the change of charge as determined by the symplectic flux across the boundary, generalizing the notion of $\SU(2)$ holonomy at the core of the LQG representation \footnote{Alternatively, one can exploit the new implementation of the corner simplicity constraints carried on below, and the ensuing appearance of new quantum numbers, to define a new spin foam model for quantum gravity.}.

In preparation for the splitting of the constraints, we gather here the Poisson brackets between the simplicity constraints and the corner phase space variables. All the brackets for this section are derived in Appendix \ref{App:simplicity-sec}. We have
\begin{subequations}
\be
\{\CC^I,\BB^J\}&=\BB^I\n^J-\CC^J\n^I-\teps^{IJ}{}_K\S^K,\label{CB}\\
\{\CC^I,\S^J\}&=\S^I \n^J+\f{1}{\beta}\teps^{IJ}{}_K\S^K,\label{CS} \\
\{\CC^I,\e_a{}^J\}&=\e_a{}^I\n^J+\f{1}{\beta}\teps^{IJ}{}_K\e_a{}^K,\label{Ce} \\
\{\CC^I,\n^J\}&=\tilde{\eta}^{IJ},\label{Cn}
\ee
\end{subequations}
where $\tilde{\eta}^{IJ}=\eta^{IJ}+\n^I\n^J$. From this, and as already anticipated, we see that $q_{ab}$, and therefore $\S^2=\beta^2 q $, are strong Dirac observables since
\be
\{\CC^I,q_{ab}\}=0,\q\q\{\CC^I,\S^2\}=0.
\ee
Now we are ready to explicitly identify the first class and the second class components of the simplicity constraints \eqref{simpl-n}. This can be done in analogy with the set of kinematical constraints \eqref{sec-con}. Indeed, since we have\footnote{This comes from the fact that
\be\la{eS}
\e_a{}^I\S_I=\f{\beta}{2}\teps_{IJK}\e_a{}^I\e_b{}^J\e_c{}^K\epsilon^{bc},
\ee
which vanishes since the indices $a,b,c$ can only take two different values.}
$\S^I\n_I=0= \e_a{}^I \S_I$, we can use the set $(\S^I,\n^I,\e^I)$ as an orthogonal basis to decompose internal vectors. In particular we can, with this basis, project components of the simplicity constraints \eqref{simpl-n}, and thereby separate them into first and second class parts.

\subsection{Second class simplicity algebra}

Let us start with the second class part of the simplicity constraints. Using the frame, we can isolate the tangential component of the simplicity constraints by considering
\be
\CC_a\coloneqq\CC_I\e_a{}^I.
\ee
The full components of the simplicity constraints can then be reconstructed from the knowledge of $(\CC\cdot\S,\CC_a)$ as
\be
\CC^I=\f{\CC\cdot\S}{\S^2}\S^I+\CC^a\e_a{}^I.  
\ee
One can now check, as in \eqref{CCa}, that the two tangential constraints form a second class pair with bracket
\be\label{CaCb}
\{\CC_a,\CC_b\}=-\f{1}{\beta}\epsilon_{ab}\big(\S^2+\BB^2\big).
 \ee
Note that the operator appearing in the right-hand side is a positive operator, which insures that $\CC_a$ are always second class. The compatibility of the Hodge dual with the bracket, which was established in the previous section, implies that
\be
\{\star\CC_a,\star\CC_b\}=\{\CC_a,\CC_b\},\q\q\{\star \CC_a,\CC_b\}+\{\CC_a,\star\CC_b\}=0.
\ee
This compatibility ensures that the holomorphic constraints $\CC^\pm_a\coloneqq\f{1}{2}(\CC_a\pm i\star\CC_a)$ commute with each other:
\be
\{\CC^\pm_a,\CC^\pm_b\}=0.
\ee
We can therefore replace the second class constraint $\CC_a=0$ by the first class condition $\CC_a^-=0$. At the quantum level this condition is imposed strongly as $\widehat \CC_a^-|\Psi\rangle=0$.
 
 \subsection{ Master constraint and semi-classical anomaly}
It is often convenient to replace the condition $\CC_a^-=0$ by the master constraint $\mathcal{M}=0$, where 
\be\label{Master}
\mathcal{M}=\CC_aq^{ab}\CC_b.
\ee 
At the classical level the condition $\CC_a^-=0$ is equivalent to the master constraint, but at the quantum level this is no longer true. Indeed, the second class nature of the constraints gives rise to an anomaly term entering the expression of the master constraint, since
\be\label{Manomaly}
\widehat{\mathcal{M}}|\Psi\rangle=(2\widehat{\CC}_a^+\widehat{q}^{ab}\widehat{\CC}_b^-+\widehat{\mathcal{A}})|\Psi\rangle.
\ee
The condition $\widehat{\CC}_a^-|\Psi\rangle=0$ leads to $\widehat{\mathcal{M}}|\Psi\rangle=\widehat{\mathcal{A}}|\Psi\rangle$, and the anomaly is
\be\la{ano}
\widehat{\mathcal{A}}=\widehat{q}^{ab}[\widehat{\CC}_a^-,\widehat{\CC}_b^+]+[\widehat{q}^{ab},\widehat{\CC}_a^+]\widehat{\CC}_b^--[\widehat{q}^{ab},\widehat{\CC}_a^-]\widehat{\CC}_b^+.
\ee
The quantum anomaly $\widehat{\mathcal{A}}$ is defined here as a quantum operator. In the semi-classical limit $\hbar\to 0$ it becomes a phase space observable $\widehat{\mathcal{A}} \to \hbar {\mathcal{A}}$ where $\mathcal{A}$ is the semi-classical anomaly.
We want to emphasize that it is possible to evaluate semi-classically this anomaly simply by replacing commutators by Poisson brackets $[\cdot,\cdot]\to i\{\cdot,\cdot\}$ in \eqref{ano}.
This will give us an idea of the phase space observable whose quantization gives the quantum anomaly.
It is one of the many examples where a proper treatment of the semi-classical analysis gives us deep insight onto the quantum theory.

In order to compute the semi-classical anomaly $\mathcal{A}$ we need to evaluate  the commutation between the holomorphic and anti-holomorphic constraints, which is related to the brackets \eqref{starCCapp} involving the Hodge dual of the constraint. We find
\be\la{C+C-}
\{\CC^-_a,\CC^+_b\}=\left(q_{ab}-\f{i}{\beta}|\S| \eps_{ab}\right)\frac{\left(\S^2+\BB^2\right)}{2i|\S|}-\frac{\CC^+_a\CC^-_b+\CC^-_a\CC^+_b}{i|\S|},
\ee
with $|\S|\coloneqq\sqrt{\S^2}=\beta\sqrt{q}$. To evaluate the anomaly we also need the bracket
\be
i \{q^{ab}, \CC_b^\pm\}=\frac{1}{|\S|}q^{ab}\big(i\star\CC_b^\pm\big)=\pm\frac{1}{|\S|}q^{ab}\CC^\pm_b.
\ee
Together this gives us the semi-classical evaluation 
\be
\mathcal{A}=  \frac{\S^2+\BB^2}{|\S|}\,,
\ee
and we expect the quantum anomaly to simply be a quantization of this operator.
Going back to the quantum discussion, let us recall that 
the master constraint program looks for a state $\Phi_0$ 
which minimizes the expectation value of $\cal M$:
\be 
\frac{\langle\Phi_0|\widehat{\cal M}|\Phi_0\rangle}{\langle\Phi_0|\Phi_0\rangle} = \mathrm{Min}_{\Phi}\left( 
\frac{\langle\Phi|\widehat{\cal M}|\Phi\rangle}{\langle\Phi|\Phi\rangle} \right)\,.
\ee
Given a Gupta--Bleuler state $\widehat{\CC}_a^-|\Psi \rangle=0$, which means that $\widehat{\cal M} |\Psi\rangle = \widehat{\cal A} |\Psi\rangle $, since $\widehat{\cal M}$ is a positive Hermitian operator  we have the inequality\footnote{This inequality follows from the Cauchy--Schwarz inequality for the states $\frac{\sqrt{\widehat{\cal M}}|\Phi_0\rangle}{\sqrt{\langle \Phi_0|\Phi_0\rangle}}$ and $ \frac{\sqrt{\widehat{\cal A}}|\Psi\rangle}{\sqrt{\langle\Psi|\Psi\rangle}}$.}
\be 
\frac{\langle\Phi_0|\widehat{\cal M}|\Phi_0\rangle}{\langle\Phi_0|\Phi_0\rangle}  \geq  
\frac{| \langle\Phi_0|\sqrt{\widehat{\mathcal{M}}}\sqrt{\widehat{\mathcal{A}}}|\Psi\rangle|^2}{|\langle\Phi_0|\Psi\rangle|^2} 
\frac{\langle\Psi|\Psi\rangle}{\langle\Psi| \widehat{\mathcal{A}}|\Psi\rangle}\,.
\ee

In order to show that the use of the master constraint is equivalent to the Gupta--Bleuler implementation of the constraints, we need to show that $\widehat{\mathcal{A}}$ weakly commutes with $\widehat{\CC}_a^-$. In this case we can simultaneously solve $\widehat{\CC}_a^-|\Psi\rangle=0$ and diagonalize the anomaly $\widehat{\mathcal{A}}|\Psi\rangle= \widehat {\mathcal{A}}_0  |\Psi\rangle$, and the previous inequality becomes
\be 
\frac{\langle\Phi_0|\widehat{\cal M}|\Phi_0\rangle}{\langle\Phi_0|\Phi_0\rangle}  \geq \mathcal{A}_0.
\ee
The minimum value is attained for $\Phi_0=\Psi$ and we see that
$\mathcal{A}_0 = \mathcal{M}_{\mathrm{min}}$.
The quantum consistency of the master constraint analysis therefore demands that the quantum anomaly operator commute with the  $\widehat{\CC}_a^-$.

We now show that this consistency condition is satisfied at the semi-classical level and that $\mathcal A$ is a Dirac observable. This follows from the brackets given in Appendix  \ref{CaSBapp}, which are
\begin{subequations}
\be
 \{\S^2 , \CC_a\}&= 2 |\S| \star \CC_a ,\la{SSC}\\
\{\BB^2 , \CC_a\}&=2 |\S| \star \CC_a,\\
\{\BB\cdot\S , \CC_a\}&=0.\label{CBBSS}
\ee
\end{subequations}
Therefore we conclude that
\be
\{\mathcal{A}, \CC_a^-\}=\frac{\BB^2}{\S^2} \CC_a^-,
\ee
which shows that the anomaly weakly Poisson-commutes with the first class constraints. This shows that one should be able to  use the master constraint to implement the second class simplicity constraints. Of course the proof here is only done at the semi-classical level.
The quantum imposition  will be carried out in \cite{Edge-Mode-IV}.

\subsection{First class simplicity algebra}

We now have to isolate the component of $\CC_I$ which is first class. By definition, this is the component which commutes strongly with the second class components $\CC_a$. There are two natural candidates, namely $\CC^2$ and $\CC\cdot \S$. Neither are suitable individually since they lead to
\be\label{C2Ca}
 \{\CC^2 , \CC_a\}
 =2 \left(1+\frac1{\beta^2}\right)|\S| \star\CC_a,
\q\q
\{\CC\cdot \S, \CC_a\}
= -\frac{2}{\beta}|\S|\star\CC_a\,,
\ee
as shown in \eqref{C2Caapp} and \eqref{CSCaapp}. This establishes however that the combination 
\be\la{diagsimp}
-\FC\coloneqq\CC^2 + \left(\beta+\beta^{-1}\right) \CC\cdot \S
\ee
commutes with $\CC_a$. Therefore, $\FC=0$ is the first class constraint we are looking for. Using the definition of $\CC^I$ we can rewrite this first class constraint as
\be\la{Cfirst}
\FC
&=(\beta^{-1} \S-\BB)\cdot(\beta \S+\BB)\cr
&=(\S^2-\BB^2) - \left(\beta-\beta^{-1}\right)  \BB\cdot \S \cr
&={\mathsf Q} + \left(\frac{{\beta}-{\beta^{-1}}}{2}\right)  \widetilde {\mathsf Q}\,.
\ee
This shows that the first class component of the simplicity constraints is a function of the $\sll(2, \C)$ Casimirs only. 
The expression  \eqref{Cfirst}  first appeared in \cite{Livine:2001jt}. 
It corresponds to the continuum version of the  ``diagonal simplicity constraint'' 
and it was studied further in \cite{Engle:2007wy, Pereira:2007nh, Livine:2007ya, Alexandrov:2010pg, Rovelli:2010ed}.
It was first observed in \cite{Engle:2007mu} that this expression corresponds to 
the first class component the discrete simplicity constraints.
The second class components were never identified in the vector formalism.
The only work that did identify a split between  first class and second class components of the simplicity constraints 
was done in the twistor formalism by Wieland \cite{Wieland:2011ru}.
However, the first class component there is not exactly the diagonal simplicity constraint and the normal is still treated kinematically. 

We can easily solve the first class simplicity constraint at the quantum level by working in a basis where the two Casimirs
$ {\mathsf Q}$ and $\widetilde {\mathsf Q}$ are diagonal.
 In preparation for the quantization of the theory, one introduces a parametrization of the Casimirs in terms of two Lorentzian weights $(k,\rho)$, with $k >0$, which are such\footnote{Explicitely we have 
 \be 
 k^2 =\frac12 \left(\sqrt{{\mathsf Q}^2+ {\widetilde{\mathsf Q}}^2}+ {\mathsf Q}\right),\q
 \rho^2=\frac12 \left(\sqrt{{\mathsf Q}^2+ {\widetilde{\mathsf Q}}^2}- {\mathsf Q}\right)
 \ee} that
 \bea
 {\mathsf Q}  =k^2-\rho^2 ,\q 
 \widetilde {\mathsf Q}  = -2k \rho  \,.
\eea
At the quantum level $(\rho, k)$ become the weights of the  unitary representations of the Lorentz algebra, and the Lorentz spin $k$ is quantized. The first class simplicity constraint $\FC=0$ is then solved by 
\be
\rho=\frac{k}{\beta},\q\mathrm{or}\q\rho=-\beta k.
\ee
The existence of two solutions is due to the fact that  the transformation $\beta\rightarrow -\beta^{-1}$ exchanging the two solutions corresponds to the duality map $\BB_{IJ}\to*\BB_{IJ}/\beta$.
This duality is broken in gravity, and only the first branch corresponds the the classical solution where $\BB=\beta^{-1}\S$. The other solution obtained after the duality transformation $\beta\to-\beta^{-1}$ is a spurious one. In fact it leads to a topological sector of the theory.

The pair $(\rho,k)$ can also be seen as the invariant Cartan weight associated with the Lorentz generators $\JJ^{IJ}$. It is clear from the algebra \eqref{PoincJJ} that the generator $\JJ_{12}$ and $\JJ_{03}$ are  the commuting Cartan elements. This means that,
given $\JJ^{IJ}$, one can always find an $\SL(2,\mathbb{C})$ transformation $g$ such that 
$(g\JJ g^{-1})^{IJ}$ is diagonal, and explicitly write
\be
{\JJ}^{IJ} = k g^{I}{}_{[1} g^{J}{}_{2]} + \rho g^I{}_{[0}g^J{}_{3]},
\q
\as\JJ^{IJ} = -\rho  g^{I}{}_{[1} g^{J}{}_{2]}+ k  g^I{}_{[0}g^J{}_{3]}\,.
\ee
The first class simplicity constraint can then be written covariantly as the simplicity condition 
$*\Sigma_\beta^{IJ}\Sigma_{\beta IJ} =0$, where  $\Sigma_\beta^{IJ}\coloneqq {\JJ}^{IJ}-\beta^{-1}*\mathsf{J}^{IJ}$. 
In this Cartan decomposition we have that 
\be 
\Sigma_\beta^{IJ} = \left(k+\frac{\rho}{\beta} \right)g^{I}{}_{[1} g^{J}{}_{2]} + \left(\rho-\frac{k}{\beta} \right) g^I{}_{[0}g^J{}_{3]}.
\ee
The first class simplicity constraint $\FC=0$ means that one of the two factors vanishes. This shows that the first class simplicity constraint is equivalent to the statement that there is a vector $N^I$ such that $\Sigma_\beta^{IJ}  N_J=0$.  The physical sector corresponds to the requirement that $N^I$ be a time-like vector. The vector $N^I$ is therefore uniquely determined by the choice of a $\JJ^{IJ}$ solution of the first class simplicity. Up to normalization we have  that $N^I=g^I{}_0$.

\subsection{Corner Dirac observables}\la{sec:Dirac-obs}

Our rewriting of the simplicity constraints does not affect the set of kinematical constraints \eqref{sec-con}. This simply follows from the fact that  $\n_a$ and $\n^2$ commute with $(\BB^I,\S^I,\e_a{}^I)$ by construction, as shown in Section \ref{sec:Bound-PS}. Overall,
this means that the first and second class simplicity constraints $(\FC,\CC_a)$ commute strongly with the kinematical constraints $(\n^2,\n_a)$.
This establishes that the full set of constraints in the initial corner phase space \eqref{poincare theta} consists of two first class constraints $(\n^2, \FC)$ and four second class constraints $(\n_a, \CC_a).$


Now that we have isolated the first class constraint $\FC$ and the second class constraints $\CC_a$ in the simplicity constraints, we can turn our attention to the corner Dirac observables. Let us recall that a physical observable $\CO$ is required to commute strongly with the second class constraints and weakly with the first class constraint, i.e.
\be
\{\CC_a,\CO\}\approx 0,\q \{\FC,\CO\}\approx0,
\ee
where the  equality $\approx0$ means that it vanishes after imposition of $\FC=0$ and $\n^2=-1$ but without using $\CC_a=0$ or $\n_a=0$.

Since $(\CC^I,\e_a{}^I)$ are Lorentz vectors, we have that $(\CC^2,\CC\cdot\S,\CC_a)$ are Lorentz scalars. It follows that they commute with $\JJ_{IJ}$, meaning that the total angular momentum is a strong Dirac observable. This is shown explicitly in Appendix \ref{Lotrentzt} as a consistency check, where we find
\be\la{CSQtilde}
\{\CC^2, \JJ_{IJ}\}=0\,,\q\{\CC\cdot\S, \JJ_{IJ}\}=0\,,\q \{\CC_a,\JJ_{IJ}\}=0\,.
\ee
For the same reason we have that
 \be\la{nSJ}
\{\n^2,\JJ_{IJ}\}=0\,,
\q \{\n_a,\JJ_{IJ}\}=0\,.
\ee
This fact shows that internal Lorentz symmetry is preserved by the imposition of the simplicity and kinematical constraints. However, the same is no longer true for the corner metric components $q_{ab}$. In fact, while we still have
\be
\{q_{ab},\n_c\}=0
\ee
by virtue of \eqref{tena}, we have that $\CC_a$ transforms as an $\mathfrak{sl}(2,\mathbb{R})$ vector, i.e.
\be
\{q_{ab},\CC_c\}=\frac{1}{\beta}(\epsilon_{ca}\CC_b+\epsilon_{cb}\CC_a) \,.
\ee
This shows that $q_{ab}$ does not commute strongly with the second class constraints. As our next goal (postponed to \cite{Edge-Mode-IV}) is to build the corner Hilbert space by means of the quantum numbers labeling the representations of the corner symmetry algebra, let us look ahead and realize that
this puzzle can easily be resolved by choosing the Gupta--Bleuler way of quantizing the second class constraints, which replaces the two real constraints $\CC_a$ by a complex one $\CC_a^-$. With this quantization method, a Dirac observable now is only required to commute weakly with $\CC_a^-$. This is easily seen to be the case since we have
\be
\{q_{ab},\CC^-_c\}=\frac{1}{\beta}(\epsilon_{ca}\CC^-_b+\epsilon_{cb}\CC^-_a)\approx 0\,.
\ee

An alternative way to proceed is to replace the second class components of the simplicity constraints with a single master constraint. In the quantum theory, and as shown in \eqref{Manomaly}, the second class nature of the constraint algebra is reflected in the fact that the zero eigenvalue does not appear in the spectrum of the master constraint operator, and a weak imposition of the constraints amounts to selecting the minimum eigenvalue instead \cite{Perez:2012wv}. The master constraint $\mathcal{M}$ was defined in \eqref{Master}. It is a Lorentz scalar and therefore commutes strongly with the first class simplicity constraint $\FC$. Moreover, we have that
\be
\{ \mathcal M, q_{cd}\}
&=2 \CC^a \{\CC_a, q_{cd}\} - \CC^a \CC^b \{q_{ab}, q_{cd} \}\cr
&=-\frac{2}{\beta}\CC^a\left(\eps_{ac}\CC_{d}+\eps_{ad}\CC_c\right)+\frac{2}{\beta}\CC^a\CC^b\left(q_{ac} \eps_{bd}+q_{bc} \eps_{ad}\right)\cr
&=0\,,
\ee
showing how the corner metric components continue to be strong Dirac observables if we rely on the master constraint approach for the second class sector of the simplicity constraints.

To summarize, we have separated the corner simplicity constraints into one first class component $\FC$ and one master constraint $\mathcal{M}$, with this latter replacing the second class components. This allowed us to show that the 6 Lorentz generators and the 3 corner metric components commute strongly with both the kinematical constraints and the simplicity constraints $(\FC,\mathcal{M})$. This therefore gives us a total of 9 observables. However, not all of them are independent. We showed in \eqref{Cfirst} that the first class component of the simplicity constraints, $\FC=0$, implies a relationship between the two $\sll(2,\C)$ Casimirs, thus reducing the number of independent Lorentz generators to 5. In addition, there is the Casimir balance equation \eqref{Sdet}. This relation further reduces the number of independent physical observables down to 7. This means that we still need to recover one of the physical degrees of freedom of  the corner phase space parametrized by \eqref{tse}, which is 8-dimensional. The missing observable is represented by the angle $\theta$ introduced in Section \ref{sec:kin-constraints}. This angle, which is conjugated to the area element as shown in \eqref{qtheta}, encodes the information about the frame field which is not captured by the spin generators and the tangential metric.

\subsection{Master constraint}

Let us finally focus on the imposition of the master constraint \eqref{Master}. At the classical level, we simply need to impose $\mathcal{M}=0$. At the quantum level however, we need to impose $\mathcal{M}=\mathcal{A}$. In preparation for the study of the quantum theory, which is the focus of the companion paper \cite{Edge-Mode-IV}, we want to establish here that the master constraint can be conveniently written as a function of the Poincar\'e spin and Lorentz weights in the form
\be\la{master rewriting}
\mathcal{M}=\frac{(\S^2-k^2)(\S^2+ \rho^2) }{\S^2}. 
\ee
In order to show this, we first use \eqref{S as ee app} to establish the identity
\be
\beta^2\e_a{}^I ( q  q^{ab})\e_b{}^J
&=\beta^2\e_a{}^I \eps^{ac} q_{cd} \eps^{bd}\e_b{}^J \cr
&=\beta^2(\e_a{}^I \eps^{ac} \e_c{}^K)\eta_{KL}(\e_d{}^L \eps^{bd}\e_b{}^J )\cr
&=\S_M\teps^{MIK}\eta_{MN}\teps^{NJL}\S_N\cr
&=\S_M\big(\tilde\eta^{IJ} \tilde\eta^{MN}-\tilde\eta^{IN}\tilde\eta^{MJ}\big)\S_N\cr
&=\tilde\eta^{IJ}\S^2-\S^I\S^J,
\ee
and then the fact that $\S^2=\beta^2q$ to write
\be
\e_a{}^Iq^{ab}\e_b{}^J=\tilde\eta^{IJ}-\f{\S^I\S^J}{\S^2}.
\ee
Contracting this with $\CC_I$ and $\CC_J$ then gives
\be
\mathcal{M}=\CC_aq^{ab}\CC_b=\BB^2 -\frac{(\BB\cdot\S)^2}{\S^2}=\S^2 - {\mathsf Q} -\frac{\widetilde{{\mathsf Q}}^2}{4 \S^2},
\ee
which can be evaluated in terms of the Lorentz weights $(k,\rho)$ to find \eqref{master rewriting}.

We therefore see that the condition $\mathcal{M}\geq 0 $ implies that  $\S^2 \geq k^2$. The weight $k$ is  the minimal admissible value for $s=|\S|$.
The condition $\mathcal{M}=0$ therefore implies that the Lorentz spin is equal to the Poincar\'e spin, i.e. $k=s$. Taking into account the first class simplicity constraint, this means that the joint solution of $\FC=0=\mathcal{M}$ is
\be\label{simpler}
\left(k=s,\rho=\frac{s}{\beta}\right),\q\mathrm{or}\q\big(k=s,\rho=-\beta s\big).
\ee 
The gravitational sector corresponds to the first branch. This establishes one of our key results, namely that the simplicity constraints imply that the value of the Lorentz Casimirs are entirely determined by the Poincar\'e spin.

\section{Symmetry breaking pattern}
\la{sec:symbreaking}

We have now arrived at a complete understanding of the role of the simplicity constraints on the corner phase space. In particular, we have shown that we still have an $\text{SL}(2,\C)^S$ symmetry even after imposing the constraints $\BB_I=\S_I/\beta$. This is in sharp contrast with the standard literature in  spin foams \cite{Freidel:2007py, Engle:2007wy,Perez:2012wv}. There, one assumes that the Lorentz symmetry generators are constrained to satisfy the discrete simplicity constraints of the form
\be\la{simp-kin}
\mathring{C}_I\coloneqq K_I-\frac{L_I}{\beta}=0, 
\ee
where $K_I=\JJ_{IJ}t^J$ are the boosts along a fixed direction $t^I$ (often taken to simply be $t^I=\delta^I_0$), and $L_I=*\JJ_{IJ}t^J$ are the rotation generators fixing this kinematical direction. Both generators satisfy $K_It^I=0=L_It^I$, and if one choses $t^I=\delta^I_0$, then $L_0=0=K_0$ and the kinematical spin and boost generators $(L_i,K_i)$ are 3 dimensional vectors with  $i=1,2,3$. These generators satisfy the commutation relations
\be\la{KL}
-i[L_I,L_J]=-\mathring\epsilon_{IJ}{}^KL_K\,,\q-i[L_I, K_J]=-\mathring\epsilon_{IJ}{}^K K_K\,,\q-i[K_I, K_J]=\mathring\epsilon_{IJ}{}^K L_K\,,
\ee
where $\mathring{\epsilon}_{IJ}{}^K\coloneqq\epsilon_{IJ}{}^{KL}t_L$. Moreover, the discrete simplicity constraints satisfy the algebra
\be
-i[\mathring{C}_I,\mathring{C}_J]=\frac{2}{\beta} \mathring\epsilon_{IJ}{}^K \mathring{C}_K+\left(1+\frac1{\beta^2} \right)\mathring\epsilon_{IJ}{}^K L_K.
\ee
We see that the structure of this algebra differs substantially from the simplicity constraint algebra (\ref{CCbra}) derived in the continuum. This is because the standard discrete analysis of the simplicity constraints differs from our continuum analysis in essential ways.

First, in the standard analysis of the spin foam simplicity constraints, one postulates implicitly that the symmetry group of BF theory, before the imposition of the simplicity constraints, is the \emph{Lorentz} group. Then the time-like vector $t^I$ is assumed to commute with the Lorentz generators, and therefore taken to be \emph{kinematical}. These conditions, in turn, imply that the simplicity constraints break the Lorentz group down to its  $\mathrm{SU}(2)$ subgroup preserving the time-like direction $t^I$. Indeed, one can check that only $L_I$ preserves the constraints and provides a Dirac observable, since we have $-i[L_I,\mathring{C}_J]=-\mathring\epsilon_{IJ}{}^K\mathring{C}_K$, while the boost operator is not a Dirac observable even weakly since
\be
-i[K_I,\mathring{C}_J]=\f{1}{\beta}\mathring\epsilon_{IJ}{}^K \left(\mathring{C}_K+ (\beta +\beta^{-1})L_K\right).
\ee
Therefore the ``usual'' spin foam simplicity constraints break the Lorentz symmetry down to an SU$(2)$ symmetry. This breaking of the Lorentz symmetry, which follows the pattern
\be
0\to H^3 \to \mathrm{SL}(2,\mathbb{C}) \to \mathrm{SU}(2) \to 0, 
\ee
with $H^3$ the Hyperbolic space, is the source of many puzzles and shortcomings of the standard treatment.

The correct continuum analysis presented here instead shows that the corner symmetry group is the \emph{Poincar\'e} group generated by $(\JJ_{IJ}, \n^K)$, before the imposition of the simplicity constraints. The presence of this elemental Poincar\'e algebra, associated entirely to the geometry itself and not to the presence of matter, is one of the most surprising features of our construction. It also establishes that the internal time-like direction $\n^I$ is a \emph{dynamical} variable, which plays the role of momenta, and as such gets rotated by the Lorentz generators. At the quantum level, this new role of the internal normal will manifest itself in the introduction of new quantum numbers entering labeling the representations of the elemental Poincar\'e algebra \footnote{In the usual Poincar\'e basis, these correspond to  the components of a spatial vector $\vec{n}\in \R^3$ labelling the positive energy states.}, as shown in \cite{Edge-Mode-IV}.
Finally, and most importantly, it shows that the simplicity constraints break the \emph{Poincar\'e} group down to its Lorentz subgroup following the pattern
\be
0\to \mathbb{R}^4 \to \mathrm{ISO}(3,1)\to \mathrm{SO}(3,1) \to  0\,.
\ee 
In this way, the imposition of the simplicity constraints does not break internal Lorentz invariance, but only breaks the translational subgroup of 
Poincar\'e. Remarkably it does so by still allowing a discrete spectra for the area operator. This resolves one of the fundamental puzzle of LQG. Let us now give a few more details concerning this Poincar\'e structure and its geometrical interpretation.

\subsection{Poincar\'e interlude}\la{sec:Poin-interlude}

It is illustrative at this point to draw a parallel between the elemental Poincar\'e structure we have discovered as the corner symmetry algebra before simplicity constraints, and its common interpretation in a physical context. Since the continuum algebra is ultra local, we can concentrate on the global structure of the algebra and ignore the $x$ dependency and the delta distribution. We are interested in the study of the elemental Poincar\'e algebra $(\JJ_{IJ},\n_{J})$ in terms of the Poincar\'e algebra of a moving particle. The commutators used in this section are given in Appendix \ref{App:Poinc}.

Let us momentarily forget about the corner variables, and consider the usual Poincar\'e generators $(J_{AB},P_C)$ and their algebra
\begin{subequations}\la{Lorentz-alg}
\be
-i[J_{IA}, J_{JB}]&=\eta_{AJ}  J_{IB} +\eta_{IB}  J_{AJ} -\eta_{IJ}  J_{AB} -\eta_{AB}  J_{IJ}\,,\\
-i[J_{IA}, P_{B}]&=P_I \eta_{AB} - P_A\eta_{IB}\,,\\
-i[P_{A}, P_{B}]&=0\,.
\ee
\end{subequations}
The Pauli--Lubanski vector $W_I$ given by
\be
W_I\coloneqq*J_{IJ}P^J
\ee
is the generator of the little group of the Poincar\'e group. The Poincar\'e algebra possesses two Casimirs: the ``mass'' squared, and the square of the Pauli--Lubanski vector defining the spin, respectively
\be
P^2=-m^2, \q W^2= m^2s(s+1).
\ee
One of the goals of this section is to establish that the Barbero--Immirzi parameter $\gamma=\beta^{-1}$ is in fact the \emph{mass} of the elemental Poincar\'e algebra, namely
\be\label{poincare mass}
m^2=\gamma^2,\q W^2 =  q \,,
\ee
where the second relation is the particle expression of the relation \eqref{Sdet}. This shows that the limit $\gamma\to 0$, which corresponds in a sense to a metric limit where the torsion is not fluctuating, corresponds in fact to a massless limit from the point of view of the elemental Poincar\'e symmetry. Moreover, since in this limit $W^2\neq0$, we expect to recover the continuous-spin representations of the Poincar\'e group \cite{Wigner:1939cj,Weinberg:1995mt,Bekaert:2005in,Schuster:2013pxj}. We will explore this massless limit in elsewhere.

Focusing here on the massive case $\gamma\neq0$, we can define the dynamical boost and spin vectors to be
\be
B_I\coloneqq\frac{J_{IJ}P^J}{m},\q S_I\coloneqq\frac{W_I}{m}=\frac{*J_{IJ}P^J}{m}.
\ee
As shown in Appendix \ref{App:Poinc}, they act on the 4-momentum as  
\be
-i[B_I,P_J]=m\left(\frac{P_IP_J}{m^2}+\eta_{IJ}\right),\q
-i[S_I,P_J]=0\,,
\ee
and they satisfy the boost-spin algebra
\begin{subequations}
\be
-mi[B_I,B_J]&=B_I P_J-B_J P_I-\eps_{IJ}{}^{KL} S_KP_L, \label{WW1}\\
-mi{[}B_I,S_J]&=S_IP_J ,\label{WW2}\\
-mi{[}S_I,S_J]&=-\eps_{IJ}{}^{KL}S_KP_L\,.\label{WW3}
\ee
\end{subequations}
This algebra was first derived by Shirokov \cite{shirokov1,shirokov2} in his study of the Poincar\'e algebra (see also \cite{Pirotte}). We see that these commutation relations recover the brackets \eqref{SS}, \eqref{BS}, and \eqref{BB} derived in the previous section upon identifying
\be
\BB_I=B_I,\q\S_I=S_I,\q\n_I=\f{P_I}{m}.
\ee
We see in particular that the internal normal is identified with the particle's 4-momentum divided by its rest mass. At the same time, we have that our Poincar\'e spin Casimir satisfies the relation \eqref{Sdet}. This suggests an interpretation of a spin network link carrying a given $\SU(2)$ irreducible representation label $s$ as a ``particle of quantum space'' carrying spin $s$ and mass $m=\gamma\hbar$. In this analogy, the time gauge corresponds to the particle's rest frame.

As explained above, the usual spin foam analysis is done introducing a decomposition of the Lorentz generators in terms of the boost and rotation generators $(K_I,L_I)=(J_{IJ}t^J,*J_{IJ}t^J)$ associated to a kinematical vector $t^I$. Using $t^I=\delta^I_0$ gives $(K_i,L_i)$. These generators are related to the dynamical boost and rotation operators by 
\be\label{BSKJ}
m S_i= P^0 L_i - (P\times K)_i, \q mB_i = P^0 K_i + (P\times L)_i, 
\ee
while $S_0 = L_iP^i$ and $B_0=K_iP^i$.

\subsection{Heisenberg frames}\la{sec:heisenberg}

We can now push even further the analysis of the elemental Poincar\'e algebra and of the geometrical nature of the BF edge modes, by going back to the precursor phase space \eqref{poincare theta} before the imposition of the kinematical constraints. For this, let us consider a particle-like parametrization of the Lorentz generators in terms of oscillators where
\be\label{JXPz}
J^{IJ}=X^IP^J-X^JP^I+ z_a{}^I \epsilon^{ab}  z_b{}^J\,.
\ee
The set $(X^I,P^I)$ corresponds to the particle's position and momentum, while $(z_a{}^I)_{a=2,3}$ are variables parametrizing the internal degrees of freedom. The non-vanishing commutation relations between these variables are
\be
-i[X^I,P^J]= \eta^{IJ},\q-i[z_a{}^I ,  z_b{}^J]=  \epsilon_{ab} \eta^{IJ}\,,
\ee
which should be put in parallel with \eqref{basic poincare brackets}. This corresponds to a decomposition of the total angular momentum in terms of the angular momentum and spin as $ J^{IJ}=L^{IJ}+S^{IJ}$, where 
\be
L^{IJ}=X^IP^J-X^JP^I,\q S^{IJ}= z_a{}^I \epsilon^{ab}  z_b{}^J.
\ee
The total angular momentum is invariant under the Heisenberg translations
\be 
X^I \to X^I + a^a   z_a{}^I + b P^I, \q z_a{}^I \to z_a{}^I + \eps_{ab} a^b P^I.
\ee
The Heisenberg translation group $H_3(\mathbb{R})$ is the 3-dimensional group generated by $P_a\coloneqq z_a{}^I P_I$ and $P^2=P^IP_I$ with commutation relations
\be
-i[P_a,P_b]=\eps_{ab} P^2, \q-i[P_a, P^2]=0.
\ee
The squared mass is therefore the central element of $H_3(\mathbb{R})$. Since this symmetry acts non-trivially on the position $X^I$, it means that fixing this symmetry amounts to choosing a position operator. Since the Heisenberg transformations also act on the spin components as
\be
S^{IJ}\to S^{IJ} - a^I P^J + a^J P^I, \q a^I\coloneqq a^a z_a{}^I,
\ee
we can chose a position operator by imposing a condition on the spin generators. The simplest way to parametrize the choice of position operator by fixing the Heisenberg frame is to chose a vector $k^I$ and to impose the condition $S^{IJ}k_J=0$. This condition breaks the $H_3(\mathbb{R})$ symmetry group down to its center, and fixes the world-line position to be
\be
X_k^I(\tau)=\frac{J^{IJ}k_J + P^I \tau }{P\cdot k}\,.
\ee
There are three natural choices for the Heisenberg frame \cite{Fleming}: the inertial frame, the rest frame, or the Newton--Wigner frame.

The \emph{inertial frame} corresponds to the choice $k^I=P^I$. This corresponds to a choice of inertial frame coordinates for which $P_a=0$. Geometrically, this symmetry breaking ensures that the momentum is normal to the sphere. In this case, the inertial position is a relativistically-invariant position denoted by $X_\text{in}^I$ and given by $X_\text{in}^I(0)=-B^I/m$. This position is non-commutative, leading to the well-known statement that it is not possible to have a sharp relativistic localisation. This corresponds implicitly to the choice we have made in our construction.

The \emph{rest frame} corresponds to the choice $k^I=t^I$, where $t^I$ is a kinematical unit time-like vector. The corresponding position is not relativistically-invariant and is also non-commutative.

Finally, the \emph{Newton--Wigner frame} is obtained with the choice $k^I=P^I+mt^I$. It leads to the only position operator which is commutative and therefore admits an accurate localisation \cite{Newton:1949cq}. This Newton--Wigner position is given by
\be\la{NW}
X^I_\text{NW}(0)
&=\f{J^{IJ}(P_J+m t_J)}{P\cdot(P+mt)}
=X^I_\text{in}(0)+\f{P^I(X_\text{in}\cdot t)}{(m-P\cdot t)}-\frac{S_\text{in}^{IJ}t_J}{(m-P\cdot t)},
\ee
where in the second equality we have chosen the inertial frame as a reference to express the total angular momentum. It can be checked that $[X^I_\text{NW},X^J_\text{NW}]=0$ which shows that it is possible to chose a commutative position operator. However, this choice breaks Lorentz invariance. It would be interesting to understand what is the  gravitational interpretation of this commutative position.
  
The total angular momentum is also invariant under the $\text{SL}(2,\mathbb{R})$ rotations $z_a{}^I\to g_a{}^b z_b{}^I$ with $ g_a{}^{c}g_b{}^{d} \eps_{cd}=\epsilon_{ab}$. These $\text{SL}(2,\mathbb{R})$ transformations are generated by the metric 
\be
q_{ab}=m\,z_a{}^Iz_b{}^J\eta_{IJ}. 
\ee
In gravitational terms, the variables of the phase space \eqref{poincare theta} are related to the parametrization \eqref{JXPz} by
\be
\X^I=-mX^I,\q\se_a{}^I=\sqrt{m}\,z_a{}^I,\q \n^I=\frac{P^I}{m},
\ee
and the kinematical constraint $\n_a=0$ is the rest frame constraint $P_a=0$ (and $\n^2=-1$ is satisfied immediately). After imposition of this constraint, the boost generator corresponds to the inertial/relativistic position operator, while the frame corresponds to the spin oscillators. Explicitly we have
\be
B^I=-m\tilde{X}^I\coloneqq-mX^I-P^I\f{(X\cdot P)}{m},
\q 
S^I=\frac{1}{2}\eps^I{}_{JKL}(z_a^J\eps^{ab} z_b^K)\frac{P^L}{m} \,,
\ee
where $\tilde{X}^I$ such that $P_I\tilde{X}^I=0$ is the  position relative to the particle's world-line. When going back to the edge mode notation this gives indeed \eqref{tse} and \eqref{Sq-def}. Finally, recalling that in this analogy $\beta$ plays the role of the inverse mass, the simplicity constraint takes the form $C^I=B^I-mS^I=0$, and therefore simply fixes the relativistic position to be proportional to the spin\footnote{Note that we work in Planck unit in this section.} as
\be\la{Cparticle}
-\tilde{X}^I=S^I.
\ee
Interestingly, such a constraint is satisfied by the endpoint of an open string, which stretches more as it spins faster \cite{Zwiebach:2004tj}.

\subsection{Area versus boosted area }\label{Boosted area}

Since we are now in a relativistic setting, there are two different notions of rotation/spin operator which arise, namely the covariant generator $\S_I=*{\JJ}_{IJ}\n^J$ and the kinematical generator $L_I=*{\JJ}_{IJ}t^J$ associated to a choice of fiducial frame. Accordingly, there are three different scalars $(\S^2,L^2,L\cdot\S)$ which we can construct and which will play a role in the quantization. The goal of this section is to understand their geometrical interpretation. While the Poincar\'e spin $|\S|$ is proportional to the area of the sphere, the other spin numbers can be understood as boosted areas. 

We have seen that $\S^I$ is a vector normal to the frame, i.e. $\e_a{}^I\S_I=0$, and that we can decompose any vector in terms of the orthonormal frame
$(\S^I,\e_a{}^I,\n^I)$. With respect to this decomposition we have that
\be
\S^I=  \S^I,\q \BB^I=-\frac{ \widetilde{\mathsf{Q}}}{2 \S^2 }\S^I+\CC^a \e_a{}^I\,,
\ee 
where we recall that $\CC^a$ is the projection of the simplicity constraint along the frame. In the previous section we have shown that the classical simplicity constraints imply the conditions
\be\label{classicalism}
-\frac{ \widetilde{\mathsf{Q}}}{2 \S^2 }=\frac{1}{\beta}, \q \CC_aq^{ab}\CC_b =  0,
\ee
and that $|\S|=\beta\sqrt{q}$ is proportional to the 2-sphere area form.

A kinematical observer is, by definition, associated to a unit time-like vector $t^I$ with $t^2=-1$. This vector $t^I$ is not part of the corner phase space, and is assumed to commute with all the other fields. Given the kinematical observer picked by $t^I$, one can define its rotation generator $L_I=*\JJ_{IJ}t^J$ and its boost generator $K_I=\JJ_{IJ} t^J$. We have shown that the Poincar\'e spin $s=|\S|$ has a clear geometrical interpretation in terms of the area operator, as indicated by the relation \eqref{Sdet}. Now what we are interested in is the geometrical interpretation of the kinematical spin $j=|L|$.

At the quantum level this kinematical spin becomes the LQG spin $j$ associated with an internal \textit{auxiliary} time-like unit vector. The generator $L^2$ is what is commonly associated to the area operator in LQG, once the time gauge is used. However, it is now clear that, in general, the internal vector $t^I$ is not aligned with the internal normal $\n^I$, but instead defines a boosted observer with respect to $\n^I$. More precisely, given $t^I$ we can define
$t_a\coloneqq\e_a{}^It_I$, which represents the pull-back of the vector $t$ on $S$, and introduce a boost angle $\eta$ such that $t^aq_{ab}t^b=(\sinh \eta)^2 $. Then we can use the basis $(\S^I,\e_a{}^I,\n^I)$ to write the decomposition
\be\la{tI}
t^I=\cosh\eta\left(\n^I\cosh{\alpha}+\f{\S^I}{|\S|}\sinh{\alpha}\right)+t^{a} \e_a{}^I\,,
\ee
where we have used that ${\S^I}/{|\S|}$ is the second internal unit normal vector to the corner surface $S$. From the  expression \eqref{J1} of the total angular momentum and its dual, we then conclude that the kinematical rotation and boosts are given by
\begin{subequations}
\be
L^I&=\n^I(\S\cdot t)-\S^I(\n\cdot t)  - \teps^I{}_{JK} t^J \BB^K , \\
K^I&=\n^I(\BB\cdot t)-\BB^I(\n\cdot t) + \teps^I{}_{JK} t^J \S^K \,.
\ee
\end{subequations}
Assuming the validity of the covariant simplicity constraints \eqref{simpl-n}, using $\teps_{IJK}\S^J\e_a{}^K= |\S| \star \e_a{}^I$ which is shown in \eqref{*eSapp}, and denoting $ \star\,t^a =q^{ab} \star t_b$,    we get 
\begin{subequations}\label{JK}
\be
L^I&=s\big(\cosh\eta \, r^I +\star\, t^a \e_a{}^I /\beta \big),\\
K^I&=\f{s}{\beta}\big( \cosh\eta \, r^I- \beta\star t^a  \e_a{}^I \big)\,,
\ee
\end{subequations}
where we recall that the Poincar\'e spin is $|\S|=s$. Here we have introduced the vector $r^I$ given by
\be
r^I\coloneqq\n^I\sinh{\alpha}+\frac{\S^I}{|\S|}\cosh{\alpha}\,,
\ee
which is such that $r^2=1$ and $\e_a{}^I r_I=0=r^It_I$. It is clear from \eqref{JK} that the covariant simplicity constraints do not imply
the validity of the kinematical ones since we then find
\be\label{ksimplicity}
K^I-\f{L^I}{\beta}=-s\left(1+\frac1{\beta^2}\right) \star t^a  \e_a{}^I.
\ee
This is another way to see that the kinematical simplicity constraints  break internal Lorentz symmetry. Furthermore, one finds that the kinematical spin $L^2$ and the projected spin $L\cdot \S$ are both bigger than the dynamical spin $\S^2$, as
\begin{subequations}
\be
L^2 &=s^2 \left(\cosh \eta^2+ \frac{t_aq^{ab} t_b}{\beta^2}\right)=s^2 \left(1+ \left(1+\frac{1}{\beta^2}\right)\sinh \eta^2\right) \geq s^2,\la{JS}\\
L\cdot \S&=-\S^2 (\n\cdot t) = s^2 \cosh \eta \cosh\alpha\geq s^2. \la{JJS}
\ee
\end{subequations}
As we will see, these inequalities are also satisfied at the quantum level.

Geometrically we can understand $|L|$ as an area element associated to the plane normal to $t^I$ and $L^I$, while $L\cdot\S /|\S|$ can be viewed as an area element associated to the plane normal to $t^I$ and $\n^I$. These 2-dimensional planes are not necessary integrable, but when they are 
$|L|$ and $L\cdot \S /|\S|$ represent boosted areas (see \cite{Rovelli:2002vp} for a discussion concerning  boosted areas in quantum gravity).
At the quantum level, the inequality \eqref{JS} can be understood as a restriction on the spin numbers
\be
j \geq s\,,
\ee
where  the spin number $j=|L|$ represents the eigenvalue of the boosted or kinematical area operator, while $s=|S|$ represents the eigenvalue of the physical area operator. 

Finally, it is also possible to establish that the kinematical area $j$ is bounded from below by the Lorentz spin $k$, i.e. that $j\geq k$. This condition can easily be derived using simple relations between the $\SL(2,\C)$ Casimirs. First, using the expressions $Q=L^2-K^2$ and $\widetilde{Q}=-2K\cdot L$ for the Casimirs, we can rewrite the simplicity constraint \eqref{simp-kin} as
\be
-\f{\widetilde{Q}}{2L^2}=\f{K\cdot L}{L^2}=\f{1}{\beta}.
\ee
Taking the square of \eqref{ksimplicity} then tell us that
\be
s^2 \left(1+\frac1{\beta^2}\right)^2t^aq_{ab}t^b=K^2-\frac{(K\cdot L)^2}{L^2} = j^2-Q-\frac{\widetilde{Q}^2}{4 j^2} \geq 0\,.
\ee
For fixed $Q=k^2-\rho^2$  and $\widetilde{Q}=-2k\rho$, we have that this is an increasing function of $j^2$ which vanishes for $j=k$. 
This means that 
$j\geq k$. This is a classical equality which is known to hold also at the quantum level.

\section{Quantization of the corner algebra }\la{sec:discretization}

Our analysis, which continues that of \cite{Edge-Mode-II}, has revealed that the corner symmetry group of tetrad gravity with Barbero--Immirzi parameter contains a factor $\SL(2,\C)^S\times\SL(2,\RR)^S_\para$, where $\SL(2,\C)^S$ is the internal Lorentz group generated by $\JJ_{IJ}$, while $\SL(2,\RR)^S_\para$ encodes the non-commutativity of the corner metric components $q_{ab}$. This product symmetry group is restricted by the fact that the Poincar\'e spin Casimir is related to the $\SL(2,\RR)_\para$ Casimir $q$ as $\S^2=\beta^2q$. As we have seen, in terms of the Poincar\'e parametrization of section \ref{sec:Poin-interlude} this means  that the Poincar\'e Casimirs $P^2=-m^2$ and $S^2$ satisfy $W^2=-P^2S^2=q$, where $m=\beta^{-1}$. We have also shown that this symmetry group descends from a double symmetry breaking pattern. First, we have $\SL(2,\RR)_\para\ltimes H_3(\mathbb{R})\to\SL(2,\RR)_\para$, which comes from the imposition of the inertial frame constraints $P_a=\se_a{}^IP_I=0$, and then we also have the Poincar\'e symmetry breaking $\mathrm{ISO}(3,1)\to\SL(2,\C)$ coming from the imposition of the simplicity constraints.

In this section we would like to provide a preliminary analysis of the quantization of the corner symmetry algebra, in order to set the stage for future work.
One of our main claims is that a theory of quantum gravity necessarily provides us with a representation of the corner symmetry group $G[S]\coloneqq\mathrm{Diff}(S)\ltimes G^S$ with $G=\SL(2,\RR)\times\SL(2,\C)$. It is therefore of utmost importance to understand what are the representations of $G[S]$, since these are the building blocks of quantum gravity.

Let us first gather some notations useful for the description of $G[S]$. Its Lie algebra, denoted $\sg[S]$, is generated by three types of generators which are densities valued in the dual $\sg^*$. We have the momentum density\footnote{In the tetrad formalism it is given by $D_a=\gamma_a{}^{IJ}(*+\beta)(e\wedge e)_{IJ}$, where $\gamma_a^{IJ}$ is the torsionless connection \cite{Edge-Mode-II}. Note that the presence of $\beta$ leads to the notion of dual diffeomorphism charges, which were studied at infinity in \cite{Godazgar:2020kqd}.} $\DD_a(x)$ generating the diffeomorphisms, the angular momentum density $\JJ^{IJ}(x)$ generating $\SL(2,\C)^S$, and the densitized\footnote{Since $\epsilon^{ab}$ is a density and $q_{ab}$ is a tensor, $\K_a{}^b$ is a matrix-valued density.} metric $\K_{a}{}^b(x)\coloneqq\beta q_{ac}\epsilon^{cb}$ generating $\SL(2,\RR)^S$. We also have the spin density $\S^I(x)=\epsilon^{IJKL}\JJ_{JK}\n_L$ generating an $\SU(2)^S$ subalgebra of $\SL(2,\C)^S$. With this we can define the smeared generators
\be\la{dens-gen}
\DD(\xi)\coloneqq\int_S\xi^a(x)\DD_a(x)\,\rd^2x,\q\JJ(\alpha)\coloneqq\f{1}{2}\int_S\JJ^{IJ}(x)\alpha_{IJ}(x)\,\rd^2x,\q\K(a)\coloneqq\f{1}{2}\int_S\K_{a}{}^b(x)a_b{}^{a}(x)\,\rd^2x,
\ee
where $\xi$ is a smooth vector field on $S$, $\alpha$ is a map $\alpha:S\to \sll(2,\C )$, and $a:S\to \sll(2,\R )$ is a map from $S$ onto symmetric traceless $2\times 2$ matrices. The quantum algebra which we are interested to represent is\footnote{We use the map $[\cdot,\cdot]_q\to i\{\cdot,\cdot\}$, as in \eqref{ano}.}
\begin{subequations}
\be
[\DD(\xi),\DD(\xi') ]_q=i\DD([\xi,\xi']_{\mathrm{Lie}}),\q[\DD(\xi),\JJ(\alpha)]_q=i\JJ({\cal L}_\xi\alpha),\q[\DD(\xi),\K(a)]_q =i \K( {\cal L}_\xi a),\\
[\JJ(\alpha), \JJ(\alpha')]_q=i\JJ([\alpha,\alpha'])\q[\K(a), \K(a')]_q=i\K([a,a'])\q[\JJ(\alpha), \K(a)]_q=0,
\ee
\end{subequations}
where $[\cdot,\cdot]_q$ denotes the quantum commutator of operators, $[\cdot ,\cdot ]_{\mathrm{Lie}}$ the Lie bracket of vector fields, ${\mathcal L}_v$ the Lie derivative, and $[\cdot,\cdot]$ the matrix commutator. In the following we will denote $(\mathsf{d}_a,\jj^{IJ}, \k_a{}^b, \s_I )$ the undensitized versions of $\DD_a=\mathsf{d}_a \sqrt{q}$, $\JJ^{IJ}=\jj^{IJ} \sqrt{q}$, $\K_a{}^b= \k_a{}^b \sqrt{q}$, and $\S_I =\s_I \sqrt{q}$.

We have seen in this work that the simplicity constraints imply that all the Casimirs for the subgroups $\SL(2,\RR)^S$, $\SU(2)^S$, and $\SL(2,\C)^S$ are proportional to the area element. This  appears in \eqref{Sdet} and \eqref{simpler}, as well as in equation (6.30) of \cite{Edge-Mode-II}. Using \eqref{Sq-def} and \eqref{QQtilde} together with the definition $q=\f{1}{2}q_{ab}q_{cd}\epsilon^{ac}\epsilon^{bd}$, the statement is that there is the following relationship among the Casimir densities:
\be
\f12 \K_a{}^b  \k_b{}^a=- {\beta^2} \sqrt{q},\q \S^I \s_I=  \beta^2 \sqrt{q},\q \f12 \JJ^{IJ}  \jj_{IJ} =(\beta^2-1) \sqrt{q},\q\f12*\JJ^{IJ} \jj_{IJ} = -2\beta \sqrt{q}\,.
\ee
These relations are the algebraic expression of the \emph{simplicity} constraints. Importantly, they imply that an irreducible and simple representation of $G^S$ is entirely determined by the choice of a measure $\mu$ which diagonalizes the area element as $ \sqrt{q}\psi  =\mu \psi$ for states in $\CH_\mu$.
The main point is that the quantization of area means that the measure of Borel sets $D_p$ inside $S$ have to be quantized. Ignoring quantization ambiguities at this stage, we have that 
\be
\mu(D_p) \simeq\frac{\ell_\text{Pl}^2\lambda_p}{\beta} , \qquad \lambda_p \in \mathbb{N},
\ee
where we have reintroduced the Planck length. The measure $\mu(D_p) $ is expected to be quantized, and belongs to a discrete set\footnote{The exact nature of the discrete set depends on the details of the quantization. We can take $\mu_p=\sqrt{\lambda_p(\lambda_p-1)} $ where $\lambda_p$ is the weight of the representation.}  which asymptotes  $\mathbb{N}$, where the asymptotic evaluation is valid for regions whose measure is large with respect to the Planck area.

\subsection{Continuous representations}\label{contrep}
 
At the quantum level, and taking also the presence of diffeomorphisms into account, we want to build a corner Hilbert space in terms of the representation states of the corner symmetry group $\mathrm{Diff}(S)\ltimes G^S$ with $G= \SL(2,\RR)\times\SL(2,\C)$. The ultimate goal is to define a quantization procedure for these infinite-dimensional algebras. We want this quantization to be \emph{local}, i.e. to assign a notion of Hilbert space $\CH(U)$ to any open subset $U\subset S$ such that we have a factorization property
\be
\CH(U\cup V)=\CH(U)\otimes\CH(V),\q\mathrm{when}\q U\cap V=\emptyset,
\ee
and a compatibility with partial order in the sense
\be
\CH(U)\subset\CH(V),\q\mathrm{when}\q U\subset V.
\ee 
The representation of the infinite-dimensional algebra also needs to be \emph{covariant}, i.e. to carry a representation of the group of diffeomorphisms of the sphere. In other words, given such a diffeomorphism $f:S\to S$, we need to find quantum operators $\CO_f:\CH(S)\to\CH(S)$ such that
\be
\CO_f\big(\CH( U)\big)\subset\CH\big(f(U)\big),
\ee
for any open subset $U\subset S$.

Even if the proper study and classification of such infinite-dimensional representations is beyond the scope of this paper, we can present some element of the underlying representation theory and give an example of a smooth representation. First, let us pick a unitary representation $V_\rho$ of the group $G$, and define the Hilbert space $\CH_\rho=L^2(S, V_\rho)$. Elements of $\CH_\rho$ are \emph{half-densities} $\psi:S\to V_\rho$ with norm
\be
\|\psi\|^2=\int_S\langle\psi(x),\psi(x)\rangle_{\rho}\,\rd^2x,
\ee
where $\langle\cdot,\cdot\rangle_\rho$ denotes the Hermitian inner product on $V_\rho$. The point-wise norm $\langle\psi(x),\psi(x)\rangle_{\rho} $ defines a density on $S$. 

Given now two group elements\footnote{We can think of  the elements $f$ and $g$ as arising from the exponentiations $f=\exp\big(i \DD(\xi)\big)$ and $g=\exp\big(i\JJ(\alpha)\big)$ or $g=\exp\big(i\K(a)\big)$.} $g\in G$ and $f\in\mathrm{Diff}(S)$, the factors $G$ and $\mathrm{Diff}(S)$ of the corner symmetry group act on $\CH_\rho$ as
\begin{subequations}
\be
(g\vartriangleright\psi) (x)&=\rho\big(g(x)\big)\psi(x),\\
(f\vartriangleright\psi) (x)&=\psi\big(f^{-1}(x)\big)\sqrt{D_f(x)},
\ee
\end{subequations}
where we have denoted $\rho:G\to\mathrm{End}(V_\rho)$, and introduced the Jacobian $D_f(x)=\mathrm{det}(\rd f^{-1})(x)$. This latter appears because the states are half-densities. This defines a local and unitary representation of $\mathrm{Diff}(S)\ltimes G^S$. We can then construct more involved representations by taking the tensor products and defining
\be
\CH^{(N)}_\brho\coloneqq\bigotimes_{i=1}^N\CH_{\rho_i}=L^2\left(S^N,V_\brho\right),
\ee
where we denote $\brho=(\rho_1,\cdots,\rho_N)$ and $V_\brho=\bigotimes_{i=1}^NV_{\rho_i}$. Elements of the tensor product Hilbert space $\CH^{(N)}_{\brho}$ are functionals $\psi(x_1,\cdots,x_N) \in V_{\brho}$, on which the action of the corner group is naturally given by
\begin{subequations}
\be
(g\vartriangleright\psi)(x_1,\cdots,x_N)&=\bigotimes_{i=1}^N\rho_i\big(g(x_i)\big)\psi(x_1,\cdots,x_N),\\
(f\vartriangleright\psi)(x_1,\cdots,x_N)&=\psi\big(f^{-1}(x_1),\cdots,f^{-1}(x_N)\big)\prod_{i=1}^N\sqrt{D_f(x_i)}.
\ee
\end{subequations}
In order to obtain an \textit{irreducible} representation of the group of diffeomorphisms, we have to restrict the states to form a representation of the symmetric group when the representation labels are identical. In other words, if $(x_1,\cdots,x_N)$ carry the same representation label (say) $\rho$, we have to impose that 
\be
\psi(x_{\sigma_1},\cdots,x_{\sigma_N})=R(\sigma)\psi(x_1,\cdots,x_N),
\ee
where $R$ is a representation of the permutation group $\sigma_N$. This representation can be chosen to be either Abelian (leading to bosonic or fermionic representations) or non-Abelian (leading to para-fermionic representations). In the 2-dimensional case, there is also the additional possibility to consider braided statistics. The choice of statistics is a central ingredient of the entropy counting formula.

We have seen that $\|\psi(x_1,\cdots,x_N)\|^2$ is a density on $S^N$ which is absolutely continuous with respect to the Lebesgue measure. However, it is important to realize that associated to an irreducible representation we have another independent measure $\mu_\brho(x)$ on the sphere, given by the diagonalisation of $\sqrt{q}(x)$. Indeed, we have seen in this work that the local Casimirs of the $\sll(2,\R)^S$ and $\sll(2,\C)^S$ algebras are  both proportional to the measure density $\sqrt{q}$. Since $\sqrt{q}$ is a Casimir operator for $G^S$, it acts diagonally on irreducible representation of $G^S$. We denote this diagonal action $\mu_\brho(x)\psi = \sqrt{q}(x)  \psi$, for all $\psi \in \CH^{(N)}_{\brho}$.

This means the choice of representation $\brho$ is characterized by a choice of measure $\mu_\brho$ which represents the value of the operator 
$\sqrt{q}$ on $\CH_\brho$. Now, because the spectrum of the area operator associated with a finite region is quantized, the measure $\mu_\brho$ is a discrete measure which is not absolutely continuous with respect to the Lebesgue measure. In fact, the measure $\mu_\brho$ resembles the mass density of a collection of 2-dimensional particles, and we can write it as
\be\la{murho}
\mu_\brho(x) =\sum_{i=1}^N\mu_i \delta^{(2)}(x-x_i),
\ee
where the individual ``masses'' are given by the value of an $\sll(2,\R)$ Casimir for the discrete series as
\be
\mu_i=\beta^{-1}\sqrt{ \lambda_i(\lambda_i-1)},\qquad \lambda_i\in \mathbb{N}.
\ee

In order to evaluate the Casimirs, it is convenient to introduce, starting from the Lie algebra-valued density $\JJ^{IJ}$, the undensitized operator $\jj^{IJ}$ such that $\JJ^{IJ}=\jj^{IJ}\sqrt{q}$. The action of the densitized operators $\jj^{I_1J_1}\cdots \jj^{I_NJ_N} \sqrt{q}$ on the elements of $\CH^{(N)}_\brho$ is then given by 
\be
\big(\jj^{I_1J_1}(x)\cdots \jj^{I_NJ_N}(x)\sqrt{q}(x)\big)\vartriangleright\psi)(x_1,\cdots,x_N)=
\sum_{i=1}^N\frac{\delta^2(x-x_i) }{\mu_i^{N-1}}\rho_i\big(\tau^{I_1J_1}\cdots \tau^{I_NJ_N}\big) \psi(x_1,\cdots,x_N),
\ee
where $\tau^{IJ}$ denotes a basis of Lorentz Lie algebra.
We can obviously do a similar construction for the $\sll(2,\R)$ generators $K_a{}^b$.
This shows that one can introduce continuous representations of the corner symmetry algebra, without the need for a discretization. We will come back to this in a future publication, and show in \cite{Edge-Mode-IV} how to construct a Fock space representation of the corner algebra. We also refer to the work \cite{Wieland:2017cmf} for a (non-covariant) Fock quantization of an infinite-dimensional corner algebra. A first instantiation of the measure \eqref{murho} has been proposed in \cite{Freidel:2019ees} through a smearing along circles around the punctures and it led to a first notion of infinitesimal diffeomorphism operator on the corner.

Finally, let us point out that these continuum representations have the interesting possibility that we can now consider the limit of large spheres as a 
thermodynamical limit where the total area and the total number of elementary  excitations are sent to infinity while keeping their density fixed. This means that the continuous representation defined here can potentially be studied in a limit where
\be
A, N\to\infty,  \qquad A/N\ \text{fixed}.
\ee

Interestingly, another continuum limit can be achieved if we send the Barbero--Immirzi parameter to zero at fixed area. This is the limit in which we recover the metric formulation of gravity. Then, in the limit $\beta\to\infty$ and $N\to\infty $ with  $N/\beta \to\bar{\rho} $, the sums become Lebesgue integrals as
\be
\int_S \mu_\rho f\,\rd^2 x=\frac1{\beta} \sum_{i=1}^N f(x_i)\sqrt{\lambda_i(\lambda_i-1)}\q \to \q\bar{\rho}  \int_S f(x) \sqrt{\lambda(x)(\lambda(x)-1)}\,\rd^2 x. 
\ee
This is consistent with the interpretation of $\beta^{-1}$ as the Poincar\'e mass. The discrete area spectrum is then interpreted as a mass gap for a gas of 2-dimensional excitations, while the total area is interpreted as the total mass of this gas. In the limit where the fundamental excitations are massless, we can have an infinite number of them with fixed density and keep the mass finite. However, for non-zero $\beta^{-1}$ we can only have a finite number of excitations. We postpone the study of these limits to future work.

\subsection{Discrete subalgebras}\la{sec:subalgebras}

As another step towards the quantization of the corner symmetry algebra $\mathfrak{g}[S]$, and in order to relate it to the quantization used in traditional LQG, we can introduce a regularization procedure and the notion of coarse-grained subalgebra. Let us focus on the Lorentz component of $\mathfrak{g}[S]$ for definiteness. Given the corner charge density $\JJ^{IJ}(x)$, we introduce its smeared version
\be
\JJ(\alpha)=\f12 \int_S\alpha_{IJ}(x)\JJ^{IJ}(x)\,\rd^2 x,
\ee
where $\alpha_{IJ}(x)$ is a 0-form symmetry transformation parameter.
 
We then start by introducing the notion of \emph{admissible partition} of the corner surface $S$. We say that $P=\{D_1,\dots,D_N\}$ is an admissible partition of $S$ if there exists a collection $D_p$ with $p=1,\cdots,N$ of measurable subsets of $S$ such that
\be 
S=\bigcup_{p=1}^ND_p,\q\mathrm{and}\q D_p\cap D_q=\emptyset\ \mathrm{when}\ p\neq q.
\ee
Let us now assume that we have a collection of closed and disjoints disks $D_p\in S$, with $p=1,\cdots,N$. We can then consider smearing parameters $\alpha$ supported only on $\cup_pD_p$ and constant on each disk. These are such that
\be
\alpha_{IJ}(x)=\sum_{p=1}^N\alpha^p_{IJ}\chi_p(x),
\ee
where
\be\la{characteristic}
\chi_p(x)= 
\begin{cases}
	1\quad\text{if}\;x\in D_p\,,\\
	0\quad\text{otherwise}\,,
\end{cases}  
\ee
are the characteristic disk functions which satisfy $\chi_p\chi_q=\delta_{pq}\chi_p$. For this choice of piecewise-constant smearing parameters we have that 
\be\la{Jp}
\JJ(\alpha)=\sum_{p=1}^N\JJ_p(\alpha),\q\text{where}\q\JJ_p(\alpha)\coloneqq {\f12}\int_{D_p}\alpha^p_{IJ}\JJ^{IJ}(x)\,\rd^2 x\,=\alpha^p_{IJ}\JJ^{IJ}_p.
\ee
These generators satisfy a discrete version of the continuum surface algebra, where the local brackets \eqref{PoincJJ} are now replaced by
\be\la{PoincJJ-reg}
\{\JJ_p(\alpha),\JJ_{p'}(\alpha')\}=\delta_{pp'}\,\JJ_p([\alpha, \alpha'])\,,
\ee
This algebra is identical to the flux algebra appearing in LQG. The difference is that it is obtained here independently of a choice of bulk discretization. It appears instead as a natural finite-dimensional subalgebra of the corner symmetry algebra. We have thus replaced the infinite-dimensional corner symmetry algebra with a finite number of copies (one for each cell) of the $\sll(2,\C)$ algebra.

While the regularization procedure \eqref{Jp} provides a well-defined starting point for quantization of the generators, things are more subtle when it comes to the Casimirs of the algebra. Indeed, in this case we are faced with products of densities, which can therefore not be integrated on a 2-dimensional surface. One possibility is to concentrate on the \textit{square root of the Casimir} instead. For instance, the smeared version of the square root of the Poincar\'e spin Casimir $\S^2$, namely
\be\la{S2con}
\sqrt{\S^2(S)}=\int_S\sqrt{\S^{I}(x)\S^J(x)\eta_{IJ}}\,\rd^2x
\ee
is well-defined at the classical level. However, the formal local expression of the corresponding operator in the integrand above is badly divergent. This is the same problem we are confronted with when quantizing the area operator in LQG. In this case, the classical quantity of interest is given by the square of the densitized triad, which becomes an operator-valued distribution in the quantum theory and makes the area operator divergent. The way to deal with the corresponding divergent operator is through a point-splitting procedure \cite{Rovelli:1994ge, Ashtekar:1996eg}. More precisely, let us use the above-introduced discs to define the discretized Poincar\'e spin Casimir
\be\la{sp}
s_p\coloneqq\sqrt{\S_p^I\S_p^J\eta_{IJ}}\,,\q\text{with}\q\S_p^I\coloneqq\int_{D_p}\S^I(x)\,\rd^2 x\,.
\ee
This is now a well-defined object in the quantum theory, as it involves a product of integrals of a single density. The quantity $\beta^{-1}s_p$ corresponds to the area of the cell $D_p$. The smeared version of the square root of the Casimir $\S^2$ can then be written as the Riemannian sum
\be\la{S2dis}
\sqrt{\S^2_N(S)}=\sum_{p=1}^N s_p\,.
\ee
Notice, however, that this regularized expression \eqref{S2dis} converges to the continuum version \eqref{S2con} only in the limit $N\rightarrow\infty$
\footnote{This is easy to see already at the classical level. For this, consider a unit 2-sphere with coordinates $\theta\in[0,\pi]$ and $\varphi\in[0,2\pi]$. Let us use the time gauge $t^I=\delta^I_0$ and fix the remaining internal gauge so that the $\su(2)$ generator $J^3$ is aligned with the orthogonal radial direction. We can then write $J_p^3=\varepsilon_\theta\varepsilon_\varphi\sin{\theta_p}$, where $\varepsilon_\theta$ and $\varepsilon_\varphi$ are the two coordinate lengths of each of the $N$ plaquettes tessellating the unit 2-sphere. These can be defined as $\varepsilon_\theta=\pi /N_\theta$ and $\varepsilon_\varphi=2\pi/N_\varphi$, where  $N_\theta$ and $N_\varphi$ are two integers such that $N_\theta N_\varphi=N$ (they label respectively the number of plaquettes in the $\theta$ and $\varphi$ direction). We can thus write the discretized surface area as
\be
\text{Ar}_N(S)=\sum_{p=1}^NJ_p^3=\sum_{p_\varphi=1}^{N_\varphi}\sum_{p_\theta=1}^{N_\theta}\varepsilon_\varphi\varepsilon_\theta\sin{\left(\frac{\pi p_\theta}{N_\theta}\right)}=\frac{2\pi^2}{N_\theta}\cot{\left(\frac{\pi}{2N_\theta}\right)}\,,
\ee
which reproduces the continuum result $\text{Ar}(S)=4\pi$ only in the limit $N_\theta\rightarrow\infty$.
}. Without the square root, the continuum analog of the sum in \eqref{S2dis} is not well defined.

However, this LQG-like regularization cannot be applied to all the Casimirs in the corner symmetry algebra. For instance, the Lorentz Casimir $\mathsf{Q}$ is not positive semi-definite, and we therefore have to follow a different strategy. Another alternative is to divide the local Casimir double density by $\sqrt{q}$ and define the smeared quantity
\be\la{Qs}
{\mathsf{Q}(S)}\coloneqq\f{1}{2}\int_S\frac{\JJ_{IJ}(x)\JJ^{IJ}(x)}{\sqrt{q}} \,\rd^2 x\,.
\ee
This can then be regularized by a Riemannian sum in terms of the discrete generators \eqref{Jp} and \eqref{sp}, namely 
\be\la{Qs2}
{\mathsf{Q}_N(S)}=\sum_{p=1}^N {\mathsf{Q}_p}\,,\quad\text{with} \quad \mathsf{Q}_p\coloneqq\frac{\beta}{s_p}\JJ_p^{IJ} \JJ_{pIJ} \,.
\ee
We therefore see that the corner observable $\mathsf{Q}$ gets discretized in terms of the ``Casimir area density'' on each cell. At the quantum level, this then introduces a dependence on the Poincar\'e spin quantum number as well, in addition to the labels of the Lorentz irreducible representations. The implications of this new regularization derived from the continuum theory will be investigated in \cite{Edge-Mode-IV}.

We can do the same analysis for  the corner metric and its $\sll(2,\mathbb{R})_\para$ algebra. Since the components of the tangential metric are 0-forms, the local symmetry generators are not densities and the prescription \eqref{Jp} needs to be revisited with a bit more care. Proceeding like in  \eqref{dens-gen}, we can define  the set of discrete tangential metric generators
\be
\K_p(a)\coloneqq \frac{\beta}{2} \int_{D_p}a_{b}{}^{a}(x)q_{ac}(x) \epsilon^{cb}\,\rd^2x\,.
\ee
The local algebra \eqref{sl2R} then yields the discrete brackets
\be\la{qpqq}
\{\K_p(a),\K_{p'}(a')\}=\delta_{pp'}\,\K_p([a,a'])\,.
\ee


The picture which emerges from this construction is that of a partitioning of space in terms of 3-dimensional bubbles, as in \cite{Freidel:2018pvm, Freidel:2019ees}. Their boundaries are tessellated by 2-dimensional cells representing interfaces between neighboring 3-dimensional bubbles. Each cell carries a representation of an $\sll(2,\C)$ algebra corresponding to Lorenz transformations, and of an $\sll(2,\RR)_\para$ algebra corresponding to area-preserving diffeomorphisms. We come back to this picture in Section \ref{sec:twisted} below.


\subsection{Inductive limit representations}

We are now going to present another construction for the representations of the group $G^S$ \cite{gelfand_1982} which follows from an inductive limit and correspond to the LQG construction. Given an admissible partition $P=\{D_1,\dots,D_N\}$ such that $S=\cup_{p=1}^ND_p$, we denote by $G_P\subset G^S$ the subset of maps $S\to G$ which are piecewise-constant on $P$. Given $g \in G_P$, we then denote $g_p\in G$ its value on $D_p$. Obviously we have an isomorphism\footnote{Note that $G^N$ denotes the product group with $N$ copies of $G$, while $G^S$ denotes the surface group of maps $S\to G$.}
\be
G_P = G^N.
\ee  
Notice now that the set of admissible partitions forms a directed poset. This means that there exists a partial order on admissible partitions, where  $P_1 \leq P_2$ iff $P_1=\cup_{p} D_p$ and $P_2=\cup_{p,q} D_{pq}$ while $D_p=\cup_q D_{pq}$. Moreover, given two admissible partitions $P_1$ and $P_2$, we can always find a third one (their intersection, or common refinement) such that  $P_1\leq P_3$ and $P_2\leq P_3$. If $P_1\leq P_2$ we can define a natural embedding
\be
I_{P_1 P_2 }: G_{P_1}\to G_{P_2},
\ee
which is such that $I_{P_1 P_3 }=I_{P_1 P_2 }I_{P_2 P_3 }$, for $P_1\leq P_2\leq P_3$. We can then finally define the group $G^S$ to be the direct limit 
\be
G^S=\varinjlim G_P.
\ee

Similarly, we can construct representations of $G^S$ using an inductive limit of representations. For this, we first chose a measure $\mu$ on the sphere $S$, and given an admissible partition $P=\cup_pD_p$ we demand that $\mu(D_p)=\lambda_p$ where $\lambda_p \in \mathbb{Z}$. Moreover, to a given $\lambda_p$ we assign a representation $\rho(\lambda_p)$ of $\SL(2,\R)\times \SL(2,\C)$ with $\SL(2,\R)$ Casimir $\mu_p=\sqrt{\lambda_p(\lambda_p-1)}$. In other words we demand the measure of $D_p$ to be the \emph{weight} of the representation $\rho(\lambda_p)$. This representation is such that its Casimirs satisfy the balance relations \eqref{Sdet} and \eqref{simpler}, which enforce that the group Casimirs are directly determined by the measure. This is nothing but the expression of the simplicity constraints. We can therefore label the simple representations by the measure of the partition and 
chose the representation of $G_P$ given by
\be
V_P(\mu) = V_{\lambda_1}\otimes \cdots \otimes V_{\lambda_N}.
\ee
In order to define the limit we then need to chose an embedding 
\be
I_{P_1P_2}: V_{P_1}   \to V_{P_2}.
\ee
This embedding follows from the repeated use of the embedding
\be
 I_{\lambda_1\lambda_2}:V_{\lambda_1+\lambda_2} \to V_{\lambda_1} \otimes V_{\lambda_2},
\ee
which can be constructed using coherent state. Indeed, since $V_\mu$ carries a discrete series representation, we can consider coherent states $|\mu,z\rangle$ which are obtained by action of group elements on the ``vacuum'' $|\mu, \mu\rangle$ which is the lowest eigenstate for the elliptic generator. The map $I_{\lambda_1\lambda_2}$ is then simply given by 
\be
I_{\lambda_1\lambda_2}\big(|\lambda_1+\lambda_2, z\rangle\big) = |\lambda_1, z\rangle\otimes |\lambda_2, z\rangle.
\ee
With these embedding maps the set $V_P$ defines a directed poset and we can consider the direct limit
\be
{\cal H}_\mu =  \varinjlim V_{P}(\mu).
\ee
The fact that the weights $\lambda_p$ are quantized means that the measure $\mu$ is a discrete measure such that $\mu(x)=\sum_{p=1}^N\lambda_p \delta^2(x-x_p)$. This shows that the representations obtained by a direct limit are more singular than the continuous representations described in section \ref{contrep}, which involved absolutely continuous measures for states. 

Notice that representations obtained by inductive limits are not differentiable, and therefore only provide representations of the subgroup $G^S$ but not of the diffeomorphism subgroup. This follows from the fact that, unlike $G^S$, $\text{Diff}(S)$ is not known to be obtainable as an inductive limit group. In particular, the inductive limit representations do not have a well-defined action of the momentum operator. We conjecture that the continuous representations defined in Section \ref{contrep} above admit a representation of the quasi-local energy, while the direct limit ones are too singular to have a well-defined energy operator. This however needs to be investigated further.

\subsection{Twisted geometries reconstruction}
\la{sec:twisted}

We have introduced in \eqref{framerot} the angle $\theta$ as the missing geometrical data necessary in order to reconstruct the 6 components of the corner coframe field $\e_a{}^I$ from the spin operator and the tangential metric components defined in \eqref{Sq-def}. We also mentioned there that this angle has an interpretation in terms of the twist angle of twisted geometries \cite{Freidel:2010aq,Freidel:2010bw}. We now want to study this in more details, and show how an extension of the discrete twisted geometry picture can be recovered from the boundary parametrization we have constructed in the continuum, by analogy with the discretization of space in terms of bubble networks introduced in \cite{Freidel:2018pvm, Freidel:2019ees}. This will allow us to introduce the notion of bulk holonomy, which also endows with a geometrical interpretation the states used to represent the corner symmetry algebra at the quantum level on a single cell decomposition of the surface.

Twisted geometries were introduced as an extension of Regge geometries, where the gluing between two neighboring polyhedra is done while relaxing a geometrical condition known as shape-matching. To understand how they are parametrized, let us consider a partition of space into flat polyhedra, along with the dual oriented graph. In this graph the vertices are dual to the polyhedra themselves, and the links are dual to the boundary faces. Each oriented link starts and ends at a vertex, respectively called source $s$ and target $t$. Let us consider a single link $e$ dual to a face. The two polyhedra (dual to) $s$ and $t$ carry their own reference frame, which induce in general two different normals to the face (dual to) $e$. Twisted geometries assign a real number $j_e\in\mathbb{R}$ to the edge, corresponding to the oriented area of the dual face, a unit normal vector $N_e^s\in\RR^3$ to the face as seen from the source polyhedron, and similarly a unit normal vector $N_e^t\in\RR^3$ to the face as seen from the target one (see Figure \ref{fig:Twisted1}). In this way, each link is assigned a triple of data $(j_e,N_e^s,N_e^t)$. Compatibility of this geometrical information then requires the existence of an $\SU(2)$ group element $g_e$ rotating one normal into the other, namely
\be\la{NR}
N_e^t=R(g_e)N_e^s\,,
\ee
with $R$ the rotation matrix in the adjoint representation. It turns out that in order to fully reconstruct from this condition \eqref{NR} the connection $g_e$ defining a notion of parallel transport between the two polyhedra, an extra angle $\theta_e\in[-\pi,\pi]$ needs to be added to the set of geometrical data assigned to the link. Such an angle encodes the component of $g_e$ corresponding to rotations along the $N_e^s$ axis. It was shown in \cite{Freidel:2010aq} that the 6-dimensional space of variables $(N_e^s, N_e^t, j_e,\theta_e)$ associated to a link defines a phase space which is symplectomorphic to $T^*\SU(2)_e$, namely to the non-gauge-invariant\footnote{For the sake of our discussion here, it is not necessary to take into account the closure constraint leading to the gauge-invariant phase space.} phase space of LQG on the link $e$. In this phase space, $(j_e,\theta_e)$ are conjugated variables. Twisted geometries represent a generalization of Regge geometries in the sense that, while the area of the shared face is the same, the two flat metrics induced on the face from the source and the target polyhedra are distinct and the geometries can therefore have a different shape. As pointed out in \cite{Haggard:2012pm}, an $\SL(2,\RR)$ transformation is needed in order to match the geometry on the face as seen from both sides. A Regge geometry is then recovered if we demand that the lengths of the edges of the face as seen from both sides to be the same. In the case of a triangulation, this amounts to fixing the  $\SL(2,\RR)$ group element between the two metrics to be trivial. The connection between the extra $\SL(2,\RR)$ transformation entering the geometrical data of twisted geometries and the corner metric algebra was first pointed out and elaborated on in \cite{Freidel:2018pvm}, although the implication of this were not not fully explored.

\begin{figure}[h!]
 \centering
 \begin{subfigure}[t]{70mm}
 \centering
\includegraphics[height=40mm]{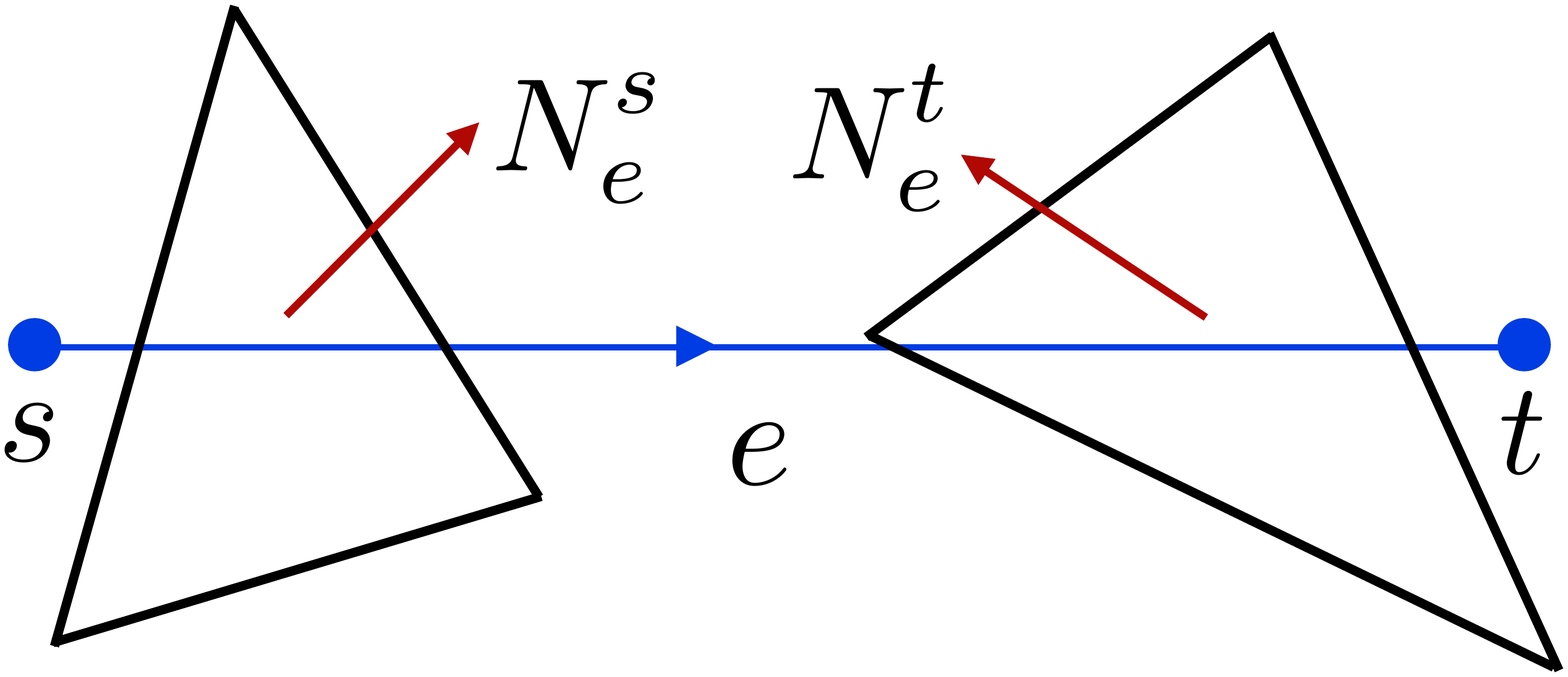}
\caption{The two induced flat metrics give the triangular faces the same area but different shapes. The gluing generates a discontinuous metric across the face.}
\la{fig:Twisted1}
\end{subfigure}
\hspace*{10.mm}
\begin{subfigure}[t]{70mm}
 \centering
\includegraphics[height=40mm]{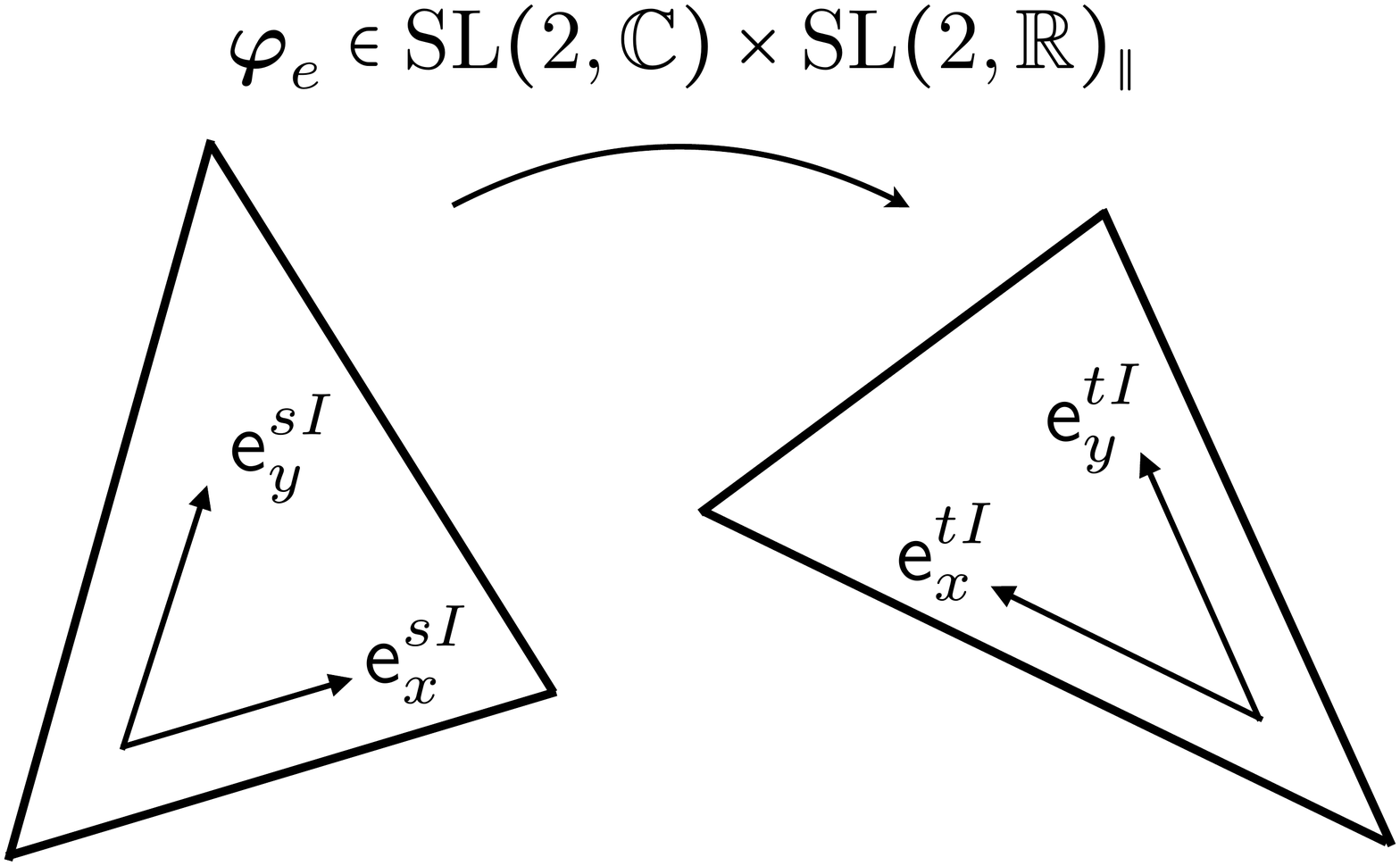}
\caption{The induced bubble geometries on the two sides can be related by a rotation generated by the group element $\boldsymbol{\varphi}_e\in \SL(2,\C)\times \SL(2, \RR)_\para$.}
\la{fig:Twisted2}
\end{subfigure}
\caption{Twisted geometry of a face and its  extension.}\la{fig:Twisted}
\end{figure}

Let us now explain how to recover a covariant extension of this discrete construction from our parametrization \eqref{Sq-def} of the corner phase space. For this we focus on a single face/patch of the regularization introduced above, that can be understood as a disk in the partition of the corner surface used for the definition of a discretization of the corner symmetry algebra in Section \ref{sec:subalgebras}. As in the analysis of \cite{Freidel:2018pvm}, we do not have to restrict to a piecewise-flat corner metric. 

Using  the definitions \eqref{sp} (and replacing the label $p$ of the face with the one of the dual edge $e$), the analog of the twisted geometry data on the dual link $e$ is now given by the set $(\S^{sI}_e/|\S|,\S^{tI}_e/|\S|,s_e,\theta_e)$. The two normals are now unit vectors in $\RR^{1,3}$, and they are related by the $\SL(2,\C)$ rotation
\be\la{glue1}
\frac{\S^{tI}_e}{|\S|}=\vphi^I_{eJ}\frac{\S^{sJ}_e}{|\S|}\,.
\ee
The twist angle $\theta_e$ necessary in order to reconstruct part of the group element $\vphi_e$ not fixed by the requirement \eqref{glue1} is given by $\theta_e\coloneqq\theta_e^t-\theta_e^s$, i.e. by the difference between the two angles parametrizing the rotational freedom \eqref{framerot} in the $\e_I^s$ and $\e_I^t$ edge mode frames respectively.\footnote{Due to a relative minus sign between the symplectic structures \eqref{TS1} associated to the source and the target faces, reflecting an orientation flip between the interior and exterior of a given bubble, we still have the conjugacy between $s_e$ and $\theta_e$.}

In addition, we also have the tangential metric data $q^e_{ab}$ associated to each link. The corresponding $\SL(2, \RR)$ group elements $\rho_e$ are the generators of the area preserving diffeomorphisms
\be\la{glue2}
{\e^t}_{a}{}^I=\rho_{ea}{}^{b} \,{\e^s}_{b}{}^I\,.
\ee
In order to elucidate the nature of the group elements $\vphi_e$ and $\rho_e$, it is useful to consider the compatibility of the frame fields induced on the face from the bulk frame fields on the two sides. This gives us also the opportunity to show how the edge modes arise from a dressing of the bulk fields when pulled back on the corner, in analogy to the treatment introduced in \cite{Donnelly:2016auv} and developed in \cite{Edge-Mode-II}. Given the bulk frames $(\tilde{e}_I^{s},\tilde{e}_I^{t})$ associated respectively to the source and the target bubbles, the edge modes $\e_a{}^I$ can be obtained from their pull back on the shared face by a rotation generated by the two group elements $\bvphi_e^s$ and $\bvphi_e^t$ as
\be
{\te^s}_{a}{}^I=(  \e^{s} \cdot \bvphi_e^s )_{a}{}^I\,,\q
{\te^t}_{a}{}^I=( \e^{t} \cdot \bvphi_e^t )_{a}{}^I \,.
\ee
The gluing of bubbles imposes the continuity of the bulk frame field across the boundary, namely the condition $\tilde e_I^{s}=\tilde e_I^{t}$ on the face, which as shown on Figure \ref{fig:Twisted2} implies the following relation between edge mode frame fields:
\be\la{glue3}
{\e^t}_{a}{}^I=({\e^s}\cdot\bvphi_e)_{a}{}^I\,\q\text{with}\q\bvphi_e=\bvphi_e^s\left(\bvphi_e^t\right)^{-1}\,.
\ee
As derived and explained in \cite{Edge-Mode-II}, the element $\bvphi_e=(\vphi_e, \rho_e)$ belongs to the group $\SL(2,\C)\times\SL(2,\RR)_\para$ and acts as with $({\e}\cdot\bvphi)_{a}{}^I = \rho_a{}^b \,\e_b{}^J \vphi_J{}^I$, which corresponds to the gluing conditions \eqref{glue1} and \eqref{glue2}. We therefore see that the edge modes provide a unified framework where the different insights about a generalized discrete geometry construction coming from \cite{Freidel:2010aq, Haggard:2012pm, Freidel:2018pvm} are clarified and implemented.

The resulting picture is that of a 3-dimensional geometry constructed by partitioning space in terms of bubbles connected to each other through interfaces represented by boundary cells. The geometrical data in the bulk of a given bubble is encoded in the edge modes living on its boundary cells. The gluing of two neighboring bubbles is encoded in a continuity condition between the two bulk coframes across the shared face. From the point of view of the edge mode geometrical data, this continuity condition translates into a transformation relation \eqref{glue3} between the edge modes induced from the bulk source frame and target frame. This transformation belongs to the corner symmetry group $\SL(2,\C)\times\SL(2, \RR)_\para$. In the quantum theory, its representation quantum numbers label all the possible configurations of quantum geometry compatible with the gluing (i.e. all the possible different ways of gluing), providing a local holographic repackaging of the fundamental degrees of freedom of quantum geometry.

Let us conclude with a remark on the nature of the two components of the corner symmetry group. The $\SL(2,\C)$ component with element $\vphi_e$ can be understood in terms of parallel transport between bubbles along the link dual to a face they share. This provides a covariant extension of the twisted geometry phase space introduced in \cite{Freidel:2010aq}, with the phase space parametrized by the set $(\S^s_{eI}/|\S|,\S^t_{eI}/|\S|,s_e,\theta_e)$ symplectomorphic to $T^*\SL(2,\C)_e$. The $\SL(2,\C)$ algebra component of this phase space corresponds to the algebra of corner Dirac observables generated by the Lorentz symmetry charges \eqref{J1}. These have associated gauge charges, vanishing on-shell due to the Gauss law, which are canonical generators of gauge transformations both in the bulk and the corner phase space, as throughly explained in \cite{Edge-Mode-II}. Note that $\SL(2,\C)$ twisted geometries have also been studied in \cite{Livine_2012}.
 
On the other hand, the $\SL(2,\RR)_\para$ component with element $\rho_e$ is more subtle, as it doesn't have an immediate interpretation in terms of a bulk holonomy. In fact, the $\SL(2, \RR)_\para$ transformations represent area-preserving diffeomorphisms of the discretized bubble surface and the associated generators, namely the metric components defined in \eqref{Sq-def}, have a purely boundary nature. What we mean with this is that, at this stage, we do not have a corresponding gauge symmetry in the bulk canonically generated by a constraint. In particular, we are lacking a bulk conservation law for the $\sll(2,\RR)_\para$ corner charges, as the Gauss law provides for the Lorentz charges. Because of this, the 4-dimensional case which we are studying differs crucially from the 3-dimensional one, where the analog of the $\SL(2, \RR)$ generators can be expressed in terms of the Lorentz generators and all the corner symmetry charges have a bulk gauge counterpart \cite{Geiller:2017whh, Freidel:2018pbr}.

\section{Conclusions}\la{sec:conclusions}

In the previous papers \cite{Edge-Mode-I, Edge-Mode-II} of this series, we have proposed to study the corner symmetry algebra of various formulations of gravity as a guiding principle towards a local holographic formulation of quantum gravity. In particular, in \cite{Edge-Mode-II} we have focused on the corner symplectic structure of the tetrad formulation of gravity with the Barbero--Immirzi parameter. We have revisited and extended the usual LQG analysis in two ways. First by relaxing the requirement of the time gauge and allowing the corner symmetry to include the Lorentz group and not only its rotation subgroup. Second by letting go of the discretization procedure and obtaining the necessary non-commutativity of boundary observables directly in a continuum formulation.
We have also  revealed an $\sll(2,\RR)_\para$ algebra associated with the corner metric, which explains the origin of the discreteness of the area spectrum, and  set the stage for the proper study, from the point of view of the corner, of the simplicity constraints. In local holography, we propose to quantize the geometry through the quantization of its corner symmetry algebra. This gives an extension of LQG, which can be viewed as a theory of the quantum boundary $\SU(2)$ fluxes. 

In the present paper, we focused on the thorough analysis of the simplicity constraints, as seen from the corner. The non-commutativity of the simplicity constraints has been the source of much confusion in the literature. These are resolved by shifting the emphasis to the corner where the non commutativity of fluxes appears naturally. Our analysis's key observation was to recognize that the internal normal field is a dynamical variable of the corner phase space and that the coframe field is non-commutative on the corner. This is encoded in the corner symplectic potential \eqref{BF potential}. It unravels the existence of an elemental Poincar\'e algebra describing the corner symmetry before the imposition of the simplicity constraints. In this picture, a suggestive particle-like description of (quantum) geometry emerges, where the internal normal plays the role of the 4-momentum, the Barbero--Immirzi parameter that of the mass, the flux that of a relativistic position, and the frame that of a spin harmonic oscillator.

The phase space structure \eqref{BF potential}  allowed us to reveal the second class nature of the corner simplicity constraints already at the classical and continuum level, and to perform a proper separation of the simplicity constraints into first and second class components. The study of this corner symplectic structure has also proven to be crucial to reconcile the imposition of the simplicity constraints with the discrete nature of the area spectrum and internal Lorentz invariance. Contrary to common claims in the literature, we have proven that the imposition of the simplicity constraints breaks down the Poincar\'e symmetry to a Lorentz symmetry. After imposing  the simplicity constraints, all Lorentz generators constitute strong Dirac observables and can be represented at the quantum level (this last point will be explicitly shown in \cite{Edge-Mode-IV}). Furthermore, the corner area generator corresponds to the Poincar\'e spin Casimir and is thus Lorentz-invariant by construction. This resolves a long standing puzzle of LQG, namely restoring compatibility between the spin foam manifestly Lorentz invariant construction in the bulk with the  Hilbert space of the canonical theory on the corner. This reinforces the conceptual and technical basis of spin foams. 

In addition to the Lorentz sector, we have established that the corner phase space is also characterized by a second set of strong Dirac observables, corresponding to the tangential metric components and satisfying an $\sll(2, \RR)_\para$ algebra. This extra symmetry algebra plays a crucial role in providing a generalized twisted geometry interpretation of a proper discretization of the corner  geometrical data. The full extent of the physical nature of these new charges will most likely require a complete quantum reconstruction of the frame field within the corner  Hilbert space carrying a representation of the $\SL(2,\mathbb{C})^S \times \SL(2,\mathbb{R})^S_\para \times \text{U}(1)^S_\para$ algebra of the corner  Dirac observables. This is the topic of a forthcoming paper in the series.

Interestingly, we have also explained how the BF phase space \eqref{BF potential} before the imposition of the simplicity constraints can itself be embedded in a larger phase space, namely \eqref{poincare theta}, which contains Poincar\'e and Heisenberg symmetries. This phase space is obtained from the BF one by relaxing what we called the  kinematical constraints. There is then a double pattern of symmetry breaking. First, the imposition of the kinematical constraints, which pick out a notion of Poincar\'e position and momentum, reduce \eqref{poincare theta} to BF theory \eqref{BF potential}, and then the simplicity constraints further reduce this latter to tetrad gravity. Again, this reveals the extent of the geometrical information which is encoded at the corner.

A powerful feature and main motivation of our program is the focus on the representations of the corner symmetry algebra and the properties of its Casimirs. Looking at it in the sense of the Kirillov orbit method \cite{Kirillov} provides a very efficient and reliable way to gain important insights into  aspects of the quantum theory while remaining in a classical framework. This pre-quantization analysis  reveals the structure of the spectra of geometrical operators even before a Hilbert space is constructed out of a given regularization procedure. 

Finally, let us end with a comment on an interesting issue, which is that of the choice of statistics for the corner excitations. This issue manifests itself most notably in the context of the LQG black hole entropy calculation, through an interplay with Chern--Simons theory (see \cite{DiazPolo:2011np} for a review), although the nature of this question is more general and far reaching. In the context of the LQG black hole entropy calculation, the punctures represent  the end points of spin network links piercing the horizon surface, where  sources of curvature and electric flux are at the same time concentrated due to the  boundary condition which is used to characterize the horizon. This discrete set of charges defined on tangential small disks around the punctures, as in the truncation \eqref{Jp}, can be understood as horizon hairs of quantum geometry contributing to the black hole entropy. It was originally argued in \cite{Krasnov:1996tb,Rovelli:1996dv} that the punctures should be considered as \textit{distinguishable}, while e.g. \cite{Ashtekar:2000eq} treats them as being \textit{indistinguihable}. In both cases though, an open question is  that of the statistics which should be assigned to these excitations. 
In fact, in the so-called ``gas of punctures'' approach, a bosonic statistics is usually assumed \cite{Ghosh:2004wq, Ghosh:2006ph, Ghosh:2011fc}. However, there  are also indications coming from alternative treatments  of the quantum horizon geometry \cite{Sahlmann:2011xu, Pithis:2014uva} that the corner punctures should obey anyonic statistics. At the same time, by taking into account an holographic degeneracy of matter states contribution to the partition function, inconsistency of distinguishability with semi-classicallity has been advocated in \cite{Ghosh:2013iwa}, where both bosonic and fermionic quantum statistics were analyzed.

It is clear that a detailed analysis of this issue requires an understanding of the action of diffeomorphisms on the boundary. This is intimately related to the issue of the quantization and representation of the continuous and infinite-dimensional corner symmetry algebra. We have presented in Section \ref{sec:discretization} preliminary ideas in this direction. In particular we have shown that it is possible to consider continuum representations acting on smooth states, and to give a representation of the corner symmetry algebra that possesses in principle an infinitesimal action of the diffeomorphism generator (see also \cite{Freidel:2019ees}) . This central feature is not available in the representations obtained by inductive limits. The technical details of this construction are left for future investigation.


 
\section*{Acknowledgement}

Research at Perimeter Institute is supported in part by the Government of Canada through the Department of Innovation, Science and Economic Development Canada and by the Province of Ontario through the Ministry of Colleges and Universities. This project has received funding from the European Union's Horizon 2020 research and innovation programme under the Marie Sklodowska-Curie grant agreement No 841923.

\appendix

\section{Corner  Poisson brackets}
\la{App:Brackets}

In this appendix we compute the various Poisson brackets used throughout Section \ref{sec:Bound-PS}. For simplicity, in all the appendices we do not include the factor $\delta^2(x,y)$ that enters the expressions since it always factors outside. We also keep the genre of the internal normal unspecified, and take $\n^2=\sigma$. The Poincar\'e potential we start with is then
\be\la{Poincare potential sigma}
\bTh_\Poin^S=\int_S\left(-\sigma\X_I\delta\n^I-\f{\beta}{2}\se_I\wedge\delta\se^I\right),
\ee
which gives the brackets
\be\la{poincare brackets sigma}
\{\X^I(x),\n^J(y)\}=-\sigma\eta^{IJ}\delta^2(x,y),
\q
\{\se_a{}^I(x),\se_b{}^J(y)\}=-\f{1}{\beta}\epsilon_{ab}\eta^{IJ}\delta^2(x,y),
\ee
and the brackets \eqref{basic poincare brackets} are recovered for $\sigma=-1$.

\subsection{Boost and frame algebra}
\la{App-Boost}

In this appendix we establish that the boost and the frame operators $(\BB_I,\e_a{}^I)$ defined in \eqref{tse} commute with the kinematical constraints $\n^2-\sigma$ and $\n_a=\se_a\cdot \n$, and we also determine their algebra. First, we define
\be
\e_a{}^I\coloneqq\se_a{}^I -\sigma  \n^I (\se_a\cdot \n),\q
\tilde\X^I\coloneqq \X^I -\sigma \n^I (\X\cdot \n).
\ee
It follows immediately from \eqref{poincare brackets sigma} that
\be
\{\tilde{\X}^I,\n^J\}=-\sigma\tilde\eta^{IJ},
\ee
where $\tilde\eta^{IJ}\coloneqq\eta^{IJ}-\sigma \n ^I \n ^J$, and therefore we get that $\tilde{\X}^I$ commutes with $\n^2-\sigma$. Then we have
\be
\{ \se_a{}^I , (\se_b\cdot \n ) \}= -\frac1{\beta} \epsilon_{ab} \n ^I,\q\q
\{ (\se_a \cdot \n ) , (\se_b\cdot \n ) \}=-\frac{\sigma}{\beta} \epsilon_{ab},
\ee
which implies that
\be\la{tena}
\{ \e_a{}^I , (\se_b\cdot \n ) \}= \{ \se_a{}^I , (\se_b\cdot \n ) \} -\sigma \n ^I \{ (\se_a \cdot \n ) , (\se_b\cdot \n ) \} = 0,
\ee
and shows that $\e_a{}^I$ commutes with $\n^2-\sigma$ (trivially) and $\se_a\cdot\n$. Moreover, it satisfies the algebra
\be\la{eeapp}
\{ \e_a{}^I ,\e_b{}^J\} =-\frac{1}{\beta}\epsilon_{ab}  \tilde\eta^{IJ}.
\ee
Next, we have the brackets
\begin{subequations}\la{Be}
\be
\{ \X^I, \e_b{}^J\}&= -\sigma \{\X^I, \n ^J (\se_b\cdot\n)\}=  \eta^{IJ} (\se_b\cdot\n) + \n ^J \se_b{}^I,\\
\{ \tilde\X^I, \e_b{}^J\}&=  \tilde \eta^{IJ} (\se_b\cdot\n)  + \n ^J \e_b{}^I,\\
\{\epsilon^{ac} \e_a{}^I (\se_c\cdot\n), \e_b{}^J \}&=-\frac{1}{\beta}\tilde \eta^{IJ} (\se_b\cdot\n),
\ee
\end{subequations}
which implies that for\footnote{Note that this is $\BB^I=\tilde{\X}^I+\beta\se^I\wedge \underline{\n}$ written on-shell of the constraint $\e_a{}^I=\se_a{}^I -\sigma  \n^I (\se_a\cdot \n)$.} $\BB^I=\tilde{\X}^I+\beta\e^I\wedge \underline{\n}=\tilde{X}^I+\beta\epsilon^{ac} \e_a{}^I (\se_c\cdot\n)$ we find
\be\label{Be bracket}
\{ \BB^I, \e_b{}^J\}&=\e_b{}^I\n^J.
\ee
One can now evaluate
\begin{subequations}
\be
\{ \tilde\X^I, (\se_b\cdot \n ) \}&= -\sigma  \se_b{}^I +  \n ^I (\se_b\cdot \n ) =-\sigma \e_b{}^I,\\
\{ \epsilon^{ac} \e_a{}^I (\se_c \cdot \n ) , (\se_b\cdot \n ) \} &=\epsilon^{ac} \{  \e_a{}^I  , (\se_b\cdot \n ) \}(\se_c \cdot \n ) + \epsilon^{ac} \e_a{}^I \{ (\se_c \cdot \n ) , (\se_b\cdot \n ) \}\cr
&=-\frac{1}{\beta}\epsilon^{ac}  \epsilon_{ab} \n ^I (\se_c \cdot \n )-\frac{\sigma}{\beta} \epsilon^{ac} \se_a{}^I \epsilon_{cb} ,\cr
&=\frac{\sigma}\beta \se_b{}^I -\frac{1}{\beta} \n ^I (\se_b \cdot \n )\cr
&=\frac{\sigma}\beta \e_b{}^I\,,
\ee
\end{subequations}
and summing the two identities we see  that $\BB^I$ is also a Dirac observable. Now, using again the notation $\e^J\wedge \underline{\n}\coloneqq \epsilon^{ab} \e_a{}^J (\se_b\cdot \n )$, let us also evaluate the brackets
\begin{subequations}
\be
\{\tilde\X^I,\tilde\X^J\}&= \tilde\X^I \n^J - \tilde\X^J \n^I,\\
\{\tilde\X^I, \e^J\wedge  \underline{\n} \}&=\sigma \e^I\wedge \se^J=  (\e^I\wedge  \underline{\n}) \n ^J +\sigma \e^I\wedge \e^J, \\
\{\e^I\wedge  \underline{\n}, \e^J\wedge  \underline{\n} \}&= -\frac{\sigma}{\beta} \e^I\wedge \e^J\,,
\ee
\end{subequations}
which imply that 
\be\la{BB2}
\{\BB^I,\BB^J\}= \BB^I \n^J - \BB^J \n^I + \sigma \beta \e^I\wedge \e^J\,.
\ee
Furthermore, we have
\be
\{\BB^I, \se_c{}^K\}&=\beta\n^J\epsilon^{ab}\{\se_{a}{}^I\se_{b}{}^J,\se^K_c\}\cr
&=-\n_J\epsilon^{ab}(\se_{a}{}^I \epsilon_{bc}\eta^{KJ}+\se_{b}{}^J\epsilon_{ac}\eta^{KI})\cr
&=\n^K\se_{c}{}^I-\n_c\eta^{KI}\,.\la{Be}
\ee
Finally we see that 
\be\la{Bnapp}
 \{\BB_I, \n^J\}
 &=\{\tilde\X_I, \n^J\}\cr
 &=  \{\X^I, \n^J\}  -\sigma \n^I \{(\X\cdot \n), \n^J\}\cr
 &=  -\sigma \eta^{IJ}+\sigma \n^I \n^J\cr
 &=-\sigma \tilde\eta^{IJ}\,,
\ee
which means once again that 
\bea\la{Beapp}
\{\BB^I, \n_a\}=0,\q \{\BB^I, \e_a{}^J\}=\e_a{}^I \n^J\,.
\eea

\subsection{Spin and frame algebra}\la{App-Spin}

Here we focus on the brackets involving the spin generator
\be
\S_I=\frac{\beta}{2} \teps_{IJK} \e_a{}^J \e_b{}^K \eps^{ab}\,. \label{defS}
\ee
We can use that 
\be\la{S as ee app}
\teps_{IJ}{}^K \S_K = \frac{\beta}{2} \teps_{IJ}{}^K\teps_{KAB}\e^A\wedge \e^{B}
= -\sigma \beta\e_I\wedge \e_{J}
\ee
to rewrite the  bracket \eqref{BB2}  as 
\be 
\{\BB^I,\BB^J\}= \BB^I \n^J - \BB^J \n^I - \teps^{IJK} \S_K=-\JJ_{IJ}\,.\la{BBapp}
\ee
The bracket among spin generators is given by 
\be\la{SSapp}
\{\S_I,\S_J\}
&=
\beta^2 \teps_{IAB} \teps_{JCD} \e_a{}^A \e_{c}{}^{C} \{\e_b{}^B,\e_{d}{}^{D}\} \epsilon^{ab}\epsilon^{cd}\cr
&=-\beta \teps_{IAB} \teps_{JCD} \eta^{BD} \e_a{}^A \epsilon^{ac} \e_{c}{}^{C} \cr
&=\sigma \beta (\eta_{IJ} \eta_{AC} -\eta_{IC}\eta_{JA}) (\e^A\wedge \e^{C})\cr
&=\sigma \beta (\e_I\wedge \e_J)\cr
&=- \teps_{IJ}{}^K \S_K.
\ee
The spin  acts on the frame by rotation, namely
\be\la{Seapp}
\{\S_I, \e_b{}^J\}=  \beta \teps_{IAB}\e_a{}^A  \{\e_{c}{}^B, \e_b{}^J\} \epsilon^{ac}=-\teps_{IAB} \e_a{}^A\tilde{\eta}^{BJ} \epsilon_{cb} \epsilon^{ac} =  \teps_{IA}{}^J \e_b^A.
\ee
To compute the bracket between the boost and spin components we use
\be
\{\tilde\X_I,\S_J\}
&=\frac{\beta}{2}\eps_{JABC}  \{ \tilde\X_I, \n^C\} (\e^A\wedge \e^{B})+\beta\eps_{JABC}\n^C\{\tilde{\X}_I,\e^A\}\wedge\e^B\cr
&=\frac{\beta}{2} \eps_{JABC}(-\sigma\delta_I^C + \n_I \n^C) (\e^A\wedge \e^{B})+\beta\eps_{JIBC}\n^C\underline{\n}\wedge\e^B \cr
&=\left(-\sigma\frac{\beta}{2}  \eps_{JABI} +  \n_I \frac{\beta}{2} \teps_{JAB}\right) (\e^A\wedge \e^{B})+\beta\teps_{JIB}\underline{\n}\wedge\e^B \cr
&=\sigma \beta \as(\e\wedge \e)_{IJ} +  \n_I \S_J+\beta\teps_{JIB}\underline{\n}\wedge\e^B \cr
&=\S_I\n_J-\S_J\n_I+\n_I\S_J+\beta\teps_{JIB}\underline{\n}\wedge\e^B \cr
&=\S_I\n_J+\beta\teps_{IJK}\e^K\wedge\underline{\n},
\ee
and
\be
\{\e_I\wedge \underline{\n}, \S_J\}
&=\eps^{ab}\big(\{\e_a{}_I, \S_J\}\n_b +  \e_a{}_I \{  \n_b, \S_J\}\big)\cr
&=\eps^{ab}\teps_{IAJ} \e_a{}^A\n_b \cr
&=-\eps_{IJK}  (\e^K\wedge\underline{\n})\,.
\ee
From this we conclude that we simply have 
\be\la{BSapp}
\{\BB_I,\S_J\}=\S_I \n _J.
\ee

\subsection{Frame algebra and Hodge star}

Here we establish various Poisson brackets involving the frame and the Hodge star. We first show
\eqref{qec}. A direct calculation gives
\be
 \beta \{\sqrt{ q }, \e_c{}^I\}
 &=\frac{\beta}{2\sqrt{ q }}   \{{ q }, \e_c{}^I\} \cr
 &= \frac{\beta}{2\sqrt{ q }} q_{ab}\eps^{aa'}\eps^{bb'} \{ q_{a'b'}, \e_c{}^I\} \cr
 &=\frac1{2\sqrt{ q }} q_{ab}\eps^{aa'}\eps^{bb'}  ( \epsilon_{c a'} e_{b'}{}^I+ \eps_{cb'} \e_{a'}{}^I)\cr
 &= \frac1{2\sqrt{ q }} (q_{cb}\eps^{bb'}  \e_{b'}{}^I
 +  q_{ac}\eps^{aa'}\e_{a'}{}^I) \cr
 &= \star \e_c{}^I. \la{detqstar}
\ee
By means of Jacobi's identity, this relation implies   that 
\bea
 \{\star\e_a{}^I,\e_d{}^J\}+ \{\e_a{}^I,\star \e_d{}^J\}
 = \{\sqrt{ q }, \{\e_a{}^I,\e_d{}^J\}\}=0.
\eea
 Using $\star^2=-1$, this also means that 
\be
\{\star\e_a{}^I,\star \e_d{}^J\}= \{\e_a{}^I,\e_d{}^J\}\,.
\ee
Moreover, we evaluate
\begin{subequations}
\be
\star q_{ab}&\coloneqq\star \e_a{}^I \e_b{}^J\eta_{IJ} = \frac{q_{ac}\eps^{cd} q_{db}}{\sqrt{ q }}= \sqrt{ q } \eps_{ab}\,,\\
q_{ab}\star &\coloneqq\e_a{}^I \star \e_b{}^J\eta_{IJ} = -\sqrt{ q } \eps_{ab}\,,\\
\star q_{ab}\star &\coloneqq\star\e_a{}^I \star \e_b{}^J\eta_{IJ} = q_{ab}\,.
\ee
\end{subequations}
We now evaluate the algebra between the frame and the dual frame. It is given by
\be
\{\star\e_a{}^I,\e_d{}^J\}
&=\left\{\frac{q_{ab} \eps^{bc} \e_c{}^I}{\sqrt{ q }}, \e_d{}^J\right\}\cr
&=- \frac{\star\e_a{}^I}{\sqrt{ q }}
\{\sqrt{ q },\e_d{}^J\} 
+ \{q_{ab} , \e_d{}^J\}\frac{ \eps^{bc} \e_c{}^I}{\sqrt{ q }}
+\frac{q_{ab} \eps^{bc} }{\sqrt{ q }} \{ \e_c{}^I , \e_d{}^J\}\cr
&=- 
\frac{\star\e_a{}^I \star\e_d{}^J }{\beta \sqrt{ q }}  
+  ( \epsilon_{d a} \e_b{}^J+ \eps_{db} \e_a{}^J)\frac{ \eps^{bc} \e_c{}^I}{\beta\sqrt{ q }}
-\frac{q_{ab} \eps^{bc} \eps_{cd} }{\beta \sqrt{ q }}  \tilde{\eta}^{IJ}\cr
&= - 
\frac{\star\e_a{}^I \star\e_d{}^J +  \e_a{}^J  \e_d{}^I}{\beta \sqrt{ q }}  
-  \epsilon_{ a d}    \frac{  \eps^{bc} \e_b{}^J \e_c{}^I}{\beta\sqrt{ q }}
+ \frac{q_{ad} }{\beta \sqrt{ q }}  \tilde{\eta}^{IJ}\cr
&= 
 \frac{q_{ad}\tilde{\eta}^{IJ} -\star\e_a{}^I\star\!\e_d{}^J -  \e_a{}^J  \e_d{}^I}{\beta \sqrt{ q }}
+ \epsilon_{ a d}    \frac{  \eps^{bc} \star\e_b{}^I \star\e_c{}^J}{\beta\sqrt{ q }}\cr
&= \frac{q_{ad}\tilde{\eta}^{IJ} -\star\e_{a}{}^J\star\!\e_{d}{}^I -  \e_{a}{}^J  \e_{d}{}^I}{\beta \sqrt{ q }}
.\label{*ee}
\ee

Furthermore, using \eqref{defS}, we have that 
\be
\teps^I{}_{JK} \S^J \e_a{}^K
&=  \frac\beta2  \teps^I{}_{JK} \teps^{J}{}_{AB} \e_b{}^A\e_c{}^B \eps^{bc} \e_a{}^K\cr
&=\frac\beta2 (\tilde\delta^I_{B}\tilde\eta_{KA}-\tilde\delta^I_{A}\tilde\eta_{KB})\e_b{}^A\e_c{}^B  \e_a{}^K  \eps^{bc}\cr
&=\frac\beta2 ( q_{ab}\e_{c}{}^I - q_{ac}\e_{b}{}^I)  \eps^{bc}\cr
&= \beta \sqrt{ q } \star \e_a{}^I\cr
&= |\S| \star \e_a{}^I
\la{*eSapp},
\ee
which is statement \eqref{*eS}.

Finally, we can prove that
\be
\teps^{IJ}{}_{K} \e_a{}^K = \frac{\star \e_a{}^I \S^J  -    \star \e_a{}^J {\S}^I}{|\S|}.
\ee
Given a vector $V^I$, we define $V_a \coloneqq V^I \e_a{}^J \eta_{IJ}$, with
 $V^a = q^{ab} V_b$. This means that 
\be
V^I =  \frac{ V\cdot \S}{\S^2} \S^I + V^a \e_a{}^I.
\ee
Now consider the vector $\teps_{IJK} V^I \e_a{}^K$ and its projections   
\be
\teps_{IJK} V^I \S^J \e_a{}^K = |\S| \star V_a,\q\q
\teps_{IJK} V^I e^J_b\e_a{}^K =-\frac1\beta V\cdot {\S} \eps_{ab}.
\ee
Therefore, we find that 
\be
\teps_{I}{}^J{}_{K} V^I \e_a{}^K = \star V_a \frac{\S^J}{|\S|}  -   \star \e_a{}^J \frac{V\cdot {\S}}{|\S|}\,,
\ee
which means that 
\be
\teps^{IJ}{}_{K} \e_a{}^K = \frac{\star \e_a{}^I \S^J  -    \star \e_a{}^J {\S}^I}{|\S|}.
\ee

\section{Algebra of simplicity constraints}
\la{App:simplicity-sec}

In this section we provide detailed evaluations of the brackets involving the full simplicity operators $\CC^I$,  their second class part $\CC_a$, and the first class part $\FC$.

\subsection{Full components}

We start by evaluating the different brackets involving the full simplicity operators $\CC^I$. First, the brackets \eqref{BBapp}, \eqref{SSapp}, \eqref{BSapp} computed above yield
\be\la{CCapp}
\{\CC^I,\CC^J\}
&=\{\BB^I,\BB^J\}-\frac1{\beta} (\{\BB^I,\S^J\}+\{\S^I,\BB^J\})+\frac1{\beta^2} \{\S^I,\S^J\}\cr
&=(\BB^{I} \n^{J}- \BB^J \n ^I) -\frac1{\beta} (\S^I\n^J-\S^J\n^I) - \left(1+ \frac1{\beta^2}\right) \teps^{IJ}{}_K \S^K,\cr
&=\CC^I\n^J-\CC^J \n^I -\left(1+ \frac1{\beta^2}\right) \teps^{IJ}{}_K \S^K.
\ee
We can also compute the various contributions. Using \eqref{BBapp} and \eqref{BSapp} we get
\be\la{CBapp}
\{\CC^I,\BB^J\}
&=\{\BB^I,\BB^J\}-\frac{1}{\beta}\{\S^I,\BB^J\}\cr
&=\BB^{I}\n^{J}- \BB^J \n^I -\teps^{IJ}{}_K \S^K+\frac{1}{\beta}\S^J \n^I\cr
&=\BB^{I}\n^{J}- \CC^J \n^I -\teps^{IJ}{}_K \S^K
\ee
which gives \eqref{CB}. Similarly, we have
\be\la{CSapp}
\{\CC^I,\S^J\}
&=\{\BB^I,\S^J\}-\frac{1}{\beta}\{\S^I,\S^J\}\cr
&= \S^{I} \n^{J}+\f{1}{\beta}\teps^{IJ}{}_K \S^K,
\ee
which is \eqref{CS}. Using \eqref{Be bracket} and \eqref{Seapp} gives
\be\la{Ceapp}
\{\CC^I, \e_a{}^J\}=\e_a{}^I\n^J+\frac{1}{\beta} \teps^{IJ}{}_K \e_a{}^K,
\ee
which is \eqref{Ce}. Finally, using the definition of $\CC^I$ and \eqref{Bnapp} gives
\be
\{\CC^I,\n^J\}=-\sigma\tilde{\eta}^{IJ},
\ee
which is \eqref{Cn} when $\sigma=-1$.

\subsection{Second class components}
\la{CaSBapp}

We can use \eqref{eeapp}, \eqref{CCapp}, and \eqref{Ceapp} to evaluate the bracket between the second class constraints as
\be
\{\CC_a,\CC_b\}
&=\e_a{}^I \e_b{}^J\{\CC_I,\CC_J\}+\{\e_a{}^I,\e_b{}^J \}\CC_I\CC_J+\e_a{}^I\{ \CC_I,\e_b{}^J \}\CC_J+\e_b{}^J \{\e_a{}^I ,\CC_J\}\CC_I\cr
&=- \left(1+\frac1{\beta^2}\right) \teps_{IJK}\e_a{}^I \e_b{}^J\S^K-\frac1\beta\epsilon_{ab} \CC^2+\frac{2}{\beta} \teps_{IJK}\e_a{}^I \e_b{}^K\CC^J\cr
&= \frac{1}{\beta}\epsilon_{ab}\left(-\left(1+\frac1{\beta^2}\right) \S^2- \CC^2-\frac{2}{\beta}\S\cdot \CC\right)\cr
&=-\frac{1}{\beta}\epsilon_{ab}\left(\S^2+\BB^2\right)\label{CCa},
\ee
where we have used that
\be\la{See}
\frac1\beta \epsilon_{ab}\S^K=\e_a{}^I \e_b{}^J\,\teps_{IJ}{}^K. 
\ee
Similarly, we can evaluate
\be\la{starCCapp}
\{\star\CC_a,\CC_b\}
&=\star\e_a{}^I\e_b{}^J\{\CC_I,\CC_J\}+\{ \star\e_a{}^I ,\e_b{}^J \}\CC_I\CC_J+ \star\e_a{}^I\{ \CC_I,\e_b{}^J \}\CC_J+\e_b{}^J \{ \star\e_a{}^I ,\CC_J\}\CC_I\cr
&=- \left(1+\frac1{\beta^2}\right) \teps_{IJK}  \star\e_a{}^I \e_b{}^J\S^K+\frac{q_{ab} \CC^2 -\star\CC_a\star\!\CC_b -  \CC_a\CC_b}{\beta \sqrt{ q }}+\frac{2}{\beta}\teps_{IJK}\star\e_a{}^I \e_b{}^K\CC^J \cr
&=\frac{q_{ab}}{\beta \sqrt{q}}\left(\left(1+\frac1{\beta^2}\right) \S^2+ \CC^2+\frac{2}{\beta}\S\cdot \CC\right)- \frac{\star\CC_a\star\!\CC_b + \CC_a\CC_b}{\beta \sqrt{q}}\cr
&=\frac{q_{ab}}{\beta \sqrt{q}}\left(\S^2+\BB^2\right)- \frac{\star\CC_a\star\!\CC_b+\CC_a\CC_b}{\beta \sqrt{q}}.
\ee
Combining these last two brackets gives
\be
\{\CC^-_a,\CC^+_b\}=\f{1}{2}\left(-\f{1}{\beta}\eps_{ab}(\S^2+\BB^2)+\f{q_{ab}}{i\beta\sqrt{q}}(\S^2+\BB^2)-\f{\star\CC_a\star\!\CC_b+\CC_a\CC_b}{i\beta \sqrt{q}}\right)
\ee
which is \eqref{C+C-} once we use $|\S|=\sqrt{\S^2}=\beta\sqrt{q}$ and $\star\CC_a\star\!\CC_b+\CC_a\CC_b=2(\CC^+_a\CC^-_b+\CC^-_a\CC^+_b)$.

We now give the brackets between the second class simplicity constraints and the various quantities appearing. Using \eqref{Seapp}, \eqref{CSapp}, and $\e_a{}^I\S_I=0$, we get
\be
\{\CC_a,\S^J\}
&=\e_{aI}\{ \CC^I,\S^J\}+\CC_I\{ \e_{a}{}^I,\S^J\}\cr
&=\e_{aI}\left(\S^I \n^J+\frac{1}{\beta} \teps^{IJ}{}_K S^K\right)-\CC_I    \teps_{JKI} \e_{a}{}^K\cr
&=-\frac{1}{\beta}  \teps^{J}{}_{IK}\e_{a}{}^I \S^K- \teps^J{}_{IK}  \e_{a}{}^I \CC^K\cr
&=\teps^J{}_{KI}\BB^K\e_{a}{}^I.\la{CaSapp}
\ee
Using \eqref{Beapp}, \eqref{CBapp}, \eqref{*eSapp}, and \eqref{eS}, we evaluate 
\be
\{\CC_a,\BB^J\}
&= \e_{aI}\{ \CC^I,\BB^J\}+\CC_I\{ \e_{a}{}^I,\BB^J\}\cr
&= \e_{aI}\left(\BB^{I} \n^{J}- \CC^J \n^I -\teps^{IJK} \S_K\right) \cr
&=\CC_a \n^{J} -\teps^J{}_{KI} \S^K\e_{a}{}^I\cr
&= \CC_a \n^{J} -|\S| \star \e_{a}{}^J.\la{CaSapp}
 \ee
Together this gives
\be
\{\CC_a,\CC^J\}=\CC_a \n^{J} -\teps^J{}_{KI}\left(\S^K+\f{1}{\beta}\BB^K\right)\e_{a}{}^I.
\ee
Using \eqref{*eSapp} again, we also obtain 
\begin{subequations}
\be
\{\CC_a,\S^2\}&=2  \{\CC_a,\S^J\} \S_J =  - \BB^K \teps_{KJI}  \S^J \e_{a}{}^I=-2 |\S|\star \CC_a\,, \\
\{\CC_a,\BB^2\}&= 2\{\CC_a,\BB^J\} \BB_J= 2(\CC_a\n^{J} - |\S|\star \e_{a}{}^J ) \BB_J = -2 |\S| \star \CC_a\,,
\ee
\end{subequations}
and
\be
\{\CC_a,\BB\cdot \S\}
&=\{\CC_a,\BB^J\} \S_J +  \BB_J \{\CC_a,\S^J\}\cr
&=(\CC_a \n^{J} -|\S| \star \e_{a}{}^J )\S_J - \BB_J\BB^K \teps_{KJI}  \S^J \e_{a}{}^I\cr
&=0.
\ee
The brackets above together give
\be\la{C2Caapp}
\frac12 \{\CC^2 , \CC_a\}
&=\CC_I\{\CC^I,\CC_a\}\cr
&=\CC_I\{\BB^I,\CC_a\}-\f{1}{\beta}\CC_I\{\S^I,\CC_a\}\cr
&=|\S|\star\CC_a+\f{1}{\beta}\CC_I\teps^I{}_{JK}\BB^J\e_a{}^K\cr
&=|\S|\star\CC_a+\f{1}{\beta^2}\CC_I\teps^I{}_{JK}\S^J\e_a{}^K\cr
&=\left(1+\frac1{\beta^2}\right)|\S|\star\CC_a,
\ee
and
\be\la{CSCaapp}
\{\CC\cdot\S,\CC_a\}
&=\CC_I\{\S^I,\CC_a\}+\{\CC^I,\CC_a\}\S_I\cr
&=-\teps^I{}_{JK}\CC_I\BB^J\e_a{}^K+\f{1}{\beta}\teps^I{}_{JK}\S_I\BB^J\e_a{}^K\cr
&=\f{2}{\beta}\teps^I{}_{JK}\S_I\BB^J\e_a{}^K\cr
&=\f{2}{\beta}\teps^I{}_{JK}\S_I\CC^J\e_a{}^K\cr
&=-\f{2}{\beta}|\S|\star\CC_a.
\ee
We finally evaluate the bracket of $\CC_a$ with the normal to find
\be
\{ \CC_a,\n^J \}
&=\e_{a}{}^I\{ \CC_I,\n^J \}\cr
&=\e_{a}{}^I(-\sigma\delta_{I}^J +  \n_I \n^J)\cr
&=-\sigma\e_{a}{}^J\,,
\ee
and with the frame to find
\be
\{ \CC_a,\e_{b}{}^J \}
&=\e_{aI}\{ \CC^I,\e_{b}{}^J \}+ \CC_I\{\e_a{}^I,\e_{b}{}^J \}\cr
&=\e_{aI}\left(\n^J\e_b{}^I+\frac{1}{\beta}  \teps^{IJ}{}_K \e_b{}^K\right) -\frac{1}\beta \epsilon_{ab}  \CC^J\cr
&=q_{ab}\n^J-\frac{1}{\beta}  \teps^{J}{}_{IK}\e_{a}{}^I \e_b{}^K-\frac{1}\beta \epsilon_{ab}  \CC^J\cr
&=q_{ab}\n^J-\frac{1}{\beta^2}  \epsilon_{ab}  \S^J-\frac{1}\beta \epsilon_{ab}  \CC^J\cr
&=q_{ab}\n^J -\frac{1}\beta \epsilon_{ab}  \BB^J\,.
\ee

\subsection{First class part}

We now evaluate the brackets of the first class simplicity constraint $\FC=-\CC^2-(\beta+\beta^{-1})\CC\cdot\S$ with $\n^I$, $\S^I$, $\BB^I$, $\CC^I$, $\CC_a$ and $\e_a{}^I$. First, we have that
\be
\{\CC\cdot\S,\n^I\} =-\sigma \S^I,\q
\{\CC^2,\n^I\}=-2\sigma\CC^I,
\ee
which implies that 
\be
\{\FC, \n^I\}= -\sigma \left(\frac2{\beta} \BB^I +\left(1-\frac1{\beta^2}\right) \S^I \right)
=-\sigma  \left(\frac2{\beta} \CC^I +\left(1+\frac1{\beta^2}\right) \S^I \right).
\ee
We also have that 
\be
\{\CC\cdot \S, \S_J\} =-\tilde\eps_{IJK} \CC^I \S^K + \S^2 \n_J,\q
\{\CC^2, \S_J\} = 2 \CC\cdot \S \n_J +\frac2{\beta^2} \tilde\eps_{IJK} \CC^I \S^K,
\ee
which gives
\be
\{\FC, \S_J\}=  -\left(1-\frac1{\beta^2}\right)
\teps_{IJK} \CC^I \S^K  + \left(\frac2{\beta} \CC\cdot \S + \left(1+\frac1{\beta^2}\right) \S^2\right) n_J.
\ee
Next we evaluate
\be
\{\CC\cdot \S, \BB_J\} =\BB\cdot \S\n_J,\q
\{\CC^2, \BB_J\} = 2 \CC\cdot \BB \n_J 
- 2 \tilde\eps_{IJK} \CC^I \S^K\,,
\ee
which gives 
\be
\{\FC, \BB_J\}=  -\frac{2}{\beta}
\teps_{IJK} \CC^I \S^K  + \left(\frac2{\beta} \CC\cdot \BB + \left(1+\frac1{\beta^2}\right) \S\cdot \BB \right) \n_J.
\ee
Furthermore, we have
\be
\{\FC, \CC_J\}=  - \frac1\beta \left(1+\frac1{\beta^2}\right)
\teps_{IJK} \CC^I \S^K  + \left(\frac2{\beta} \CC^2 + \left(1+\frac1{\beta^2}\right) \S\cdot \CC \right) \n_J.
\ee
We now evaluate the bracket of $\FC$ with $ \e_a{}^I$. For this we use
\be
\{\CC^2 , \e_a{}^J\}= 2\CC_a \n^J +\frac2{\beta} \teps_{I}{}^J{}_K   \CC^I \e_a{}^K,
\ee
and
\be
\{\CC\cdot \S , \e_a{}^J\}
&=\S_I\{\CC^I , \e_a{}^J\}+\CC_I\{\S^I , \e_a{}^J\}\cr
&= \S_I\left(\e_a{}^I \n^J+\frac{1}{\beta} \teps^{IJ}{}_K \e_a{}^K\right) - \CC_I \left(\teps^{IJ}{}_K \e_a{}^K\right)\cr
&= \teps^{IJ}{}_K \left( \frac{\S_I}{\beta} - \CC_I\right) \e_a{}^K\,,\la{CSeapp}
\ee
from which we get
\be
\{\FC, \e_{aJ}\}
&=\frac2\beta\CC_a \n_J+\frac2{\beta^2} \teps_{IJK}   \CC^I \e_a{}^K+\left(1+\frac1{\beta^2}\right) \teps_{IJK} \left( \frac{\S^I}{\beta} - \CC^I\right) \e_a{}^K\cr
&=\frac2\beta\CC_a \n_J+\left(\frac1{\beta^2}-1\right) \teps_{IJK} \CC^I \e_a{}^K+\frac1\beta \left(1+\frac1{\beta^2}\right)\teps_{IJK} \S^I \e_a{}^K\,.
\ee
As a consistency check we finally verify that 
\be
\{\FC,\CC_a\}
&=\{\FC, \e_{aJ}\} \CC^J + \{\FC, \CC_J\} \e_a{}^J\cr
&=\frac1\beta \left(1+\frac1{\beta^2}\right)\teps_{IJK} \S^I \CC^J  \e_a{}^K   - \frac1\beta \left(1+\frac1{\beta^2}\right)
\teps_{IJK} \CC^I \e_a{}^J \S^K\cr
&=0.
\ee

\subsection{Lorentz transformation}
\label{Lotrentzt}

The bracket between the Lorentz generators and the simplicity constraints can be computed as
\be
\sigma\{\CC_I,\JJ_{JK}\} =\,& \{\CC_I,\BB_{J}\} \n_{K}
+\BB_J\{\CC_I,\n_{K}\}
- \{\CC_I,\BB_K\} \n_J -\BB_K\{\CC_I,\n_J\}
-\teps_{JK}{}^L\{\CC_I, \S_L \}-\eps_{JK}{}^{LA}\{\BB_I, \n_A\} \S_L
\cr
=\,&  
 \BB_{I} \n_{J} \n_{K}- \BB_J \n_I \n_{K} -\teps_{IJ}{}^A \S_A \n_{K} +\frac{1}{\beta}\S_J \n_I  \n_{K}
 \cr
 &+\BB_J(-\sigma\eta_{IK} +  \n_I \n_K)\cr
 &-\left( \BB_{I} \n_{K} \n_{J}- \BB_K \n_I \n_{J} -\teps_{IK}{}^A \S_A \n_{J} +\frac{1}{\beta}\S_K \n_I  \n_{J}\right)
\cr
&-\BB_K(-\sigma\eta_{IJ} +  \n_I \n_J)\cr
&-  \teps_{JK}{}^L (\S_I \n_L+\frac{1}{\beta} \teps_{ILA} S^A)\cr
&+\eps_{JK}{}^{LA}(\sigma\eta_{IA}-\n_I\n_A) \S_L
\cr
=\,&-\teps_{IJ}{}^A \S_A \n_{K}+\frac{1}{\beta}\S_J \n_I  \n_{K}
-\sigma\eta_{IK}\BB_J\cr
&+\teps_{IK}{}^A \S_A \n_{J} -\frac{1}{\beta}\S_K \n_I  \n_{J}
+\sigma\eta_{IJ}\BB_K\cr
&+\frac1\beta  \teps_{JK}{}^L\teps_{IAL} S^A\cr
&-\sigma \eps_{IJK}{}^{L}\S_L -\teps_{JK}{}^L\n_I\S_L \cr
=\,&-\teps_{IJ}{}^A \S_A \n_{K}+\frac{1}{\beta}\S_J \n_I  \n_{K}
-\sigma\eta_{IK}\BB_J\cr
&+\teps_{IK}{}^A \S_A \n_{J} -\frac{1}{\beta}\S_K \n_I  \n_{J}
+\sigma\eta_{IJ}\BB_K\cr
&-\frac1\beta  (\sigma\eta_{JI}\S_K- \n_I\n_J\S_K -\sigma\eta_{KI}\S_J+\n_I\n_K \S_J)\cr
&-\sigma \eps_{IJK}{}^{L}\S_L -\teps_{JK}{}^L\n_I\S_L \cr
=\,&\sigma(\eta_{IJ}\CC_K-\eta_{IK}\CC_J)\,,
\la{JC-app}
\ee
where we have used
\be
\teps_{JK}{}^L\teps_{IAL} \S^A
&=-\sigma (\tilde\eta_{JI}\tilde\eta_{KA}-\tilde\eta_{JA}\tilde\eta_{KI}) \S^A\cr
&=-\sigma (\tilde\eta_{JI}\eta_{KA}-\eta_{JA}\tilde\eta_{KI}) \S^A\cr
&=-\sigma\eta_{JI}\S_K+ \n_I\n_J\S_K +\sigma\eta_{KI}\S_J-\n_I\n_K \S_J\,.
\ee
By means of \eqref{JC-app}, it follows that
\be
\{\CC_I,\JJ_{JK}\}\S^I
&=\CC_K\S_J-\CC_J\S_K\cr
&=\BB_K\S_J-\BB_J\S_K\,.
\ee
Moreover,
\be
\sigma \{\S_I,\JJ_{JK}\}
&=\{\S_I,\BB_{J}\} \n_{K}- \{\S_I,\BB_K\} \n_J-\teps_{JK}{}^L\{\S_I, \S_L \}\cr
&=-  \S_J \n_I\n_{K} +\S_K \n_I\n_J+\teps_{JK}{}^L \teps_{ILA} S^A\cr
&=-  \S_J \n_I\n_{K} +\S_K \n_I\n_J
+\sigma\eta_{JI}\S_K- \n_I\n_J\S_K -\sigma\eta_{KI}\S_J+\n_I\n_K \S_J\cr
&=\sigma\eta_{JI}\S_K-\sigma\eta_{KI}\S_J\,,
\ee
from which
\be\la{nSJ}
\n^I\{\S_I,\JJ_{JK}\}=\n_J \S_K-\n_K\S_J
\ee
and
\be
\CC^I\{\S_I,\JJ_{JK}\}=\BB_J\S_K-\BB_K\S_J\,.
\ee
Therefore,
\be\la{CSJ}
\{\CC\cdot\S,\JJ_{JK}\}=\{\CC_I,\JJ_{JK}\}\S^I+\CC^I\{\S_I,\JJ_{JK}\}=0.
\ee
from which \eqref{CSQtilde} follows, and
\be
\{\n\cdot\S,\JJ_{JK}\}
&=\{\n_I,\JJ_{JK}\}\S^I+\n^I\{\S_I,\JJ_{JK}\}\cr
&=\sigma \{\n_I,\BB_J\}\n_K\S^I-\sigma\{\n_I,\BB_K\}\n_J\S^I+\n_J\S_K-\n_K \S_J\cr
&= (\eta_{IJ} \n_K-\eta_{IK}\n_J)\S^I+\n_J\S_K-\n_K \S_J\cr
&=0.
\ee

Furthermore, we have
\be
\sigma\{\e_{aI}, \JJ_{JK}\}
&=\{\e_{aI},\BB_J\}\n_K-\{\e_{aI},\BB_K\}\n_J
-\teps_{JK}{}^L\{\e_{aI},\S_L\}\cr
&=- \n_I \n_K\e_a{}_J+\n_I\n_J \e_{aK}
+\teps_{JK}{}^L \teps_{LAI} \e_a{}^A\cr
&=- \n_I \n_K\e_a{}_J+\n_I\n_J \e_{aK}
+\sigma (\tilde\eta_{JI}\tilde\eta_{KA}-\tilde\eta_{JA}\tilde\eta_{KI})\e_a{}^A\cr
&=- \n_I \n_K\e_a{}_J+\n_I\n_J \e_{aK}\cr
&\phantom{=\ }+\sigma \left((\eta_{JI}-\sigma \n_I\n_J)(\eta_{KA}-\sigma\n_K\n_A)-(\eta_{JA}-\sigma\n_J\n_A)(\eta_{KI}-\sigma\n_I\n_K)\right)\e_a{}^A\cr
&=\sigma \left(\eta_{JI}(\eta_{KA}-\sigma\n_K\n_A)\e_a{}^A-(\eta_{JA}-\sigma\n_J\n_A)\eta_{KI}\e_a{}^A\right)\cr
&=\sigma(\eta_{IJ}\e_{aK}-\eta_{IK}\e_{aJ})\,,
\ee
from which we get
\be\la{CaJ}
\{\CC_a,\JJ_{JK}\}
&=\CC^I\{\e_{aI}, \JJ_{JK}\}+\{\CC_I,\JJ_{JK}\}\e_{aI}\cr
&=\CC_J\e_{aK}
-\CC_K\e_{aJ}+\CC_K\e_{aJ}-\CC_J\e_{aK}\cr
&=0\,.
\ee
Lastly, let us verify that
\be
\{\n_a,\JJ_{JK}\}
&=\n^I\{\se_{aI},\JJ_{JK}\}+\se_{a}{}^I\{\n_I,\JJ_{JK}\}\cr
&=\n_J \se_{aK}-\n_K \se_{aJ}+\n_K \se_{aJ}-\n_J \se_{aK}\cr
&=0\,.
\ee

\section{Poincar\'e algebra}\la{App:Poinc}

Here we consider the Poincar\'e generators $J_{AB}$ and $P_C$ satisfying the algebra
\begin{subequations}
\be
-i[J_{IA},J_{JB}]&=\eta_{AJ} J_{IB} +\eta_{IB} J_{AJ} -\eta_{IJ} J_{AB} -\eta_{AB} J_{IJ}\,,\\
-i[J_{IA},P_B]&=P_I \eta_{AB} - P_A\eta_{IB}\,\\
-i[P_A,P_B]&=0.
\ee
\end{subequations}
The duality transformation $*J_{IJ} \coloneqq \frac12 \epsilon_{IJ}{}^{KL} J_{KL}$ satisfies the following  compatibility conditions  with the commutator:
\be
[*J_{IA},*J_{JB}]=-[J_{IA},J_{JB}],\q[*J_{IA},J_{JB}] =[J_{IA},*J_{JB}].
\ee
We now define the boost and spin generators to respectively be
\be
B_I \coloneqq\f{J_{IJ}P^J}{m}\,,\q S_I\coloneqq\f{*J_{IJ}P^J}{m}\,,
\ee
which means that 
\begin{subequations}
\be
P^2 J_{IJ}&=m\big(B_I P_J-B_J P_I- \teps_{IJ}{}^K S_K\big)\,,\\
P^2 \as{J}_{IJ}&=m\big(S_I P_J-S_J P_I+ \teps_{IJ}{}^K B_K\big)\,,
\ee
\end{subequations}
with $\teps_{IJ}{}^K \coloneqq\eps_{IJ}{}^{KL}P_L$ and $P^2=-m^2$. The boost and spin generators are such that 
\be
-mi[B_I, P_J]= P_IP_J - P^2 \eta_{IJ},\q
-i[S_I, P_J]=0,
\ee
and they satisfy the boost-spin algebra
\begin{subequations}
\be
-mi[B_I,B_J]&=B_I P_J-B_J P_I-\teps_{IJ}{}^K S_K,\\
-mi[B_I,S_J]&=S_I P_J ,\\
-mi[S_I,S_J]&=-\teps_{IJ}{}^K S_K.
\ee
\end{subequations}
Denoting the simplicity constraint by
\be
C_I= B_I -\frac1{\beta} S_I\,,
\ee
we get 
\be
-mi[C_I,C_J]= C_IP_J-C_JP_I- \frac{(1+\beta^2)}{\beta^2} \teps_{IJ}{}^K S_K\,.
\ee
This follows from the explicit computations
\be
-mi[B_I,B_J]
&=\f{1}{m}\big([J_{IA},J_{JB}]P^AP^B + J_{JB} [J_{IA},P^B ] P^A- J_{IA}[J_{JB}, P^A] P^B\big),\cr
&=P_J B_{I} - P_I B_{J}  -\f{P^2 J_{IJ}}{m}+P_I B_J + \f{P^2J_{IJ}}{m}  - P_J B_I +\f{P^2J_{IJ}}{m}\cr
&=\f{P^2 J_{IJ}}{m}\cr
&=B_I P_J-B_J P_I - \teps_{IJ}{}^K S_K\,,
\ee
\be
-mi[B_I,S_J]
&=\f{1}{m}\big([J_{IA},\as{J}_{JB}]P^AP^B + \as{J}_{JB} [J_{IA},P^B ] P^A- J_{IA}[\as{J}_{JB}, P^A] P^B\big),\cr
&=P_J S_I - P_I S_J  - \f{P^2 \as{J}_{IJ}}{m} +  P_I S_J + \f{P^2\as{J}_{IJ}}{m}  \cr
&=P_J S_I\,,
\ee
and
\be
-mi[S_I,S_J]
&=\f{1}{m}\big([\as{J}_{IA},\as{J}_{JB}]P^AP^B + \as{J}_{JB} [\as{J}_{IA},P^B ] P^A- \as{J}_{IA}[\as{J}_{JB}, P^A] P^B\big),\cr
&=\f{P^2 J_{IJ}}{m} -P_J B_{I} + P_I B_{J}  ,\cr
&=-\teps_{IJ}{}^K S_K\,.
\ee
We can also write the algebra for the kinematical boost and rotation generators 
\be
K_I = J_{IJ} t^J, \qquad L_I=*J_{IJ}t^J, 
\ee
where $t^J$ is a kinematical vector. The calculations are similar but simpler since $t^J$ commutes with $J_{IJ}$. Denoting $\mathring{\epsilon}_{IJ}{}^K=\eps_{IJ}{}^{KL}t_L$ as in the main text, we have the rotation commutators
\be
-i[L_I,L_J]
&=[\as{J}_{IA},\as{J}_{JB}]t^At^B ,\cr
&=t^2 J_{IJ} +  t_I K_{J}-t_J K_{I}  ,\cr
&=-\mathring{\epsilon}_{IJ}{}^K L_K\,,
\ee 
the boost commutators
\be
-i[K_I,K_J]
&=[J_{IA},J_{JB}]t^At^B ,\cr
&=t_J K_{I} - t_I K_{J}  -t^2 J_{IJ},\cr
&=\mathring{\epsilon}_{IJ}{}^K L_K\,,
\ee
and the mixed commutators
\be
-i[K_I,L_J]
&=[J_{IA},\as{J}_{JB}]P^AP^B ,\cr
&=t_J L_I - t_I L_J  - t^2 \as{J}_{IJ}, \cr
&=-\mathring{\epsilon}_{IJ}{}^K K_K\,.
\ee

\bibliographystyle{bib-style2}
\bibliography{Biblio.bib}

\end{document}

%% file: Macro.tex
\usepackage{caption}
\usepackage{currvita}
\usepackage{alphabeta}

\newcommand{\la}{\label}
\newcommand{\bee}{\nopagebreak[3]\begin{equation*}}
\newcommand{\be}{\begin{equation}}
\newcommand{\ee}{\end{equation}}

\newcommand{\eee}{\end{equation*}}
\newcommand{\beq}{\begin{eqnarray}}
\newcommand{\eeq}{\end{eqnarray}}
\newcommand{\bea}{\begin{eqnarray}}
\newcommand{\eea}{\end{eqnarray}}
\newcommand{\baa}{\nopagebreak[3]\begin{eqnarray*}}
\newcommand{\eaa}{\end{eqnarray*}}

\def\be#1\ee{\begin{align}#1\end{align}}
\def\nn{\nonumber}
\def\q{\qquad}
\def\f{\frac}
\def\eps{\epsilon}
\def\teps{\tilde{\epsilon}}

\def\ip{\lrcorner\,}
\def\pa{\partial}

\def\as{{\ast}}

\def\para{\text{\tiny{$\parallel$}}}
\def\Poin{{\mathrm{PH}}}

\def\rd{\mathrm{d}}

\def\un{\underline{n}}

\def\vphi{\varphi}
\def\bvphi{\boldsymbol{\varphi}}
\def\bd{\boldsymbol{\delta}}

\def\brho{{\boldsymbol\rho}}

\newcommand{\R}{\mathbb{R}}
\newcommand{\RR}{\mathbb{R}}
\newcommand{\C}{\mathbb{C}}

\def\BF{\mathrm{BF}}

\def\ECH{\mathrm{ECH}}
\def\GR{\mathrm{GR}}

\def\BF{\mathrm{BF}}
\def\ts{\tilde{s}}
\def\tb{\tilde{b}}
\def\tS{\tilde{S}}
\def\tB{\tilde{B}}

\def\tE{\tilde{E}}

\def\tK{\tilde{K}}

\def\bTh{\boldsymbol\Theta}
\def\jj{\mathsf{j}}
\def\k{\mathsf{k}}
\def\K{\mathsf{K}}

\def\FC{\mathcal{C}}

\def\CC{\mathsf{C}}
\def\DD{\mathsf{D}}

\def\CH{\mathcal{H}}

\def\CO{\mathcal{O}}

\def\n{\mathsf{n}}
\def\s{\mathsf{s}}
\def\JJ{\mathsf{J}}
\def\jj{\mathsf{j}}
\def\ext{\mathrm{ext}}
\def\qq{q}

\def\SU{\text{SU}}

\def\SL{\text{SL}}
\def\su{\mathfrak{su}}

\def\sll{\mathfrak{sl}}
\def\sg{\mathfrak{g}}

\def\e{\mathsf{e}}
\def\se{\mathsf{z}}
\def\te{\tilde{e}}

\def\BB{\mathsf{B}}
\def\S{\mathsf{S}}

\def\X{\mathsf{X}}

\numberwithin{equation}{section}

%% file: Gravity-Edge-Modes-3.bbl
\providecommand{\href}[2]{#2}\begingroup\raggedright\begin{thebibliography}{100}

\bibitem{Edge-Mode-I}
L.~Freidel, M.~Geiller and D.~Pranzetti, \emph{{Edge modes of gravity. Part I.
  Corner potentials and charges}},
  \href{http://dx.doi.org/10.1007/JHEP11(2020)026}{\emph{JHEP} {\bfseries 11}
  (2020) 026}, [\href{https://arxiv.org/abs/2006.12527}{{\ttfamily
  2006.12527}}].

\bibitem{Edge-Mode-II}
L.~Freidel, M.~Geiller and D.~Pranzetti, \emph{{Edge modes of gravity. Part II.
  Corner metric and Lorentz charges}},
  \href{http://dx.doi.org/10.1007/JHEP11(2020)027}{\emph{JHEP} {\bfseries 11}
  (2020) 027}, [\href{https://arxiv.org/abs/2007.03563}{{\ttfamily
  2007.03563}}].

\bibitem{Donnelly:2016auv}
W.~Donnelly and L.~Freidel, \emph{{Local subsystems in gauge theory and
  gravity}}, \href{http://dx.doi.org/10.1007/JHEP09(2016)102}{\emph{JHEP}
  {\bfseries 09} (2016) 102},
  [\href{https://arxiv.org/abs/1601.04744}{{\ttfamily 1601.04744}}].

\bibitem{Peld_n_1994}
P.~Peld{\'a}n, \emph{{Actions for Gravity, with Generalizations: A Review}},
  \href{http://dx.doi.org/10.1088/0264-9381/11/5/003}{\emph{Class. Quant.
  Grav.} {\bfseries 11} (May, 1994) 1087--1132}.

\bibitem{BarroseSa:2000vx}
N.~Barros~e Sa, \emph{{Hamiltonian analysis of general relativity with the
  Immirzi parameter}},
  \href{http://dx.doi.org/10.1142/S0218271801000858}{\emph{Int. J. Mod. Phys.
  D} {\bfseries 10} (2001) 261--272},
  [\href{https://arxiv.org/abs/gr-qc/0006013}{{\ttfamily gr-qc/0006013}}].

\bibitem{Alexandrov:2011ab}
S.~Alexandrov, M.~Geiller and K.~Noui, \emph{{Spin Foams and Canonical
  Quantization}}, \href{http://dx.doi.org/10.3842/SIGMA.2012.055}{\emph{SIGMA}
  {\bfseries 8} (2012) 055}, [\href{https://arxiv.org/abs/1112.1961}{{\ttfamily
  1112.1961}}].

\bibitem{Perez:2012wv}
A.~Perez, \emph{{The Spin Foam Approach to Quantum Gravity}},
  \href{http://dx.doi.org/10.12942/lrr-2013-3}{\emph{Living Rev. Rel.}
  {\bfseries 16} (2013) 3}, [\href{https://arxiv.org/abs/1205.2019}{{\ttfamily
  1205.2019}}].

\bibitem{Engle:2007qf}
J.~Engle, R.~Pereira and C.~Rovelli, \emph{{Flipped spinfoam vertex and loop
  gravity}},
  \href{http://dx.doi.org/10.1016/j.nuclphysb.2008.02.002}{\emph{Nucl. Phys.}
  {\bfseries B798} (2008) 251--290},
  [\href{https://arxiv.org/abs/0708.1236}{{\ttfamily 0708.1236}}].

\bibitem{Freidel:2007py}
L.~Freidel and K.~Krasnov, \emph{{A New Spin Foam Model for 4d Gravity}},
  \href{http://dx.doi.org/10.1088/0264-9381/25/12/125018}{\emph{Class. Quant.
  Grav.} {\bfseries 25} (2008) 125018},
  [\href{https://arxiv.org/abs/0708.1595}{{\ttfamily 0708.1595}}].

\bibitem{Engle:2007wy}
J.~Engle, E.~Livine, R.~Pereira and C.~Rovelli, \emph{{LQG vertex with finite
  Immirzi parameter}},
  \href{http://dx.doi.org/10.1016/j.nuclphysb.2008.02.018}{\emph{Nucl. Phys.}
  {\bfseries B799} (2008) 136--149},
  [\href{https://arxiv.org/abs/0711.0146}{{\ttfamily 0711.0146}}].

\bibitem{0264-9381-4-5-011}
Y.~N. Obukhov, \emph{The palatini principle for manifold with boundary},
  {\emph{Classical and Quantum Gravity} {\bfseries 4} (1987) 1085}.

\bibitem{bianchi2012horizon}
E.~Bianchi and W.~Wieland, \emph{Horizon energy as the boost boundary term in
  general relativity and loop gravity},  2012.

\bibitem{Bodendorfer:2013hla}
N.~Bodendorfer and Y.~Neiman, \emph{{Imaginary action, spinfoam asymptotics and
  the `transplanckian' regime of loop quantum gravity}},
  \href{http://dx.doi.org/10.1088/0264-9381/30/19/195018}{\emph{Class. Quant.
  Grav.} {\bfseries 30} (2013) 195018},
  [\href{https://arxiv.org/abs/1303.4752}{{\ttfamily 1303.4752}}].

\bibitem{Alexandrov:2002br}
S.~Alexandrov and E.~R. Livine, \emph{{SU(2) loop quantum gravity seen from
  covariant theory}},
  \href{http://dx.doi.org/10.1103/PhysRevD.67.044009}{\emph{Phys. Rev.}
  {\bfseries D67} (2003) 044009},
  [\href{https://arxiv.org/abs/gr-qc/0209105}{{\ttfamily gr-qc/0209105}}].

\bibitem{Alexandrov:2007pq}
S.~Alexandrov, \emph{{Spin foam model from canonical quantization}},
  \href{http://dx.doi.org/10.1103/PhysRevD.77.024009}{\emph{Phys. Rev.}
  {\bfseries D77} (2008) 024009},
  [\href{https://arxiv.org/abs/0705.3892}{{\ttfamily 0705.3892}}].

\bibitem{Alexandrov:2008da}
S.~Alexandrov, \emph{{Simplicity and closure constraints in spin foam models of
  gravity}}, \href{http://dx.doi.org/10.1103/PhysRevD.78.044033}{\emph{Phys.
  Rev.} {\bfseries D78} (2008) 044033},
  [\href{https://arxiv.org/abs/0802.3389}{{\ttfamily 0802.3389}}].

\bibitem{Gielen:2010cu}
S.~Gielen and D.~Oriti, \emph{{Classical general relativity as BF-Plebanski
  theory with linear constraints}},
  \href{http://dx.doi.org/10.1088/0264-9381/27/18/185017}{\emph{Class. Quant.
  Grav.} {\bfseries 27} (2010) 185017},
  [\href{https://arxiv.org/abs/1004.5371}{{\ttfamily 1004.5371}}].

\bibitem{Baratin:2010wi}
A.~Baratin and D.~Oriti, \emph{{Group field theory with non-commutative metric
  variables}},  [\href{https://arxiv.org/abs/1002.4723}{{\ttfamily
  1002.4723}}].

\bibitem{Baratin:2011tx}
A.~Baratin and D.~Oriti, \emph{{Quantum simplicial geometry in the group field
  theory formalism: reconsidering the Barrett-Crane model}},
  \href{http://dx.doi.org/10.1088/1367-2630/13/12/125011}{\emph{New J. Phys.}
  {\bfseries 13} (2011) 125011},
  [\href{https://arxiv.org/abs/1108.1178}{{\ttfamily 1108.1178}}].

\bibitem{Wieland_2014}
W.~M. Wieland, \emph{{A new action for simplicial gravity in four dimensions}},
  \href{http://dx.doi.org/10.1088/0264-9381/32/1/015016}{\emph{Class. Quant.
  Grav.} {\bfseries 32} (2015) 015016},
  [\href{https://arxiv.org/abs/1407.0025}{{\ttfamily 1407.0025}}].

\bibitem{Wieland_2017}
W.~Wieland, \emph{{Discrete gravity as a topological field theory with
  light-like curvature defects}},
  \href{http://dx.doi.org/10.1007/JHEP05(2017)142}{\emph{JHEP} {\bfseries 05}
  (2017) 142}, [\href{https://arxiv.org/abs/1611.02784}{{\ttfamily
  1611.02784}}].

\bibitem{Wieland:2017zkf}
W.~Wieland, \emph{{New boundary variables for classical and quantum gravity on
  a null surface}},
  \href{http://dx.doi.org/10.1088/1361-6382/aa8d06}{\emph{Class. Quant. Grav.}
  {\bfseries 34} (2017) 215008},
  [\href{https://arxiv.org/abs/1704.07391}{{\ttfamily 1704.07391}}].

\bibitem{Wieland:2017cmf}
W.~Wieland, \emph{{Fock representation of gravitational boundary modes and the
  discreteness of the area spectrum}},
  \href{http://dx.doi.org/10.1007/s00023-017-0598-6}{\emph{Annales Henri
  Poincare} {\bfseries 18} (2017) 3695--3717},
  [\href{https://arxiv.org/abs/1706.00479}{{\ttfamily 1706.00479}}].

\bibitem{Bodendorfer_2013}
N.~Bodendorfer, T.~Thiemann and A.~Thurn, \emph{{New variables for classical
  and quantum gravity in all dimensions: II. Lagrangian analysis}},
  \href{http://dx.doi.org/10.1088/0264-9381/30/4/045002}{\emph{Class. Quant.
  Grav.} {\bfseries 30} (Jan, 2013) 045002}.

\bibitem{Bodendorfer:2013jba}
N.~Bodendorfer, T.~Thiemann and A.~Thurn, \emph{{New Variables for Classical
  and Quantum Gravity in all Dimensions V. Isolated Horizon Boundary Degrees of
  Freedom}},
  \href{http://dx.doi.org/10.1088/0264-9381/31/5/055002}{\emph{Class. Quant.
  Grav.} {\bfseries 31} (2014) 055002},
  [\href{https://arxiv.org/abs/1304.2679}{{\ttfamily 1304.2679}}].

\bibitem{Bodendorfer:2013sja}
N.~Bodendorfer, \emph{{Black hole entropy from loop quantum gravity in higher
  dimensions}},
  \href{http://dx.doi.org/10.1016/j.physletb.2013.09.043}{\emph{Phys. Lett. B}
  {\bfseries 726} (2013) 887--891},
  [\href{https://arxiv.org/abs/1307.5029}{{\ttfamily 1307.5029}}].

\bibitem{Freidel:2010aq}
L.~Freidel and S.~Speziale, \emph{{Twisted geometries: A geometric
  parametrisation of SU(2) phase space}},
  \href{http://dx.doi.org/10.1103/PhysRevD.82.084040}{\emph{Phys. Rev. D}
  {\bfseries 82} (2010) 084040},
  [\href{https://arxiv.org/abs/1001.2748}{{\ttfamily 1001.2748}}].

\bibitem{Freidel:2010bw}
L.~Freidel and S.~Speziale, \emph{{From twistors to twisted geometries}},
  \href{http://dx.doi.org/10.1103/PhysRevD.82.084041}{\emph{Phys. Rev. D}
  {\bfseries 82} (2010) 084041},
  [\href{https://arxiv.org/abs/1006.0199}{{\ttfamily 1006.0199}}].

\bibitem{Rovelli:1994ge}
C.~Rovelli and L.~Smolin, \emph{{Discreteness of area and volume in quantum
  gravity}}, \href{http://dx.doi.org/10.1016/0550-3213(95)00150-Q,
  10.1016/0550-3213(95)00550-5}{\emph{Nucl. Phys.} {\bfseries B442} (1995)
  593--622}, [\href{https://arxiv.org/abs/gr-qc/9411005}{{\ttfamily
  gr-qc/9411005}}].

\bibitem{Ashtekar:1996eg}
A.~Ashtekar and J.~Lewandowski, \emph{{Quantum theory of geometry. 1: Area
  operators}},
  \href{http://dx.doi.org/10.1088/0264-9381/14/1A/006}{\emph{Class. Quant.
  Grav.} {\bfseries 14} (1997) A55--A82},
  [\href{https://arxiv.org/abs/gr-qc/9602046}{{\ttfamily gr-qc/9602046}}].

\bibitem{Freidel:2015gpa}
L.~Freidel and A.~Perez, \emph{{Quantum gravity at the corner}},
  \href{http://dx.doi.org/10.3390/universe4100107}{\emph{Universe} {\bfseries
  4} (2018) 107}, [\href{https://arxiv.org/abs/1507.02573}{{\ttfamily
  1507.02573}}].

\bibitem{Freidel:2016bxd}
L.~Freidel, A.~Perez and D.~Pranzetti, \emph{{Loop gravity string}},
  \href{http://dx.doi.org/10.1103/PhysRevD.95.106002}{\emph{Phys. Rev.}
  {\bfseries D95} (2017) 106002},
  [\href{https://arxiv.org/abs/1611.03668}{{\ttfamily 1611.03668}}].

\bibitem{Freidel:2019ofr}
L.~Freidel, E.~R. Livine and D.~Pranzetti, \emph{{Kinematical Gravitational
  Charge Algebra}},
  \href{http://dx.doi.org/10.1103/PhysRevD.101.024012}{\emph{Phys. Rev.}
  {\bfseries D101} (2020) 024012},
  [\href{https://arxiv.org/abs/1910.05642}{{\ttfamily 1910.05642}}].

\bibitem{Hasiewicz:1990xc}
Z.~Hasiewicz, J.~Kowalski-Glikman, J.~Lukierski and J.~van Holten,
  \emph{{{BRST} Formulation of the Gupta-bleuler Quantization Method}},
  \href{http://dx.doi.org/10.1063/1.529161}{\emph{J. Math. Phys.} {\bfseries
  32} (1991) 2358--2364}.

\bibitem{Kalau:1991jp}
W.~Kalau, \emph{{On Gupta-Bleuler quantization of systems with second class
  constraints}}, \href{http://dx.doi.org/10.1142/S0217751X93000163}{\emph{Int.
  J. Mod. Phys. A} {\bfseries 8} (1993) 391--406}.

\bibitem{Plebanski:1977zz}
J.~F. Plebanski, \emph{{On the separation of Einsteinian substructures}},
  \href{http://dx.doi.org/10.1063/1.523215}{\emph{J. Math. Phys.} {\bfseries
  18} (1977) 2511--2520}.

\bibitem{Capovilla:1991kx}
R.~Capovilla, T.~Jacobson and J.~Dell, \emph{{A Pure spin connection
  formulation of gravity}},
  \href{http://dx.doi.org/10.1088/0264-9381/8/1/010}{\emph{Class. Quant. Grav.}
  {\bfseries 8} (1991) 59--73}.

\bibitem{Capovilla:1991qb}
R.~Capovilla, T.~Jacobson, J.~Dell and L.~Mason, \emph{{Selfdual two forms and
  gravity}}, {\emph{Class. Quant. Grav.} {\bfseries 8} (1991) 41--57}.

\bibitem{Obukhov:1996rb}
Y.~Obukhov and S.~Tertychny, \emph{{Vacuum Einstein equations in terms of
  curvature forms}},
  \href{http://dx.doi.org/10.1088/0264-9381/13/6/025}{\emph{Class. Quant.
  Grav.} {\bfseries 13} (1996) 1623--1640},
  [\href{https://arxiv.org/abs/gr-qc/9603040}{{\ttfamily gr-qc/9603040}}].

\bibitem{Capovilla:2001zi}
R.~Capovilla, M.~Montesinos, V.~Prieto and E.~Rojas, \emph{{BF gravity and the
  Immirzi parameter}},
  \href{http://dx.doi.org/10.1088/0264-9381/18/5/101}{\emph{Class. Quant.
  Grav.} {\bfseries 18} (2001) L49--L52},
  [\href{https://arxiv.org/abs/gr-qc/0102073}{{\ttfamily gr-qc/0102073}}].

\bibitem{Reisenberger:1996pu}
M.~P. Reisenberger and C.~Rovelli, \emph{{*Sum over surfaces* form of loop
  quantum gravity}},
  \href{http://dx.doi.org/10.1103/PhysRevD.56.3490}{\emph{Phys. Rev.}
  {\bfseries D56} (1997) 3490--3508},
  [\href{https://arxiv.org/abs/gr-qc/9612035}{{\ttfamily gr-qc/9612035}}].

\bibitem{Reisenberger:1997sk}
M.~P. Reisenberger, \emph{{A Lattice world sheet sum for 4-d Euclidean general
  relativity}},  [\href{https://arxiv.org/abs/gr-qc/9711052}{{\ttfamily
  gr-qc/9711052}}].

\bibitem{Barrett:1997gw}
J.~W. Barrett and L.~Crane, \emph{{Relativistic spin networks and quantum
  gravity}}, \href{http://dx.doi.org/10.1063/1.532254}{\emph{J. Math. Phys.}
  {\bfseries 39} (1998) 3296--3302},
  [\href{https://arxiv.org/abs/gr-qc/9709028}{{\ttfamily gr-qc/9709028}}].

\bibitem{Baez:1997zt}
J.~C. Baez, \emph{{Spin foam models}},
  \href{http://dx.doi.org/10.1088/0264-9381/15/7/004}{\emph{Class. Quant.
  Grav.} {\bfseries 15} (1998) 1827--1858},
  [\href{https://arxiv.org/abs/gr-qc/9709052}{{\ttfamily gr-qc/9709052}}].

\bibitem{Markopoulou:1997wi}
F.~Markopoulou and L.~Smolin, \emph{{Causal evolution of spin networks}},
  \href{http://dx.doi.org/10.1016/S0550-3213(97)00488-4}{\emph{Nucl. Phys. B}
  {\bfseries 508} (1997) 409--430},
  [\href{https://arxiv.org/abs/gr-qc/9702025}{{\ttfamily gr-qc/9702025}}].

\bibitem{Freidel:1998pt}
L.~Freidel and K.~Krasnov, \emph{{Spin foam models and the classical action
  principle}}, {\emph{Adv. Theor. Math. Phys.} {\bfseries 2} (1999)
  1183--1247}, [\href{https://arxiv.org/abs/hep-th/9807092}{{\ttfamily
  hep-th/9807092}}].

\bibitem{Barrett:1999qw}
J.~W. Barrett and L.~Crane, \emph{{A Lorentzian signature model for quantum
  general relativity}},
  \href{http://dx.doi.org/10.1088/0264-9381/17/16/302}{\emph{Class. Quant.
  Grav.} {\bfseries 17} (2000) 3101--3118},
  [\href{https://arxiv.org/abs/gr-qc/9904025}{{\ttfamily gr-qc/9904025}}].

\bibitem{Livine:2001jt}
R.~E. Livine and D.~Oriti, \emph{{Barrett-Crane spin foam model from
  generalized BF type action for gravity}},
  \href{http://dx.doi.org/10.1103/PhysRevD.65.044025}{\emph{Phys. Rev.}
  {\bfseries D65} (2002) 044025},
  [\href{https://arxiv.org/abs/gr-qc/0104043}{{\ttfamily gr-qc/0104043}}].

\bibitem{Reisenberger:1996ib}
M.~P. Reisenberger, \emph{{A Left-handed simplicial action for Euclidean
  general relativity}},
  \href{http://dx.doi.org/10.1088/0264-9381/14/7/012}{\emph{Class. Quant.
  Grav.} {\bfseries 14} (1997) 1753--1770},
  [\href{https://arxiv.org/abs/gr-qc/9609002}{{\ttfamily gr-qc/9609002}}].

\bibitem{DePietri:1998hnx}
R.~De~Pietri and L.~Freidel, \emph{{so(4) Plebanski action and relativistic
  spin foam model}},
  \href{http://dx.doi.org/10.1088/0264-9381/16/7/303}{\emph{Class. Quant.
  Grav.} {\bfseries 16} (1999) 2187--2196},
  [\href{https://arxiv.org/abs/gr-qc/9804071}{{\ttfamily gr-qc/9804071}}].

\bibitem{Freidel:1999rr}
L.~Freidel, K.~Krasnov and R.~Puzio, \emph{{BF description of
  higher-dimensional gravity theories}}, {\emph{Adv. Theor. Math. Phys.}
  {\bfseries 3} (1999) 1289--1324},
  [\href{https://arxiv.org/abs/hep-th/9901069}{{\ttfamily hep-th/9901069}}].

\bibitem{Buffenoir:2004vx}
E.~Buffenoir, M.~Henneaux, K.~Noui and P.~Roche, \emph{{Hamiltonian analysis of
  Plebanski theory}},
  \href{http://dx.doi.org/10.1088/0264-9381/21/22/012}{\emph{Class. Quant.
  Grav.} {\bfseries 21} (2004) 5203--5220},
  [\href{https://arxiv.org/abs/gr-qc/0404041}{{\ttfamily gr-qc/0404041}}].

\bibitem{Alexandrov:2008fs}
S.~Alexandrov and K.~Krasnov, \emph{{Hamiltonian Analysis of non-chiral
  Plebanski Theory and its Generalizations}},
  \href{http://dx.doi.org/10.1088/0264-9381/26/5/055005}{\emph{Class. Quant.
  Grav.} {\bfseries 26} (2009) 055005},
  [\href{https://arxiv.org/abs/0809.4763}{{\ttfamily 0809.4763}}].

\bibitem{Alexandrov:2006wt}
S.~Alexandrov, E.~Buffenoir and P.~Roche, \emph{{Plebanski theory and covariant
  canonical formulation}},
  \href{http://dx.doi.org/10.1088/0264-9381/24/11/003}{\emph{Class. Quant.
  Grav.} {\bfseries 24} (2007) 2809--2824},
  [\href{https://arxiv.org/abs/gr-qc/0612071}{{\ttfamily gr-qc/0612071}}].

\bibitem{Alexandrov:2010pg}
S.~Alexandrov, \emph{{The new vertices and canonical quantization}},
  \href{http://dx.doi.org/10.1103/PhysRevD.82.024024}{\emph{Phys. Rev.}
  {\bfseries D82} (2010) 024024},
  [\href{https://arxiv.org/abs/1004.2260}{{\ttfamily 1004.2260}}].

\bibitem{Alexandrov:2010un}
S.~Alexandrov and P.~Roche, \emph{{Critical Overview of Loops and Foams}},
  \href{http://dx.doi.org/10.1016/j.physrep.2011.05.002}{\emph{Phys. Rept.}
  {\bfseries 506} (2011) 41--86},
  [\href{https://arxiv.org/abs/1009.4475}{{\ttfamily 1009.4475}}].

\bibitem{Anza:2014tea}
F.~Anz{\`a} and S.~Speziale, \emph{{A note on the secondary simplicity
  constraints in loop quantum gravity}},
  \href{http://dx.doi.org/10.1088/0264-9381/32/19/195015}{\emph{Class. Quant.
  Grav.} {\bfseries 32} (2015) 195015},
  [\href{https://arxiv.org/abs/1409.0836}{{\ttfamily 1409.0836}}].

\bibitem{Alexandrov:2001wt}
S.~Alexandrov, \emph{{Choice of connection in loop quantum gravity}},
  \href{http://dx.doi.org/10.1103/PhysRevD.65.024011}{\emph{Phys. Rev.}
  {\bfseries D65} (2002) 024011},
  [\href{https://arxiv.org/abs/gr-qc/0107071}{{\ttfamily gr-qc/0107071}}].

\bibitem{Alexandrov:2002xc}
S.~Alexandrov, \emph{{Hilbert space structure of covariant loop quantum
  gravity}}, \href{http://dx.doi.org/10.1103/PhysRevD.66.024028}{\emph{Phys.
  Rev.} {\bfseries D66} (2002) 024028},
  [\href{https://arxiv.org/abs/gr-qc/0201087}{{\ttfamily gr-qc/0201087}}].

\bibitem{Engle:2007uq}
J.~Engle, R.~Pereira and C.~Rovelli, \emph{{The loop-quantum-gravity
  vertex-amplitude}},
  \href{http://dx.doi.org/10.1103/PhysRevLett.99.161301}{\emph{Phys. Rev.
  Lett.} {\bfseries 99} (2007) 161301},
  [\href{https://arxiv.org/abs/0705.2388}{{\ttfamily 0705.2388}}].

\bibitem{Alesci:2007tx}
E.~Alesci and C.~Rovelli, \emph{{The complete LQG propagator: I. Difficulties
  with the Barrett-Crane vertex}},
  \href{http://dx.doi.org/10.1103/PhysRevD.76.104012}{\emph{Phys. Rev.}
  {\bfseries D76} (2007) 104012},
  [\href{https://arxiv.org/abs/0708.0883}{{\ttfamily 0708.0883}}].

\bibitem{Conrady:2008mk}
F.~Conrady and L.~Freidel, \emph{{On the semiclassical limit of 4d spin foam
  models}}, \href{http://dx.doi.org/10.1103/PhysRevD.78.104023}{\emph{Phys.
  Rev.} {\bfseries D78} (2008) 104023},
  [\href{https://arxiv.org/abs/0809.2280}{{\ttfamily 0809.2280}}].

\bibitem{Barrett:2009gg}
J.~W. Barrett, R.~J. Dowdall, W.~J. Fairbairn, H.~Gomes and F.~Hellmann,
  \emph{{Asymptotic analysis of the EPRL four-simplex amplitude}},
  \href{http://dx.doi.org/10.1063/1.3244218}{\emph{J. Math. Phys.} {\bfseries
  50} (2009) 112504}, [\href{https://arxiv.org/abs/0902.1170}{{\ttfamily
  0902.1170}}].

\bibitem{Rovelli:2010wq}
C.~Rovelli, \emph{{A new look at loop quantum gravity}},
  [\href{https://arxiv.org/abs/1004.1780}{{\ttfamily 1004.1780}}].

\bibitem{Livine:2007ya}
E.~R. Livine and S.~Speziale, \emph{{Consistently Solving the Simplicity
  Constraints for Spinfoam Quantum Gravity}},
  \href{http://dx.doi.org/10.1209/0295-5075/81/50004}{\emph{EPL} {\bfseries 81}
  (2008) 50004}, [\href{https://arxiv.org/abs/0708.1915}{{\ttfamily
  0708.1915}}].

\bibitem{Engle:2007mu}
J.~Engle and R.~Pereira, \emph{{Coherent states, constraint classes, and area
  operators in the new spin-foam models}},
  \href{http://dx.doi.org/10.1088/0264-9381/25/10/105010}{\emph{Class. Quant.
  Grav.} {\bfseries 25} (2008) 105010},
  [\href{https://arxiv.org/abs/0710.5017}{{\ttfamily 0710.5017}}].

\bibitem{Pereira:2007nh}
R.~Pereira, \emph{{Lorentzian LQG vertex amplitude}},
  \href{http://dx.doi.org/10.1088/0264-9381/25/8/085013}{\emph{Class. Quant.
  Grav.} {\bfseries 25} (2008) 085013},
  [\href{https://arxiv.org/abs/0710.5043}{{\ttfamily 0710.5043}}].

\bibitem{Rovelli:2010ed}
C.~Rovelli and S.~Speziale, \emph{{Lorentz covariance of loop quantum
  gravity}},
  \href{http://dx.doi.org/10.1103/PhysRevD.83.104029}{\emph{Phys.Rev.}
  {\bfseries D83} (2011) 104029},
  [\href{https://arxiv.org/abs/1012.1739}{{\ttfamily 1012.1739}}].

\bibitem{Ding:2010ye}
Y.~Ding and C.~Rovelli, \emph{{Physical boundary Hilbert space and volume
  operator in the Lorentzian new spin-foam theory}},
  \href{http://dx.doi.org/10.1088/0264-9381/27/20/205003}{\emph{Class. Quant.
  Grav.} {\bfseries 27} (2010) 205003},
  [\href{https://arxiv.org/abs/1006.1294}{{\ttfamily 1006.1294}}].

\bibitem{Livine:2007vk}
E.~R. Livine and S.~Speziale, \emph{{A new spinfoam vertex for quantum
  gravity}}, \href{http://dx.doi.org/10.1103/PhysRevD.76.084028}{\emph{Phys.
  Rev.} {\bfseries D76} (2007) 084028},
  [\href{https://arxiv.org/abs/0705.0674}{{\ttfamily 0705.0674}}].

\bibitem{Dupuis:2010iq}
M.~Dupuis and E.~R. Livine, \emph{{Revisiting the Simplicity Constraints and
  Coherent Intertwiners}},  [\href{https://arxiv.org/abs/1006.5666}{{\ttfamily
  1006.5666}}].

\bibitem{Dupuis:2011wy}
M.~Dupuis, L.~Freidel, E.~R. Livine and S.~Speziale, \emph{{Holomorphic
  Lorentzian Simplicity Constraints}},
  \href{http://dx.doi.org/10.1063/1.3692327}{\emph{J. Math. Phys.} {\bfseries
  53} (2012) 032502}, [\href{https://arxiv.org/abs/1107.5274}{{\ttfamily
  1107.5274}}].

\bibitem{Wieland:2011ru}
W.~M. Wieland, \emph{{Twistorial phase space for complex Ashtekar variables}},
  \href{http://dx.doi.org/10.1088/0264-9381/29/4/045007}{\emph{Class. Quant.
  Grav.} {\bfseries 29} (2012) 045007},
  [\href{https://arxiv.org/abs/1107.5002}{{\ttfamily 1107.5002}}].

\bibitem{Speziale_2012}
S.~Speziale and W.~M. Wieland, \emph{Twistorial structure of loop-gravity
  transition amplitudes},
  \href{http://dx.doi.org/10.1103/physrevd.86.124023}{\emph{Phys. Rev.}
  {\bfseries D86} (Dec, 2012) }.

\bibitem{Dittrich:2008ar}
B.~Dittrich and J.~P. Ryan, \emph{Phase space descriptions for simplicial 4d
  geometries},
  \href{http://dx.doi.org/10.1088/0264-9381/28/6/065006}{\emph{Class. Quant.
  Grav.} {\bfseries 28} (Feb, 2011) 065006}.

\bibitem{Dittrich:2010ey}
B.~Dittrich and J.~P. Ryan, \emph{Simplicity in simplicial phase space},
  \href{http://dx.doi.org/10.1103/physrevd.82.064026}{\emph{Phys. Rev.}
  {\bfseries D82} (Sep, 2010) }.

\bibitem{Freidel:2019ees}
L.~Freidel, E.~R. Livine and D.~Pranzetti, \emph{{Gravitational edge modes:
  from Kac--Moody charges to Poincar{\'e} networks}},
  \href{http://dx.doi.org/10.1088/1361-6382/ab40fe}{\emph{Class. Quant. Grav.}
  {\bfseries 36} (2019) 195014},
  [\href{https://arxiv.org/abs/1906.07876}{{\ttfamily 1906.07876}}].

\bibitem{Edge-Mode-IV}
L.~Freidel, M.~Geiller and D.~Pranzetti, \emph{{Edge modes of gravity - IV:
  Corner Hilbert space}},  [\href{https://arxiv.org/abs/2009.xxxxx}{{\ttfamily
  2009.xxxxx}}].

\bibitem{Wigner:1939cj}
E.~P. Wigner, \emph{{On Unitary Representations of the Inhomogeneous Lorentz
  Group}}, \href{http://dx.doi.org/10.2307/1968551}{\emph{Annals Math.}
  {\bfseries 40} (1939) 149--204}.

\bibitem{Weinberg:1995mt}
S.~Weinberg, \emph{{The Quantum theory of fields. Vol. 1: Foundations}}.
\newblock Cambridge University Press, 2005.

\bibitem{Bekaert:2005in}
X.~Bekaert and J.~Mourad, \emph{{The Continuous spin limit of higher spin field
  equations}},
  \href{http://dx.doi.org/10.1088/1126-6708/2006/01/115}{\emph{JHEP} {\bfseries
  01} (2006) 115}, [\href{https://arxiv.org/abs/hep-th/0509092}{{\ttfamily
  hep-th/0509092}}].

\bibitem{Schuster:2013pxj}
P.~Schuster and N.~Toro, \emph{{On the Theory of Continuous-Spin Particles:
  Wavefunctions and Soft-Factor Scattering Amplitudes}},
  \href{http://dx.doi.org/10.1007/JHEP09(2013)104}{\emph{JHEP} {\bfseries 09}
  (2013) 104}, [\href{https://arxiv.org/abs/1302.1198}{{\ttfamily 1302.1198}}].

\bibitem{shirokov1}
M.~I. Shirokov, \emph{A group theoretical considertion of the basis of
  relativistic quantum mechanics i:. the general properties of the
  inhomogeneous lorentz group}, {\emph{Soviet Physics-JETP 6} {\bfseries
  6(33),} ((1958)) 665, 673}.

\bibitem{shirokov2}
M.~I. Shirokov, \emph{A group theoretical considertion of the basis of
  relativistic quantum mechanics ii:. classification of the irreducible
  representations of the inhomogeneous lorentz group}, {\emph{Soviet
  Physics-JETP 6} {\bfseries 6(33)} ((1958)) 919, 929}.

\bibitem{Pirotte}
C.~Pirotte, \emph{{M\'ethode de Shirokov et alg\`ebres de spin du groupe de
  Poincar\'e}}, {\emph{Physica} {\bfseries 63} (1973) 373--383}.

\bibitem{Fleming}
G.~N. Fleming, \emph{Covariant position operators, spin, and locality},
  {\emph{Physical Review B} {\bfseries 137} (1965) 188--197}.

\bibitem{Newton:1949cq}
T.~Newton and E.~P. Wigner, \emph{{Localized States for Elementary Systems}},
  \href{http://dx.doi.org/10.1103/RevModPhys.21.400}{\emph{Rev. Mod. Phys.}
  {\bfseries 21} (1949) 400--406}.

\bibitem{Zwiebach:2004tj}
B.~Zwiebach, \emph{{A first course in string theory}}.
\newblock Cambridge University Press, 2006.

\bibitem{Rovelli:2002vp}
C.~Rovelli and S.~Speziale, \emph{{Reconcile Planck scale discreteness and the
  Lorentz-Fitzgerald contraction}},
  \href{http://dx.doi.org/10.1103/PhysRevD.67.064019}{\emph{Phys. Rev. D}
  {\bfseries 67} (2003) 064019},
  [\href{https://arxiv.org/abs/gr-qc/0205108}{{\ttfamily gr-qc/0205108}}].

\bibitem{Godazgar:2020kqd}
H.~Godazgar, M.~Godazgar and M.~J. Perry, \emph{{Hamiltonian derivation of dual
  gravitational charges}},  [\href{https://arxiv.org/abs/2007.07144}{{\ttfamily
  2007.07144}}].

\bibitem{Freidel:2018pvm}
L.~Freidel and E.~R. Livine, \emph{{Bubble networks: framed discrete geometry
  for quantum gravity}},
  \href{http://dx.doi.org/10.1007/s10714-018-2493-y}{\emph{Gen. Rel. Grav.}
  {\bfseries 51} (2019) 9}, [\href{https://arxiv.org/abs/1810.09364}{{\ttfamily
  1810.09364}}].

\bibitem{gelfand_1982}
{I. M. Gel'fand, M. I. Graev, I. N. Bernstein, V. A. Ponomarev, S. I. Gel'fand,
  A. M. Vershik}, \emph{Representation Theory: Selected Papers}.
\newblock London Mathematical Society Lecture Note Series. Cambridge University
  Press, 1982,
  \href{http://dx.doi.org/10.1017/CBO9780511629310}{10.1017/CBO9780511629310}.

\bibitem{Haggard:2012pm}
H.~M. Haggard, C.~Rovelli, W.~Wieland and F.~Vidotto, \emph{{Spin connection of
  twisted geometry}},
  \href{http://dx.doi.org/10.1103/PhysRevD.87.024038}{\emph{Phys. Rev. D}
  {\bfseries 87} (2013) 024038},
  [\href{https://arxiv.org/abs/1211.2166}{{\ttfamily 1211.2166}}].

\bibitem{Livine_2012}
E.~R. Livine, S.~Speziale and J.~Tambornino, \emph{Twistor networks and
  covariant twisted geometries},
  \href{http://dx.doi.org/10.1103/physrevd.85.064002}{\emph{Physical Review D}
  {\bfseries 85} (Mar, 2012) }.

\bibitem{Geiller:2017whh}
M.~Geiller, \emph{{Lorentz-diffeomorphism edge modes in 3d gravity}},
  \href{http://dx.doi.org/10.1007/JHEP02(2018)029}{\emph{JHEP} {\bfseries 02}
  (2018) 029}, [\href{https://arxiv.org/abs/1712.05269}{{\ttfamily
  1712.05269}}].

\bibitem{Freidel:2018pbr}
L.~Freidel, F.~Girelli and B.~Shoshany, \emph{{2+1D Loop Quantum Gravity on the
  Edge}}, \href{http://dx.doi.org/10.1103/PhysRevD.99.046003}{\emph{Phys. Rev.
  D} {\bfseries 99} (2019) 046003},
  [\href{https://arxiv.org/abs/1811.04360}{{\ttfamily 1811.04360}}].

\bibitem{Kirillov}
A.~A. Kirillov, \emph{Lectures on the Orbit Method}, vol.~64 of \emph{Graduate
  Studies in Mathematics}.
\newblock American Mathematical Society, 2004.

\bibitem{DiazPolo:2011np}
J.~Diaz-Polo and D.~Pranzetti, \emph{{Isolated Horizons and Black Hole Entropy
  In Loop Quantum Gravity}},
  \href{http://dx.doi.org/10.3842/SIGMA.2012.048}{\emph{SIGMA} {\bfseries 8}
  (2012) 048}, [\href{https://arxiv.org/abs/1112.0291}{{\ttfamily 1112.0291}}].

\bibitem{Krasnov:1996tb}
K.~V. Krasnov, \emph{{Counting surface states in the loop quantum gravity}},
  \href{http://dx.doi.org/10.1103/PhysRevD.55.3505}{\emph{Phys. Rev. D}
  {\bfseries 55} (1997) 3505--3513},
  [\href{https://arxiv.org/abs/gr-qc/9603025}{{\ttfamily gr-qc/9603025}}].

\bibitem{Rovelli:1996dv}
C.~Rovelli, \emph{{Black hole entropy from loop quantum gravity}},
  \href{http://dx.doi.org/10.1103/PhysRevLett.77.3288}{\emph{Phys. Rev. Lett.}
  {\bfseries 77} (1996) 3288--3291},
  [\href{https://arxiv.org/abs/gr-qc/9603063}{{\ttfamily gr-qc/9603063}}].

\bibitem{Ashtekar:2000eq}
A.~Ashtekar, J.~C. Baez and K.~Krasnov, \emph{{Quantum geometry of isolated
  horizons and black hole entropy}}, {\emph{Adv. Theor. Math. Phys.} {\bfseries
  4} (2000) 1--94}, [\href{https://arxiv.org/abs/gr-qc/0005126}{{\ttfamily
  gr-qc/0005126}}].

\bibitem{Ghosh:2004wq}
A.~Ghosh and P.~Mitra, \emph{{An Improved lower bound on black hole entropy in
  the quantum geometry approach}},
  \href{http://dx.doi.org/10.1016/j.physletb.2005.05.003}{\emph{Phys. Lett. B}
  {\bfseries 616} (2005) 114--117},
  [\href{https://arxiv.org/abs/gr-qc/0411035}{{\ttfamily gr-qc/0411035}}].

\bibitem{Ghosh:2006ph}
A.~Ghosh and P.~Mitra, \emph{{Counting black hole microscopic states in loop
  quantum gravity}},
  \href{http://dx.doi.org/10.1103/PhysRevD.74.064026}{\emph{Phys. Rev. D}
  {\bfseries 74} (2006) 064026},
  [\href{https://arxiv.org/abs/hep-th/0605125}{{\ttfamily hep-th/0605125}}].

\bibitem{Ghosh:2011fc}
A.~Ghosh and A.~Perez, \emph{{Black hole entropy and isolated horizons
  thermodynamics}},
  \href{http://dx.doi.org/10.1103/PhysRevLett.107.241301}{\emph{Phys. Rev.
  Lett.} {\bfseries 107} (2011) 241301},
  [\href{https://arxiv.org/abs/1107.1320}{{\ttfamily 1107.1320}}].

\bibitem{Sahlmann:2011xu}
H.~Sahlmann, \emph{{Black hole horizons from within loop quantum gravity}},
  \href{http://dx.doi.org/10.1103/PhysRevD.84.044049}{\emph{Phys. Rev. D}
  {\bfseries 84} (2011) 044049},
  [\href{https://arxiv.org/abs/1104.4691}{{\ttfamily 1104.4691}}].

\bibitem{Pithis:2014uva}
A.~G. Pithis and H.-C. Ruiz~Euler, \emph{{Anyonic statistics and large horizon
  diffeomorphisms for Loop Quantum Gravity Black Holes}},
  \href{http://dx.doi.org/10.1103/PhysRevD.91.064053}{\emph{Phys. Rev. D}
  {\bfseries 91} (2015) 064053},
  [\href{https://arxiv.org/abs/1402.2274}{{\ttfamily 1402.2274}}].

\bibitem{Ghosh:2013iwa}
A.~Ghosh, K.~Noui and A.~Perez, \emph{{Statistics, holography, and black hole
  entropy in loop quantum gravity}},
  \href{http://dx.doi.org/10.1103/PhysRevD.89.084069}{\emph{Phys. Rev. D}
  {\bfseries 89} (2014) 084069},
  [\href{https://arxiv.org/abs/1309.4563}{{\ttfamily 1309.4563}}].

\end{thebibliography}\endgroup
